\documentclass[12pt]{article}

\usepackage[dvips]{epsfig}
\usepackage{amsmath,amssymb,amsfonts,enumerate,bbm}
\usepackage{amsthm}
\usepackage{caption}
\usepackage{url}
\usepackage{fullpage}
\usepackage{ifthen}
\usepackage{graphicx}
\usepackage{subfigure}
\usepackage{wrapfig}
\usepackage{color,soul}

\long\def\commabs #1\commabsend{}
\long\def\commful #1\commfulend{}
\long\def\comment #1\commentend{}

\newtheorem{theorem}{Theorem}[section]
\newtheorem{lemma}[theorem]{Lemma}
\newtheorem{observation}[theorem]{Observation}
\newtheorem{corollary}[theorem]{Corollary}

\newtheorem{claim}[theorem]{Claim}
\newtheorem{fact}[theorem]{Fact}
\newtheorem{definition}[theorem]{Definition}
\def\deg{\mbox{\tt deg}}

\def\LCA{\mbox{\tt LCA}}

\def\Cost{\mbox{\tt Cost}}

\newcommand{\dist}{\mbox{\rm dist}}

\def\inline#1:{\par\vskip 7pt\noindent{\bf #1:}\hskip 10pt}
\def\Proof{\par\noindent{\bf Proof:~}}
\def\blackslug{\hbox{\hskip 1pt \vrule width 4pt height 8pt
    depth 1.5pt \hskip 1pt}}
\def\QED{\quad\blackslug\lower 8.5pt\null\par}
\def\LastE{\mbox{\tt LastE}}

\def\NSource{\mathcal{K}}

\def\FTMBFS{\mbox{\tt FT-MBFS}}

\def\FTBFS{\mbox{\tt FT-BFS}}

\def\dnsitem{\vspace{-4pt}\item}
\def\dnsparagraph#1{\par\vspace{2pt}\noindent{\em #1}}

\def\Cost{\mbox{\tt Cost}}

\def\UncoverPairs{\mathcal{UP}}
\def\Add{\mbox{\tt Add}}
\def\keps{K_{\epsilon}}
\begin{document}

%\begin{titlepage}

%\def\thepage{}

%%%%%%%%%%%%%%%%%%%%%%%%%%%%%%%%%%%%%%%%%%%%%%%%%%%%%%%%%%%%%%%%%%%%%%%%
\title{Fault Tolerant BFS Structures: \\ A Reinforcement-Backup Tradeoff}

\author{
Merav Parter
\thanks{Department of Computer Science and Applied Mathematics.
The Weizmann Institute of Science, Rehovot, Israel.
E-mail: {\tt \{merav.parter,david.peleg\}@ weizmann.ac.il}.
Supported in part by the Israel Science Foundation (grant 894/09),
and the I-CORE program of the Israel PBC and ISF (grant 4/11).}
\thanks{Recipient of the Google European Fellowship in distributed computing;
research is supported in part by this Fellowship.}
\and
David Peleg $^*$
}

\maketitle

\begin{abstract}
This paper initiates the study of fault resilient network structures that mix
two orthogonal protection mechanisms:
(a) {\em backup}, namely, augmenting the structure with many (redundant)
low-cost but fault-prone components, and
(b) {\em reinforcement}, namely, acquiring high-cost but fault-resistant
components.
To study the trade-off between these two mechanisms in a concrete setting,
we address the problem of designing
a $(b,r)$ {\em fault-tolerant} BFS (or $(b,r)$ \FTBFS\ for short) structure,
%an $\epsilon$ {\em fault-tolerant} BFS (or $\epsilon$ \FTBFS\ for short) structure,
namely, a subgraph $H$ of the network $G$ consisting of two types of edges:
a set $E' \subseteq E$ of $r(n)$ fault-resistant {\em reinforcement} edges,
which are assumed to never fail,
%where $|E'|=O(n^{1-\epsilon})$,
and a (larger) set $E(H) \setminus E'$ of $b(n)$ fault-prone {\em backup} edges,
such that subsequent to the failure of a single fault-prone backup edge
$e \in E \setminus E'$, the surviving part
%$H'$
of $H$ still contains an BFS spanning tree for (the surviving part of) $G$,
satisfying
$\dist(s,v,H\setminus \{e\}) \leq \dist(s,v,G\setminus \{e\})$
for every $v \in V$ and $e \in E \setminus E'$.
We establish the following tradeoff between $b(n)$ and $r(n)$:
For every real $\epsilon \in [0,1]$, if $r(n) = {\tilde\Theta}(n^{1-\epsilon})$,
then $b(n) = {\tilde\Theta}(n^{1+\epsilon})$ is necessary and sufficient.
More specifically, it was shown in \cite{PPFTBFS13} that for $\epsilon=1$,
\FTBFS\ structures (with no reinforced edges) require
$\Theta(n^{3/2})$ edges, and this number of edges is sufficient.
At the other extreme, if $\epsilon=0$, then $n-1$ reinforced edges
are sufficient with no need for backup.
Here, we present a polynomial time algorithm that given an undirected graph
$G=(V,E)$, a source vertex $s$ and a real $\epsilon \in [0,1]$,
constructs a $(b(n),r(n))$ \FTBFS\ with $r(n) = O(n^{1-\epsilon})$ and
$b(n) = O(\min\{1/\epsilon \cdot n^{1+\epsilon} \cdot \log n, n^{3/2}\})$.
We complement this result by providing a nearly matching lower bound, showing
that there are $n$-vertex graphs for which any $(b(n),r(n))$ \FTBFS\ structure
requires $\Omega(\min\{n^{1+\epsilon}, n^{3/2}\})$ backup edges when $r(n)=\Omega(n^{1-\epsilon})$ edges are reinforced.
\end{abstract}

\section{Introduction}
\dnsparagraph{Background and Motivation.}
%The concept of survivable networks is among the most frequently recurring one in the problem of designing communication networks. Survivability is defined by the capability of a given system to fulfill its mission in the presence of failures, where the term mission refers to a wide range of requirements or goals.
%For example, the survivability of a communication network implies that the network remains functional (e.g., connected) when links or nodes fail.
%
%In the general setting of \emph{Network Design} problems, the goal is to find a \emph{minimum cost} subgraph satisfying certian connectivity requirements between vertices. This captures a wide variety of classical problems such as \mbox{\tt Minimum Cost Flow}, \mbox{\tt Minimum Stiener Tree}, and many more.
%\textbf{MP: we may start from the next paragraph.}
Modern day communication networks support a variety of logical structures and services, and depend on their undisrupted operation. Following the immense recent advances in telecommunication networks, the explosive growth of the Internet, and our increased dependence on these infrastructures, guaranteeing the survivability of communication networks has become a major objective in both practice and theory. An important aspect of this objective is \emph{survivable network design}, namely, the design of low cost high resilience networks that satisfy certain desirable {\em performance requirements} concerning, e.g., their connectivity, distance or capacity. Our focus here, however, is not on planning survivable networks ``from scratch'', but rather on settings where an initially existing infrastructure needs to be improved and optimized.

Our interest in this paper is in exploring a natural ``quality vs. quantity'' tradeoff in survivable network design. Designers and manufactures often face the following design choice when dealing with ensuring product reliability. One option is to invest heavily in the quality and resilience of the various components of the product, making them essentially failure-free. An alternative option is to use unreliable but cheap components, and ensure the reliability of the whole product by employing redundancy, namely,  including several ``copies" of each component in the design, so that the failure of one component will not disable the operation.
%For example, whereas in the past, the common trend was to design expensive and reliable washing machines, nowadays these machines are often built cheaper but less reliable.

In the context of survivable network design, where the goal is to overcome link disconnections, the ``quantity-based'' approach to survivability relies on adding to the network many inexpensive (but failure-prone) {\em backup} links, counting on redundancy to provide resilience and guarantee the desired performance requirements in the presence of failures. In contrast, a ``quality-based'' approach may rely on {\em reinforcing} some of the network links, and thus making them failure-resistant (but expensive), counting on these links to ensure the performance requirements. Clearly, these two approaches address two different and \emph{orthogonal} factors affecting the survivability of a network: the \emph{topology}, e.g., the presence of redundant alternate paths, and the \emph{reliability} of individual network components. We would like to study the tradeoff betwen these two factors in various survivable network design problems.

Towards exploring this tradeoff, we consider the following ``mixed'' model. Assume that the existing infrastructure consists of a given fixed set $V$ of vertices and a collection $E$ of existing links, and it is required to decide, for each link, among the following three choices: (a) discard the link (in which case it will cost us nothing), (b) purchase it as is (at some low cost $B$), or (c) ``reinforce'' it (at some high cost $R$), making it failure-resilient.
The existing initial graph $G(V,E)$ provides a baseline for comparison, in the sense that if we decide on the conservative approach of making no changes, namely, purchasing all the links of the existing network $G$ ``as is'' (at a cost of $B\cdot |E|$), then the performance properties that can be guaranteed in the presence of failures are those of the existing $G$. An alternative baseline is obtained by the opposite extreme, namely, basing the design on selecting the smallest subgraph $H$ of $G$ that satisfies the desired performance requirements in the absence of failures, and reinforcing all its links, thus ensuring this performance level.

Unfortunately, both of these two extremes might be too costly. Hence, constructing a survivable subnetwork with a \emph{limited} budget introduces a tradeoff between backup and reinforcement and the system designer is faced with a choice:
%A high \emph{quality} link is expensive but may reduce the \emph{total} number of links in the network, as there is no longer need to supply alternative routes upon its failing.
reinforcing just a few of the links may potentially lead to considerable savings, by allowing one to discard many of the ordinary backup edges and still obtain the same performance properties.
\begin{wrapfigure}{r}{0.2\textwidth}
  \begin{center}
    \includegraphics[width=0.20\textwidth]{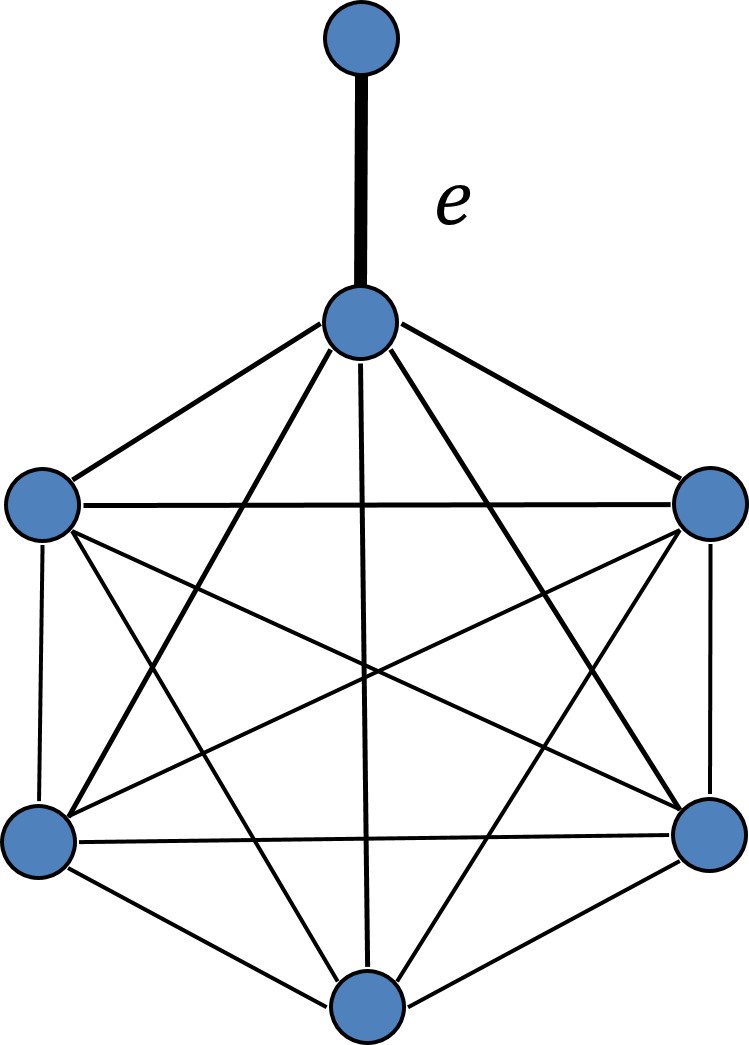}
  \end{center}
\end{wrapfigure}
To illustrate this point, consider for example an $n$-vertex network consisting of a single vertex $s$ connected via a single edge $e$ to an $n-1$-vertex clique (see figure). The edge connectivity of this network is 1, as the removal of $e$ disconnects the graph.
Hence the conservative approach of keeping all existing edges
% , according to the traditional topology-based view,
leaves this network with a low level of survivability.
In contrast, in a mixed model allowing also reinforcements, it is sufficient to reinforce a single edge, namely, $e$,
%(e.g., make this link wired instead of wireless)
in order to obtain a high level of survivability, even by purchasing only a fraction of the edges of the clique.
%Hence, according to the mixed backup-reinforcement view, the network is survivable.

\dnsparagraph{Our Contributions.}
To initiate the study of the tradeoff between reinforcement and backup in survivable network design, we consider in this paper the concrete problem of designing (in the mixed model) a {\em fault-tolerant Breadth-First structure} (or \FTBFS\ for short), namely, a subnetwork that preserves distances with respect to a given source vertex $s$ in the presence of an edge failure. Formally, given a network $G(V,E)$ and a source vertex $s$ in $G$, a $(b(n),r(n))$-\FTBFS\ is a subgraph $H$ of $G$ consisting of two types of edges: a set $E' \subseteq E$ of $r(n)$ fault-resistant {\em reinforcement} edges, which are assumed to never fail, and a (larger) set $E(H) \setminus E'$ of $b(n)$ fault-prone {\em backup} edges, such that subsequent to the failure of a single fault-prone backup edge $e \in E \setminus E'$, the surviving part of $H$ still contains an BFS spanning tree for (the surviving part of) $G$, satisfying $\dist(s,v,H\setminus \{e\}) \leq \dist(s,v,G\setminus \{e\})$ for every $v \in V$ and $e \in E \setminus E'$. We establish the following tradeoff between $b(n)$ and $r(n)$: For every real $\epsilon \in [0,1]$, if $r(n) = {\tilde\Theta}(n^{1-\epsilon})$,
then $b(n) = {\tilde\Theta}(\min\{n^{1+\epsilon}, n^{3/2}\})$ is necessary and sufficient.

It was shown in \cite{PPFTBFS13}, that for $\epsilon=1$, \FTBFS\ structure requires $\Theta(n^{3/2})$ edges. In the other extreme case of $\epsilon=0$, by reinforcing all the edges of the BFS tree, no backup is needed. This result can also be interpreted in the following manner. Let $B$ (resp., $R$) be the cost of a backup (resp., reinforced) edge. The total cost of a $(b,r)$ \FTBFS\ structure is given by $B \cdot b(n)+R \cdot r(n)=\widetilde{O}(n^{1-\epsilon}\cdot B+n^{1+\epsilon}\cdot R)$. Hence, the minimum cost $(b,r)$ \FTBFS\ obtained by taking $\epsilon=\widetilde{O}(\log(R/B)/\log n)$.

We complement the upper bound construction of $(r,b)$ \FTBFS\ structures by presenting a nearly matching lower bound. We show
that there are $n$-vertex graphs for which any $(b(n),r(n))$ \FTBFS\ structure for $r(n)=\Omega(n^{1-\epsilon})$ requires $\Omega(\min\{n^{1+\epsilon}, n^{3/2}\})$ backup edges.
In our lower bound constructions, we also consider a generalized structure referred to as a
$(b,r)$ {\em fault-tolerant multi-source BFS tree}, or {\em \FTMBFS\ tree} for short,
aiming to provide a $(b,r)$ \FTBFS\ structure at each source vertex $s\in S$ for some subset of sources $S\subseteq V$.
We show that a $(b,r)$ \FTMBFS\ structure for $r(n)=\Omega(|S|^{\epsilon}\cdot n^{1-\epsilon})$ requires
$b(n)=\Omega(\min\{\sqrt{|S|}\cdot n^{3/2},|S|^{1-\epsilon} \cdot n^{1+\epsilon}\})$ edges for every $\epsilon \in (0,1]$.

\noindent {\em Techniques and proof outline.}
Studying $(b,r)$ \FTBFS\ structures significantly differs from their standard \FTBFS\ counterparts (for $r(n)=0$) in both the upper and lower bounds.
Let $\pi(s,v)$ be an $s-v$ shortest-path in $G$. The initial structure consists of the BFS tree $T_0=\bigcup_{v \in V}\pi(s,v)$. It is then augmented by adding to it the last edges of some carefully chosen replacement-paths.
For an edge $e \in \pi(s,v)$, a replacement path $P_{v,e}$ is \emph{new-ending} path if its last edge was not present in the structure when the path was selected by the algorithm. A new-ending replacement path $P_{v,e}$ has the following structure. It consists of a prefix of $\pi(s,v)$ followed by a detour $D$ avoiding the failing edge $e$ and joining the $\pi(s,v)$ path at the terminal $v$.
%
%Our algorithm exploits the structure of new-ending replacement paths to construct $(b,r)$-\FTBFS\ structures. Essentially, a key section in our analysis concerns with collecting the last edges from a subset of new-ending replacement paths. The analysis then bounds the size of the structure as well as the number of edges that are not protected in the given structure.
%
An essential component in our analysis deals with the detour segment of the single failure replacement paths.
The analysis of \FTBFS\ structure \cite{PPFTBFS13}
focused on a \emph{single} terminal $v$ and showed that it has $O(\sqrt{n})$ new-ending replacement-paths (with distinct last edges). The current setting of $(b,r)$ \FTBFS\ structures is more involved and requires studying the interactions between detours of \emph{different} vertices.
In particular, the current construction has two simultaneous objectives: minimizing the number of backup edges in the structure as well as selecting at most $r(n)$ reinforced edges. In other words, when constructing a $(b,r)$ \FTBFS\ structure with $o(n^{3/2})$ edges, one has the privilege of discarding the protection against the failure of $r(n)$ edges,
% Consequently, the remaining $r(n)$ unprotected edges
which are reinforced.

The upper bound of \cite{PPFTBFS13} was achieved by analyzing the interactions between the detours of $s-v$ new-ending replacement-paths $P_{v,e_i}$ and $P_{v,e_{j}}$ for some $e_i,e_j \in \pi(s,v)$. It was shown that upon a proper construction of the replacement-paths, these detours are vertex disjoint\footnote{if their last edge is distinct}, except for the common endpoint $v$, and hence these detours are vertex-consuming, which enables bounding their number. In contrast, studying $(b,r)$ structures requires understanding the interaction between detours of \emph{distinct} terminals. These detours may overlap and are not necessarily vertex disjoint, hence bounding their number calls for new techniques.

Our key observation is that the interactions (referred hereafter as \emph{interference}) between detours can be roughly classified into two types depending on the relation between the edges protected by the corresponding detours. Each of these interference types gives raise to unique structural characterization and volume constraints that enable us to bound the cardinality of their corresponding paths. The first type of interference concerns $P_{v,e}$ and $P_{t,e'}$ paths both whose failing edges $e$ and $e'$ occur below the least common ancestor of $v$ and $t$ in the BFS tree $T_0$. We show that adding the last edges of $O(n^{\epsilon})$ such replacement-paths protecting against the failure of the deepest edges of each $s-v$ path is sufficient, i.e., it leaves no unprotected edge (that is protected by a replacement-path of this type) in the structure. We then turn to consider the second type of interference, where at least one of the faulty edges, say $e$, occurs above the least common ancestor of $v$ and $t$. Analyzing the interaction between detours that protect edges on the \emph{same} shortest-path turns out to be more involved. Our technique is based on the \emph{heavy-path-decomposition} procedure of Sleator and Tarjan \cite{DCOMP81} (slightly adapted by Baswana and Khanna \cite{BS10}), applied on $T_0$. This decomposition is obtained by $O(\log n)$ recursive calls on partial trees $T' \subseteq T_0$, where each recursive call results with a collection of paths in $T_0$ whose edges appear in the $s-v$ shortest-paths of a \emph{distinct} set of vertices. The advantage of this approach
%in our setting
is that equipped with our interference classification, the analysis is reduced to solving the subproblem (i.e., designing the $(b,r)$ structure) for the case where the failing events are restricted to a given path $\psi\subseteq T_0$ in the tree-decomposition (a similar approach is taken in \cite{BS10} for a different problem). In other words, when handling the second type of interference, there is an \emph{independence} between the tree-decomposition paths $\psi_i,\psi_j \subseteq T_0$ that were generated at the same level of the recursion.
% Following sentence not too important, hide if short in space
Since there are $O(\log n)$ recursion levels, summing over all levels increases our bounds by a logarithmic factor.
The final structure $H$ is then given by the union of the substructures for each of the paths in the tree-decomposition\footnote{The paths of the tree-decomposition do not cover all the edges in the BFS tree, however, the remaining uncovered edges can be handled directly by adding the last edges of the corresponding replacement-paths.}.
By collecting the last edges of carefully selected replacement-paths protecting the failures on $\psi$, for every path $\psi$ in the tree-decomposition, it is then shown that there are $\widetilde{O}(n^{1-\epsilon})$ unprotected edges in the structure.

Turning to the lower bound, $(b,r)$ \FTBFS\ structures for large $b(n)$ and $r(n)$ values require a more delicate construction when compared to standard \FTBFS\ structures. The design of the lower bound graph is governed by  two opposing forces whose balance is to be found. Specifically, since detours are vertex consuming, to end up with a dense structure with many backup edges, the detours (and as a result also the shortest-paths) of many vertices should \emph{collide}. For instance, in the lower bound construction of \FTBFS\ structures, the $s-v$ shortest-path of $\Theta(n)$ vertices is the \emph{same}. In other words, a large number of backup edges implies packing many shortest-paths and detours efficiently. Since the lower bound construction of \cite{PPFTBFS13} involved only $\Theta(\sqrt{n})$ edges on the $s-v$ shortest-paths, a new approach is needed when trying to maximize the number of reinforced edges in the structure to $O(n^{1-\epsilon})$ for $\epsilon \in (0,1/2)$.
In particular, large reinforcement forces the construction to distribute the vertices on \emph{distinct} shortest-paths so as to increase the number of edges that have large cost and hence should be reinforced. Our construction then finds the fine balance between these forces, matching our upper bounds up to logarithmic factors.
\dnsparagraph{Related Works.}
To the best of our knowledge, this paper is the first to study the backup - reinforcement tradeoff in survivable nework design for $(b,r)$ \FTBFS\ structures.
%DP: I am not sure it is safe to claim that this is the first appearance of the general concept of the mixed model.

The question of designing sparse \FTBFS\ structures (without link reinforcement) has been studied in \cite{PPFTBFS13}, using the notion of \emph{replacement paths}. For a source node $s$, a target node $v$ and an edge $e\in G$, a \emph{replacement path} is the shortest $s-v$ path $P_{v,e}$ that does not go through $e$. An \FTBFS\ structure consists of the collection of all $P_{v,e}$ replacement paths for every target $v \in V$ and edge $e \in E$.
It is shown in \cite{PPFTBFS13} that for every graph $G$ and source node $s$ there exists a (polynomial time constructible) \FTBFS\ structure $H$ with $O(n^{3/2})$ edges. This result was complemented by a matching lower bound
showing that for every sufficiently large integer $n$, there exist
an $n$-vertex graph $G$ and a source node $s \in V$, for which every \FTBFS\ structure is of size $\Omega(n^{3/2})$. Hence the insistence on {\em exact} distances makes \FTBFS\ structures significantly denser (hence expensive) compared their fault-prone counterparts (namely, BFS trees). This last observation motivates the idea of studying the mixed model and makes \FTBFS\ structures an attractive platform for studying the backup-reinforcement tradeoff.
%$(b,r)$ \FTBFS\ structures, in order to allow the design of a sparse subgraph with $b(n)< n^{3/2}$ edges and $r(n)$ reinforced (fault-resistant) edges.
%In fact, the starting point for studying $(r,b)$ \FTBFS\ structure is the lower bound construction of \cite{PPFTBFS13}. In this lower-bound graph, there are $\Omega(n^{3/2})$ backup edges that are required to be added to any \FTBFS\ structure to protect the failing of $\Theta(\sqrt{n})$ edges out of the $n-1$ edges in the original BFS tree. This raises the question of whether   reinforcement of some carefully chosen $\Theta(n^{1/2-\epsilon})$ BFS tree edges results in a much denser structure for some $\epsilon \in (0,1/2]$. In this paper, we answer this question in the affirmative and present a construction for $(b(n),r(n))$ \FTBFS\ structure with $r(n) = \widetilde{O}(n^{1-\epsilon})$ and
%$b(n) = \widetilde{O}(\min\{1/\epsilon \cdot n^{1+\epsilon} \cdot \log n, n^{3/2}\})$.

The notion of {\em \FTBFS}\ trees is also closely related to the \emph{single-source replacement paths} problem, studied in \cite{GW12}. That problem requires to compute the collection $\mathcal{P}(s)$ of all $s-t$ replacement paths $P_{t,e}$ for every $t \in V$ and every failed edge $e$ that appears on the $s-t$ shortest-path in $G$. The vast literature on \emph{replacement paths} (cf. \cite{BK09,GW12,RTREP05,TZ05,WY10}) focuses on \emph{time-efficient} computation of the these paths as well as on their efficient maintenance in data structures (a.k.a {\em distance oracles}).

Constructions of sparse fault tolerant \emph{spanners} for $\mathbb{R}^d$ Euclidean space were studied in \cite{CZ03,LNS98,L99}. Algorithms for constructing sparse edge and vertex fault tolerant spanners for arbitrary undirected weighted graphs were presented in \cite{CLPR09-span,DK11}.
%
% More detailed history of F-T spanners:
%The question of whether it is possible to construct a sparse fault tolerant \emph{spanner} for an arbitrary undirected weighted graph, raised in \cite{CZ03}, was answered in the affirmative in \cite{CLPR09-span}, presenting algorithms for constructing an $f$-vertex fault tolerant $(2k-1)$-spanner of size $O(f^2 k^{f+1} \cdot n^{1+1/k}\log^{1-1/k}n)$ and an $f$-edge fault tolerant $2k-1$ spanner of size $O(f\cdot n^{1+1/k})$ for a graph of size $n$. A randomized construction attaining an improved tradeoff for vertex fault-tolerant spanners was shortly afterwards presented in \cite{DK11}, yielding (with high probability) for every graph $G = (V,E)$, odd integer $s$ and integer $f$, an $f$-vertex fault-tolerant $s$-spanner with $O\left(f^{2-\frac{2}{s+1}}n^{1+\frac{2}{s+1}}\log{n}\right)$ edges. This should be contrasted with the best stretch-size tradeoff currently known for non-fault-tolerant spanners \cite{TZ01}, namely, $2k-1$ stretch with ${\tilde O}(n^{1+1/k})$ edges. Fault tolerant spanners for the $d$-dimensional Euclidean case were studied in \cite{CZ03,LNS98,L99}.
Note, however, that the use of costly link reinforcements for attaining fault-tolerance in spanners is less attractive than for \FTBFS\ structures, since the cost of adding fault-tolerance via backup edges (in the relevant complexity measure) is often low (e.g., merely polylogarithmic in the graph size $n$), hence the gains expected from using reinforcement are relatively small.
%In contrast, in the case of  \FTBFS\ structures, the insistence on exact distances makes them significantly denser compared their fault-prone counterparts (i.e., BFS trees). Hence, \FTBFS\ structure provides a good platform for studying the backup-reinforcement tradeoff.
%, with the hope that reinforcement of bounded number of edges gives raise to sparser structures.

Constructions of edge fault-tolerant spanners with \emph{additive} stretch
%resilient to edge failures
are given in \cite{FTAdd12}, and the case of single vertex fault has been recently studied in \cite{PDISC14}.
%
%Additional previous work on sparse / compact fault-tolerant structures and services concerned structures that are {\em distance-preserving} (i.e., dealing with distances, shortest paths or shortest routes), {\em global} (i.e., centered on ``all-pairs'' variants), and {\em approximate} (i.e., settling for near optimal distances) \cite{Bern10}, such as {\em spanners}, {\em distance oracles} and {\em compact routing schemes} \cite{ABLP-89:stoc}.
\dnsparagraph{Discussion.}
The presented mixed model implicitly reflects the notion of \emph{economy of scale}, stating that the cost per unit decreases with the amount of purchased units. Indeed, economy of scale has been incorporated explicitly in many network design tasks by using a \emph{sublinear} cost function. These scenarios are known in the literature as \emph{buy-at-bulk} problems \cite{BAUYBULKAA,SALBBUY97,BBUYCHE10} in which capacity is sold with a ``volume discount": the more capacity is bought, the cheaper is the price per unit of bandwidth.

Economy of scale arises implicitly in $(b,r)$ \FTBFS\ structures. The main objective in constructing  $(b,r)$ \FTBFS\ is to minimize the number of backup edges $b(n)$ subject to a bound limit on the number of reinforced edges $r(n)=O(n^{1-\epsilon})$. We say that a vertex $v \in V$ \emph{uses} an edge if the edge is on the $s-v$ shortest path in $G$ (for the sake of discussion, assume that all shortest paths in $G$ are unique). The \emph{cost} of an edge $e$, $\Cost(e)$, is the number of backup edges required to be added to the structure upon its failing (to provide alternative shortest-paths in the surviving network). Since reinforcement is expensive, it is beneficial to reinforce an edge that has many users (i.e., appears on many $s-v$ shortest paths). The intuition behind this observation is that an edge $e$ with many users may cause a considerable damage upon failing, hence its cost $\Cost(e)$ should be large. Whereas in the traditional backup mechanism $\Cost(e)$ scales with the number of users, in the reinforcement mechanism the reinforcement cost of each edge is fixed and independent of the number of users. Therefore, for certain $B/R$ ratios, the cost of reinforcement becomes sublinear with the number of users.
A similar intuition arises in the \emph{single sink rent-or-buy} problem \cite{RBMS}.
%The input for this problem is an undirected graph $G=(V,E)$ with an edge cost $c_e\geq 0$ for every $e \in E$, a source vertex $s$, a set of terminals $R \subseteq V$, and a parameter $M\geq 1$. The output is a subgraph $H'$ connecting all terminals to the root. To build these paths, one can buy an edge $e$ at cost $M \cdot c_e$, and once bought, any terminal can use this edge. Alternately, one can rent an edge $e$ at cost $c_e$, but then it is necessary to pay the rental cost for each terminal that uses this edge. The goal is to find a feasible subgraph of minimum cost. Renting (respectively, buying) an edge is the informal analogue to reinforcement (resp., backup), in the sense that buying / backup costs are proportional to the number of users and renting / reinforcement costs are fixed and hence may provide a sublinear cost for a large number of users (hence illustrating the economy of scale).

The problem considered here, namely, the construction of $(r,b)$ \FTBFS\ structure, deviates from the traditional rent-or-buy and buy-at-bulk network design tasks in the sense that it concerns a fault resilient distance preserving (and not merely connectivity preserving) structure. In addition, whereas most network design tasks are stated as combinatorial optimization problems and their solution employs polyhedral combinatorics, the goal of the current paper is to establish a \emph{universal} bound on the tradeoff between backup and reinforcement.

This paper aims at establishing universal lower and upper bounds for $(b,r)$ \FTBFS\ structures. In particular, although the universal upper bound is nearly tight (upto logarithmic factors), our upper bound constructions might be far from optimal in some instances (see the example of Fig. 5 in \cite{PPFTBFSArxiv13}).
This motivates the study of $(b,r)$ \FTBFS\ structures from the combinatorial optimization point of view.
Specifically, two natural optimization problems can be defined within this context. The first (resp., second) formulation aims at optimizing the number of backup edges $b(n)$ (resp., reinforced edges $r(n)$) subject to a given bound on the number of reinforced edges (resp., backup edges). For example, when $r(n)=\widetilde{O}(n^{1-\epsilon})$, we know that for every input graph $G$ and source vertex $s \in V$, one can construct a structure with $b(n)=\widetilde{O}(n^{1+\epsilon})$ backup edges, and there are graphs for which $b(n)=\Omega(n^{1+\epsilon})$ backup edges are essential. Yet, there are graphs for which $b(n)=O(n)$ backup edges are sufficient and using the current construction is too wasteful.

Aside from optimization tasks for $(r,b)$ \FTBFS\ structures,
the presented reinforcement-backup tradeoff can be studied in a more generalized setting. In fact, it can be integrated into a large collection of survivability network design tasks.
We hope that this work will pave the way for studying this setting, leading to new theoretical tools and techniques as well as to a better understanding of fault resilient structures.

\section{Preliminaries and Notation}

Let $E(v,G)=\{(u,v) \in E(G) ~\mid~ u \in V\}$ be the set of edges incident to the vertex $v$ in the graph $G$ and let $\deg(v,G)=|E(v,G)|$ denote the degree of $v$
in $G$. When the graph $G$ is clear from the context,
we may simply write $\deg(v)$ and $E(v)$.
For a subgraph $G'=(V', E') \subseteq G$
(where $V' \subseteq V$ and $E' \subseteq E$)
and a pair of vertices $u,v \in V$, let $\dist(u,v, G')$ denote the
shortest-path distance in edges between $u$ and $v$ in $G'$.
For a path $P=[u_1, \ldots, u_k]$, let $\LastE(P)$ denote the last edge of $P$,
let $|P|$ denote the length of $P$ in edges, i.e., $k-1$, and let $P[u_i, u_j]$ be the subpath of $P$
from $u_i$ to $u_j$. For paths $P_1$ and $P_2$
where the last vertex of $P_1$ equals the first vertex of $P_2$,
let $P_1 \circ P_2$ denote
the path obtained by concatenating $P_2$ to $P_1$.
Throughout, the edges of these paths are considered to be directed
away from the source $s$. Given an $s-t$ path $P$ and an edge
$e=(u,v) \in P$, let $\dist(s, e, P)$ be the distance (in edges) between $s$
and $e$ on $P$.
%In addition,
For an edge $e=(u,v)\in T_0$, define
$\dist(s,e)=i$ if $\dist(s,u,G)=i-1$ and $\dist(s,v,G)=i$.
For a subset $V' \subseteq V$, let $G(V')$ be the subgraph of $G$ induced by $V'$.
Let $\LCA(u,v)$ be the least common ancestor of $u$ and $v$ in $T_0$.

For vertices $u, v \in V$ and subgraph $G' \subseteq G$, let $SP(u, v, G')$ be the collection of all $s-v$ shortest-path in $G'$, i.e, $|P|=\dist(s,v, G')$ for every $P \in SP(s, v, G')$.
For a positive weight assignment $W: E(G) \to \mathbb{R}_{>0}$, let $SP(s, v, G',W)$ be the collection of $s-v$ shortest-paths in $G'$ according to the weights of $W$.
In this paper, the weight assignment $W$ is chosen as to
guarantee the uniqueness the shortest-paths in every $G' \subseteq G$. That is $W$ is used to break to shortest-path ties in $G'$ in a consistent manner. In such a case, we override notation and let $SP(s, v, G',W) \in SP(s,v,G')$ be the unique $s-v$ shortest-path in $G'$ with the weights of $W$. Given a source vertex $s$ and target vertex $v$, let $\pi(s,v)=SP(s, v, G,W)$ be the unique $s-v$ path in $G$ according to $W$. Define $T_0(s)=\bigcup_{v \in V}\pi(s,v)$ as the  BFS tree rooted at $s$. When the source $s$ is clear from the context, we simply write $T_0$.

For a vertex $v$ and an edge $e$, each path in $SP(s, v, G \setminus \{e\})$ is referred to as a \emph{replacement-path}.
%
%an $s-v$ path $P_{v,e}$ is a \emph{replacement-path} if $P_{v,e}$ is in $SP(s, v, G \setminus \{e\})$.
%
Note that if $e \notin \pi(s,v)$, then $\pi(s,v)$ is a replacement path as it appears in $SP(s, v, G \setminus \{e\})$.
A vertex $w$ is a \emph{divergence point} of the $s-v$ paths  $P_1$ and $P_2$ if $w \in P_1 \cap P_2$ but the next vertex $u$ after $w$ (i.e., such that $u$ is closer to $v$) in the path $P_1$ is not in $P_2$.

\dnsparagraph{$\epsilon$ \FTBFS\ and protected edges.}
For a subgraph $H \subseteq G$ and and a source vertex $s$, an edge $e$ is \emph{protected} in $H$ if $\dist(s,v,H \setminus \{e\})=\dist(s,v,G \setminus \{e\})$ for every $v \in V$ and otherwise it is \emph{unprotected}. In other words, the edge $e$ is protected if for every vertex $v$, $H$ contains at least one replacement path $P_{v,e} \in SP(s,v,G \setminus\{e\})$.
\begin{definition}[$\epsilon$ \FTBFS]
\label{def:epsbfstree}
For every real $\epsilon \in [0,1]$, a subgraph $H \subseteq G$ is an $\epsilon$ \FTBFS\ with respect to $s$, if it contains $O(n^{1-\epsilon})$ unprotected edges. That is, there exists a subset of $O(n^{1-\epsilon})$ edges $E'$ such that
$\dist(s,v, H\setminus \{e\})=\dist(s,v, G\setminus \{e\})$ for every $v \in V$ and $e \in E(G)\setminus E'$.
Alternately, an $\epsilon$ \FTBFS\ $H$ can be thought of as a $(b,r)$ \FTBFS\ taking $E'$ to be the set of $r(n)=O(n^{1-\epsilon})$ reinforcement edges and $E(H)\setminus E'==O(b(n))$ to be the backup edges.  
\end{definition}
Note that in the context of the reinforcement-backup model, unprotected edges are viewed as edges that should be reinforced
in the structure, since by definition, in the $(b,r)$ \FTBFS\ structure, all backup edge are \emph{protected}, and unprotected edges are not allowed to exist.

We now define a more refined notion of protected edges that is determined by the existence of the \emph{last edges} of the replacement-paths in the subgraph $H$ (instead of requiring the existence of the entire replacement path in $H$).
Given a subgraph $H \subseteq G$, we say that the edge $e$ is $v$-\emph{last-unprotected} in $H$ if there exists no replacement path $P_{v,e} \in SP(s,v,G \setminus \{e\})$ whose last edge $\LastE(P_{v,e})$ is in $H$, otherwise the edge is $v$-\emph{last-protected}.
An edge $e$ is \emph{last-unprotected} in $H$, if there exists at least one vertex $v \in V$ for which $e$ is $v$-last-unprotected, otherwise it is \emph{last-protected}.
Note that the notion of protected edge refers to the case where every vertex $v$ has at least one $s-v$ replacement-path protecting against the failing of $e$ in $H$. In contrast, the notion of last-protected edges refers to the existence of the last edge of these replacement-path (and not the entire path) in the subgraph $H$.
The next observation relates the properties of ``last-protected" and ``protected". %Due to lack of space, missing proofs are deferred to Appendix \ref{app:missproof}.
\begin{observation}
\label{obs:lastedge}
If $e$ is last-protected in $H$, then $e$ is protected, i.e., $\dist(s,v, H \setminus \{e\})=\dist(s,v, G \setminus \{e\})$,
$\forall$
%for every
$v \in V$.
\end{observation}
\Proof
Let $e$ be a last-protected edge in $H$. Assume towards contradiction that the claim does not hold and let
$$BV=\{v \mid \dist(s,v, H \setminus \{e\}) >  \dist(s,v,G \setminus \{e\})\}$$
be the set of ``bad vertices," namely, vertices for which the $s-v$ shortest path distance in $H \setminus \{e\}$ is greater than that in $G\setminus \{e\}$.
(By the contradictory assumption, it holds that $BV\ne \emptyset$.)
For every vertex $v$, let $P^*_{v,e} \in SP(s,v,G \setminus \{e\})$ be a replacement-path satisfying that $\LastE(P^*_{v,e}) \in H$, (since $e$ is $v$-last-protected for every $v$, this path exists).
Define
$BE(v)=P^*_{v,e} \setminus E(H)$ to be the set of ``bad edges,''
namely, the set of $P^*_{v,e}$ edges that are missing in $H$.
By definition, $BE(v) \neq \emptyset$ for every bad vertex $v \in BV$.
Let $\widetilde{d}(v)=\max_{e \in BE(v)}\{\dist(s,e,P^*_{v,e})\}$ be the maximal depth
of a missing edge in $BE(v)$, and let $DM(v)$ denote that ``deepest
missing edge'' for $v$, i.e., the edge $e$ on $P^*_{v,e}$ satisfying
$\dist(s,e,P^*_{v,e})=\widetilde{d}(v)$.
Finally, let $v' \in BV$ be the vertex that minimizes $\widetilde{d}(v)$,
and let $e_1=(x,y) \in BE(v')$ be
the deepest missing edge on $P^*_{v',e}$, namely, $e_1=DM(v')$.
Note that $e_1$ is the {\em shallowest} ``deepest missing edge''
over all bad vertices $v \in BV$.
Let $P_1=P^{*}_{s, y, e}$, $P_2=P^{*}_{s, y, e}[s,y]$ and
$P_3=P^{*}_{s, v', e}[y, v']$.
Note that since $v' \in BV$, it follows that also $y \in BV$.
(Otherwise, if $y \notin BV$, then any $s-y$ shortest-path
$P' \in SP(s, y, H \setminus \{e\})$, where $|P'|=|P^*_{y,e}|$, can be appended to $P_3$ resulting in
$P''=P' \circ P_3$
such that (1) $P'' \subseteq H\setminus \{e\}$ and (2)
$|P''|=|P'|+|P_3|=|P_2|+|P_3|=|P^{*}_{s,v',e}|$, contradicting the fact that
$v' \in BV$.)
We conclude that $y \in BV$.
Finally, note that $\LastE(P_1) \in H$ by definition, and therefore
the deepest missing edge of $y$ must be shallower, i.e.,
$\widetilde{d}(y)<\widetilde{d}(v')$. However, this is in contradiction to our choice
of the vertex $v'$. The lemma follows.
\QED
%}%\APPENDLASTEDGE
\section{Algorithm}
\label{sec:uni_upb}
In this section, we describe a construction of an $\epsilon$ \FTBFS\ subgraph $H$ containing $O(1/\epsilon' \cdot \log n \cdot n^{1+\epsilon'})$ edges where
$\epsilon'=\epsilon+\log (\log n/\epsilon)/\log n$ for every $\epsilon \in (0,1]$.
In the next section we prove the following.
\begin{theorem}
\label{thm:square}
For every input $n$-vertex graph $G$, source vertex $s$ and real $\epsilon \in (0,1]$, there exists an $\epsilon$ \FTBFS\ $H \subseteq G$ with $O(\min\{1/\epsilon' \cdot \log n \cdot n^{1+\epsilon'},n^{3/2}\})$ edges. Hence, in particular, one can construct a $(b,r)$ \FTBFS\ structure with $b(n) = O(\min\{1/\epsilon \cdot n^{1+\epsilon} \cdot \log n, n^{3/2}\})$ and $r(n) = O(1/\epsilon \cdot n^{1-\epsilon}\cdot \log n)$.
\end{theorem}
By \cite{PPFTBFS13}, there exists a polynomial time algorithm for constructing \FTBFS\ structures with $O(n^{3/2})$ edges, hence the claim holds trivially for $\epsilon \geq 1/2$, and to establish the theorem, it remains to consider the case where $\epsilon \in (0,1/2)$. To deal with this case, we next describe an explicit construction for an $\epsilon$ \FTBFS\ $H$ which is then analyzed in the following subsection.

\subsection{Phase (S0): Preprocessing}
\dnsparagraph{Algorithm \mbox{\tt Pcons} for constructing the replacement-paths.}
The goal of the preprocessing phase (S0) is to define a function $\mathcal{RP}:(V \times E) \to E$ that maps each vertex-edge pair $\langle v,e \rangle$ to a replacement path $P_{v,e} \subseteq E$. These paths will be used in the main construction. Let $T_0$ be a BFS tree rooted at $s$ in $G$.
Algorithm \mbox{\tt Pcons}  iterates over every vertex $v \in V$ and every edge $e \in \pi(s,v)$. For a given pair $\langle v,e \rangle$, the algorithm first tests if there exists an $s-v$ replacement path whose last edge is already in $T_0$.
Let $G'(v)=(G \setminus E(v,G))\cup E(v,T_0)$.
Now, if $\dist(s,v, G'(v)\setminus \{e\})=\dist(s,v,G \setminus \{e\})$, then let $\mathcal{RP}(\langle v,e \rangle)=P_{v,e}=SP(s,v, G'(v)\setminus \{e\},W)$.
Else, (i.e., the replacement-path $P_{v,e}$ must include a new last edge that is not in $T_0$), the algorithm attempts to select the $s-v$ replacement-path whose divergence point from $\pi(s,v)$ is as close to $s$ as possible.
Specifically, let $\pi(s,v)=[u_0=s, u_1, \ldots, u_k=v]$ and $e=(u_{i},u_{i+1})$. For every $j \in \{0, \ldots, i\}$, define $G_j(v)=G \setminus V(\pi(u_j,u_k))\cup \{u_j,u_k\}$. Note that $e \notin G_j(v)$. Define $j^*$ as the minimal index $j$ satisfying that $\dist(s,v, G_j(v))=\dist(s,v, G \setminus \{e\})$ and let $P_{v,e}=SP(s,v, G_{j*}(v)\setminus \{e\},W)$.
%Throughout, we refer to the replacement-paths collection as multi-sets. That is, even if $P_{v,e_i}=P_{v,e_j}$ for $e_i\neq e_j$, these paths are considered as ``different" paths. Spefifically, the paths $P_{v,e}$ are identified by their unique key $\langle v,e\rangle$.

\dnsparagraph{Replacement-path classification.}
A replacement-path $P=P_{v,e}$ is \emph{new-ending} if its last edge is not in $T_0$. A vertex-edge pair $\langle v,e \rangle$ is \emph{uncovered} if its replacement path $P_{v,e}$ is new-ending.
\begin{wrapfigure}{r}{0.2\textwidth}
  \begin{center}
    \includegraphics[width=0.50\textwidth]{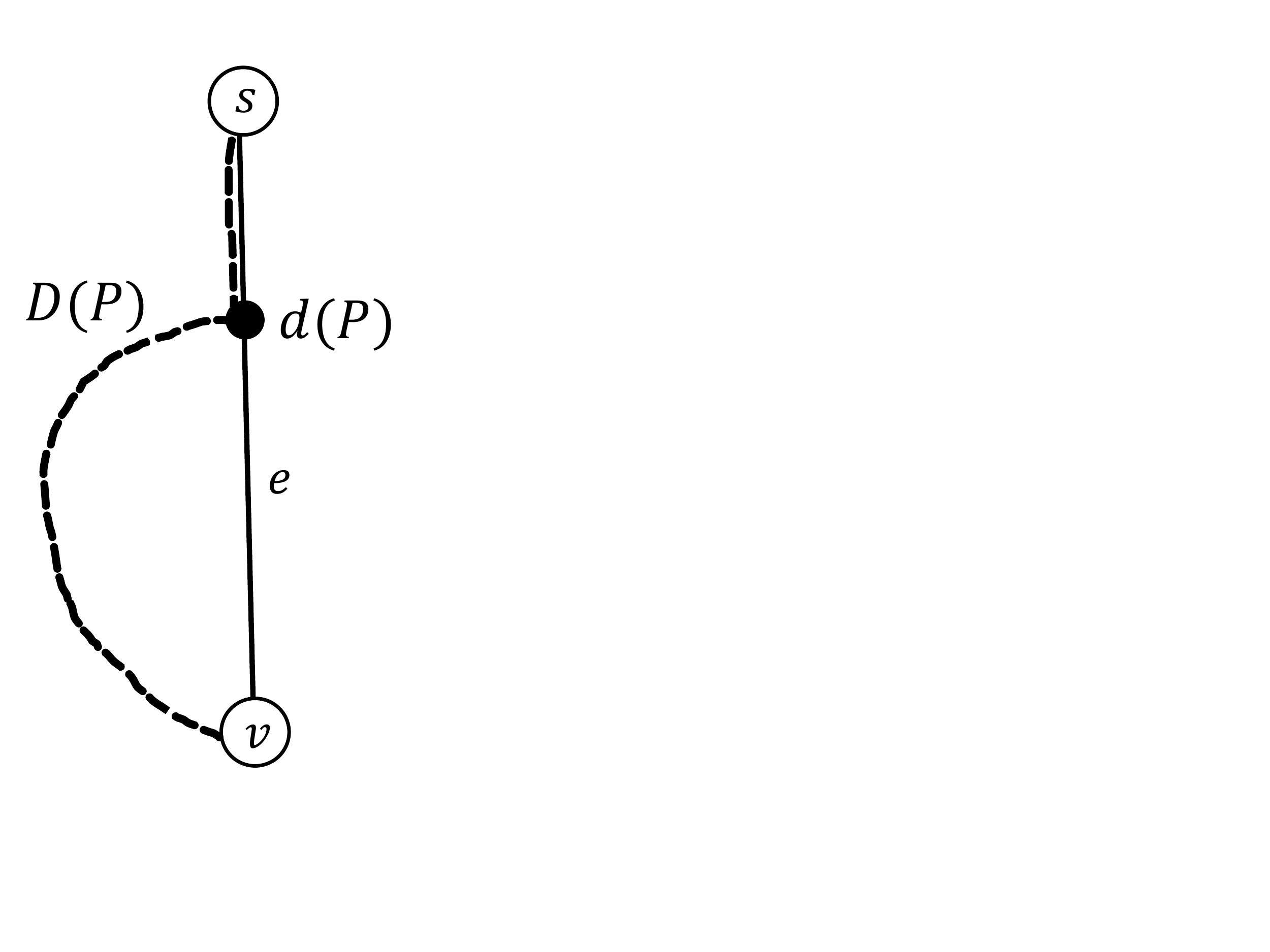}
  \end{center}
\end{wrapfigure}
The following observation follows immediately by the construction of Alg. \mbox{\tt Pcons}.
\begin{observation}
\label{obs:detour_exist}
Consider a new-ending path $P=P_{v,e}$. Let $d(P)$ be the first divergence point of $P$ and $\pi(s,v)$. Then $P$ can be decomposed into $P=\pi(s,d(P)) \circ D(P)$ where $D(P)=P \setminus E(\pi(s,v))$, referred to as the \emph{detour segment}, departs from $\pi(s,v)$ at $d(P)$ and returns only at $v$, i.e.,
$D(P)=P[d(P),v]$ and $\pi(s,v)$ are vertex disjoint besides the common endpoints $d(P)$ and $v$.
\end{observation}

Let $\UncoverPairs=\{\langle v,e \rangle ~\mid~ \LastE(P_{v,e})\notin T_0\}$ be the collection of all uncovered vertex-edge pairs.
Let $\UncoverPairs(v)=\{\langle w,e \rangle \in \UncoverPairs ~\mid~ w=v\}$ be the uncovered pairs of $v$ (hence, $\UncoverPairs=\bigcup_{v \in V}\UncoverPairs(v)$).
%
%Note it might be the case where $\LastE(P_i)=\LastE(P_j)$ for some two distinct paths $P_i,P_j \in \UncoverPairs$ and also that $P_{v,e_i}=P_{v,e_j}$ for $e_i,e_j \in \pi(s,v)$.
%
%\subsection{Analysis}
%For a tree $T'$ and a leaf vertex $v'$, denote by the
%\emph{suffix-path} of $v$ the maximal suffix of the $\pi(s,v)$ paths whose all vertices have degree at most $2$.
%
%
%
%For every edge $e$, let $\mathcal{P}(e)=\{P_{t,e} ~\mid~ t \in V, P_{t,e} \in \UncoverPairs\}$ be the collection of new-ending paths protecting against the failing of $e$, let $E(e)=\bigcup_{P \in \mathcal{P}(e)} \LastE(P)$ and finally, let $\Val(e)=|E(e)|$ be the number of new edges needed to protect the failing of the edge $e$.
\par
\begin{wrapfigure}{r}{0.2\textwidth}
  \begin{center}
    \includegraphics[width=0.5\textwidth]{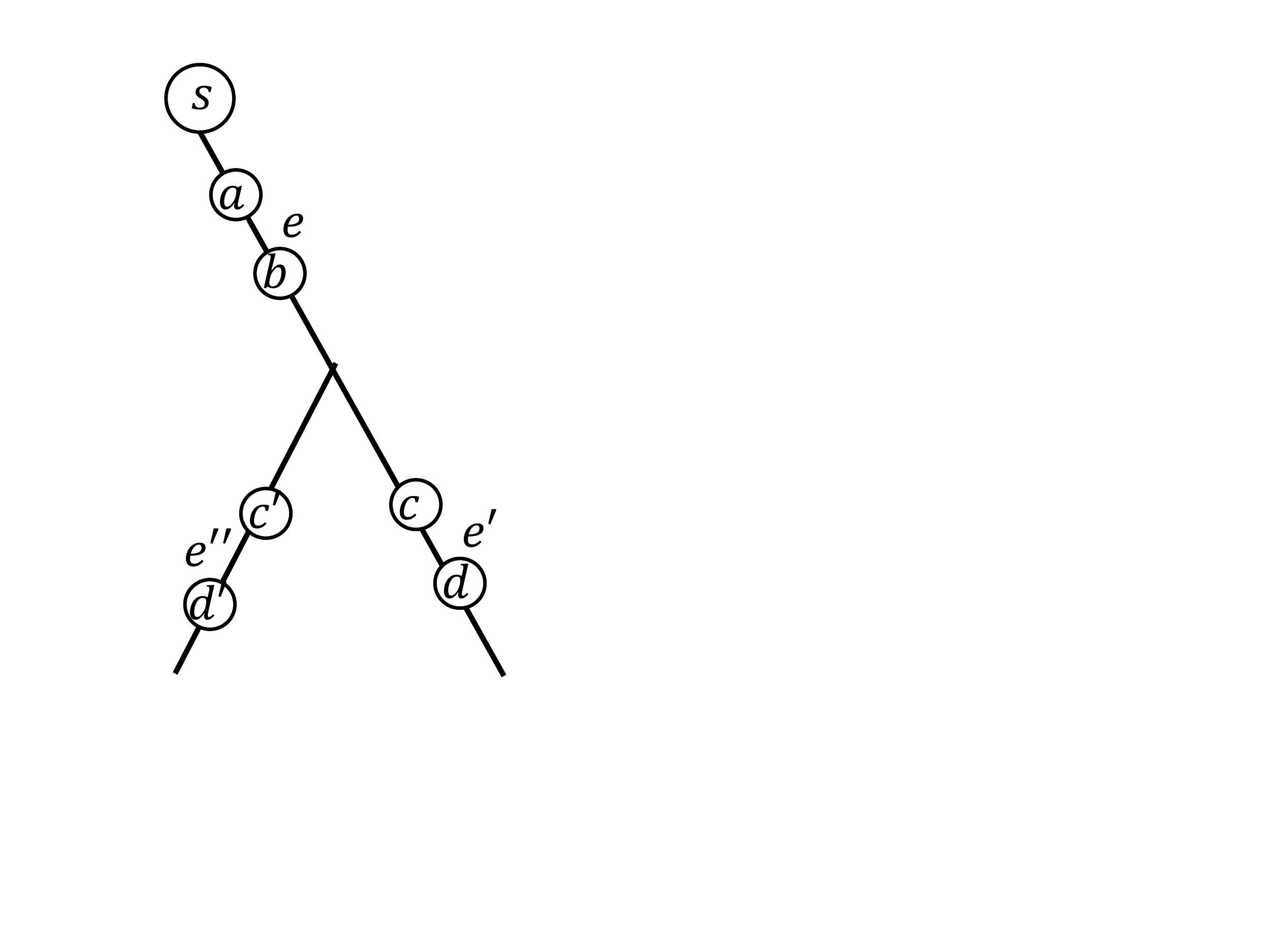}
    %\caption{ Relations between tree edges: $e \sim e'$ and $e' \not\sim e''$.}
  \end{center}
\end{wrapfigure}
Throughout, we consider the edges of $T_0$ to be directed away from $s$, hence referring to the edge $e=(x,y) \in T_0$ implies that $\dist(s,x,G)<\dist(s, y,G)$.

The following definitions are key to in our construction and the subsequent analysis. For two tree edges $e=(a,b),e'=(c,d) \in T_0$, we say that $e\sim e'$ if $\LCA(b,d) \in \{b,d\}$, i.e., $e,e' \in \pi(s,v)$ for $v=\{b,d\} \setminus \{\LCA(b,d)\}$, otherwise  $e\not\sim e'$. (In the figure, $e \sim e'$ and $e' \not\sim e''$.)
%For a new-ending path $P=P_{v,e}$ such that $\langle v,e\rangle \in \UncoverPairs$, let $D(P)$ be the detour segment of $P$ (by Obs. \ref{obs:detour_exist} such exists).
In our construction, we may impose an ordering on a subset of $v$'s uncovered pairs.
For a given subset of $v'$s pairs $\Add(\mathcal{P},v)=\{\langle v,e_1\rangle, \ldots, \langle v,e_k\rangle\} \subseteq \UncoverPairs(v)$, let $\overrightarrow{\mathcal{P}}(v)$ be ordered in increasing distance of $e_i$ from $v$, i.e., $\overrightarrow{\mathcal{P}}(v)=\{
\langle v,e_{i_1}\rangle, \ldots, \langle v,e_{i_k}\rangle\}$
where $\dist(v, e_{i_1},\pi(s,v))<\ldots <\dist(v, e_{i_k},\pi(s,v))$.

\dnsparagraph{$(\sim)$-interference and  $(\not\sim)$-interference.}
The paths $P=P_{v,e}, P'=P_{t,e'}$ for  $\langle v,e \rangle,\langle t,e' \rangle \in \UncoverPairs$  and $v \neq t$ \emph{interfere} with each other if their detours intersect at some vertex $z$ internal to both, i.e.,
%that is, if
\begin{equation}
\label{eq:interfere}
V(D(P))\cap V(D(P')) \nsubseteq \{d(P),d(P'),v,t\}~.
\end{equation}
Note that according to this definition, interference is \emph{symmetric}, i.e., if $P$ interferes with $P'$ then $P'$ interferes with $P$ as well.
For every uncovered pair $\langle v,e \rangle \in \UncoverPairs$, denote the set of pairs $\langle t,e' \rangle$ whose corresponding path $P'=P_{t,e'}$ interferes with $P=P_{v,e}$ by
$$\mathcal{I}(\langle v,e \rangle)=\{\langle t,e' \rangle \in \UncoverPairs ~\mid~t \neq v, P_{t,e'} \mbox{~and~} P_{v,e} \mbox{~satisfy~ Eq. (\ref{eq:interfere})}\}.$$

\begin{figure}[h!]
\begin{center}
\includegraphics[scale=0.31]{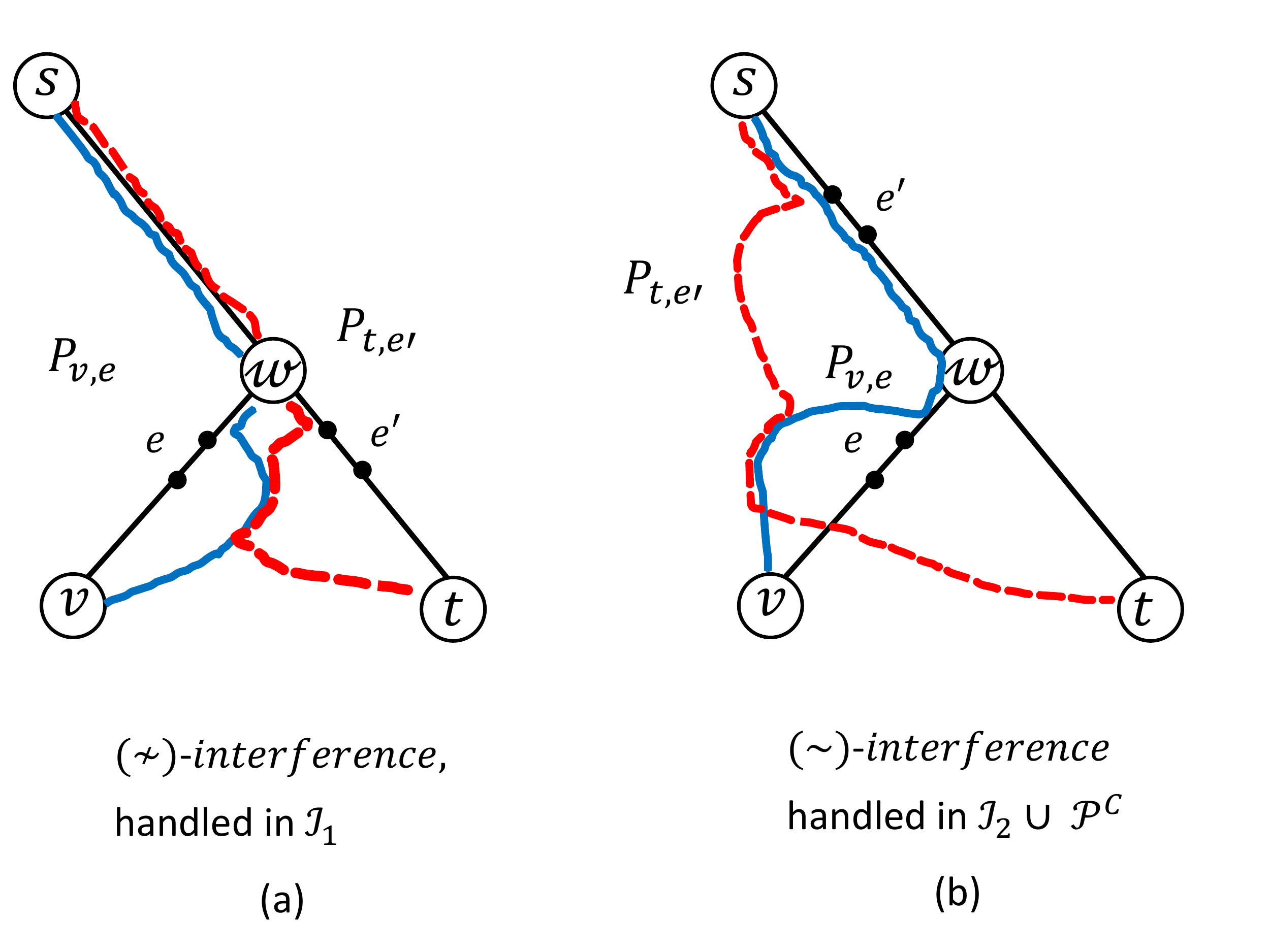}
\caption{ Illustration of the two types of interference. The replacement paths $P_{v,e}$ (solid) and $P_{t,e'}$ are
(a) $(\not\sim)$ interfering as $e \not\sim e'$, and (b) $(\sim)$ interfering as $e\sim e'$.
\label{fig:interpaths}}
\end{center}
\end{figure}

Our construction is heavily based on distinguishing between two types of interference, depending on the relation of the two failing edges protected by the interfered paths. In particular, if the interfering paths $P_{v,e}$ and $P_{t,e'}$ satisfy that $e \not\sim e'$, then we call it $(\not\sim)$-\emph{interference}, and if $e \sim e'$, then it is $(\sim)$-\emph{interference}. For an illustration, see Fig. \ref{fig:interpaths}.
%%%%%%%%%%%%%%%%%%%
Let $\mathcal{I}^{\not\sim}(\langle v,e \rangle)=\{\langle t,e' \rangle \in \mathcal{I}(P) ~\mid~ e\not\sim e'\}$ be the set of pairs whose corresponding paths $(\not\sim)$-\emph{interfere} with $P_{v,e}$.

A given subset of uncovered pairs $\mathcal{P}' \subseteq \UncoverPairs$ is called a $(\sim)$-\emph{set} if $\mathcal{I}^{\not\sim}(\langle v,e \rangle)\cap \mathcal{P}' = \emptyset$ for \emph{every} $\langle v,e \rangle \in \mathcal{P}'$. Otherwise, it is called $(\not\sim)$-\emph{set}. In other words, in a $(\sim)$-set there is no $(\not\sim)$-interference between any pair of paths.

\subsection{The main construction}
Let us start with an overview of the main construction phases.
The initial structure $H$ contains $T_0$. In phases (S1) and (S2), we add backup edges to $H$ corresponding to last edge of the new ending replacement paths $P_{v,e}$ so that eventually the set of $T_0$ edges unprotected by $H$ is bounded by $O(1/\epsilon \cdot n^{1-\epsilon}\cdot \log n)$.
(These edges will have to be reinforced; all other edges of $H$ will be taken as backup edges.)
The high level idea of our main construction is as follows. First, we divide the uncovered pairs $\UncoverPairs$ into two sub-sets $(\not\sim)$-set $\mathcal{I}_1$ and a $(\sim)$-set $\mathcal{I}_2$, by letting
$$\mathcal{I}_1=\{\langle v,e \rangle \in \UncoverPairs ~\mid~\mathcal{I}^{\not\sim}(\langle v,e \rangle) \neq \emptyset\} \mbox{~~and~~} \mathcal{I}_2=\UncoverPairs  \setminus \mathcal{I}_1.$$
%(Note that it might be the case that the same path $P_{v,e_i}=P_{v,e_j}$ appears both in $\mathcal{I}_1$ and in $\mathcal{I}_2$. For example, if $e_i$ appears above $e_j$ on $P_{v,e}$ and $P_{v,e_i}=P_{v,e_j}$ interferes with $P_{t,e'}$ such that (1) $e_i$ is above $\LCA(v,t)$ on $\pi(s,v)$ and (2) $e_j$ (resp., $e'$) appears below $\LCA(v,t)$ on $\pi(s,v)$ (resp., $\pi(s,t)$).)
%This is why its important to recognize the paths $P_{v,e}$ by the key of the pair $\langle v,e\rangle$ and not according to the structure of the paths as the same path may be classified differently depending on its key.

Phase (S1) starts by setting the first $(\sim)$-set to be $\mathcal{P}^C_0=\mathcal{I}_2$.
Then, Phase (S1) employs an iterative process of $\keps=O(1/\epsilon)$ iterations. Each of these iterations does the following.
For every vertex $v \in V$,
the algorithm repeatedly adds the $\lceil n^{1/\epsilon} \rceil$ distinct last edges of the remaining $s-v$ replacement-paths of the uncovered pairs in $\mathcal{I}_1$ protecting the deepest edges on $\pi(s,v)$.
In addition, each such iteration $i$ may yield an additional $(\sim)$-set, $\mathcal{P}^C_i$, which would be handled in Phase (S2).  Thus, at the end of Phase (S1), we have at most $O(1/\epsilon)$ such $(\sim)$-sets $\mathcal{P}^C_i$ that \emph{partially} cover the pairs of $\mathcal{I}_1$. The last edges of the replacement-paths of the pairs in $\mathcal{I}_1$ that are not covered by the $(\sim)$-sets are added to $H$.  Phase (S2) of the algorithm is then devoted for considering the $(\sim)$-sets (i.e., $\mathcal{I}_2$ and the $O(1/\epsilon)$  additional $(\sim)$-sets that were created in Phase (S1)). For each such $(\sim)$-set $\mathcal{P'}$ and for every vertex $v$, the algorithm adds a collection of
$O(n^{\epsilon} \cdot \log n)$ backup edges corresponding to the last edges of $s-v$ replacement-paths of the pairs in $\mathcal{P'}$. The analysis shows that after Phase (S2), for each of the $O(1/\epsilon)$ $(\sim)$-sets $\mathcal{P}^C_i$ the number of edges protected by replacement paths corresponding to the pairs collection $\mathcal{P}^C_i$ that are still unprotected by $H$ is $O(n^{1-\epsilon}\cdot \log n)$. Hence,
%implying that
overall there are at most $O(1/\epsilon \cdot n^{1-\epsilon}\cdot \log n)$ edges that are still unprotected by $H$. Those edges will have to be reinforced.
\par We now describe the algorithm in detail.

\dnsparagraph{Phase (S1): Handling the $(\not\sim)$-set $\mathcal{I}_1$.}
%
%
%
%\paragraph{Bounding the class $\mathcal{I}_1$.}
%For every path $P \in \mathcal{I}_1$, let $\mathcal{I}^{\not\sim}(P)=\{ P' \in \mathcal{I}_1 \cap \mathcal{I}(P) ~\mid~ F(P')\not\sim F(P)\}$ be the set of interfered paths to $P$ satisfying that
%the two protected edges are not on the same shortest-path in $T_0$.
%

\begin{figure}[h!]
\begin{center}
\includegraphics[scale=0.35]{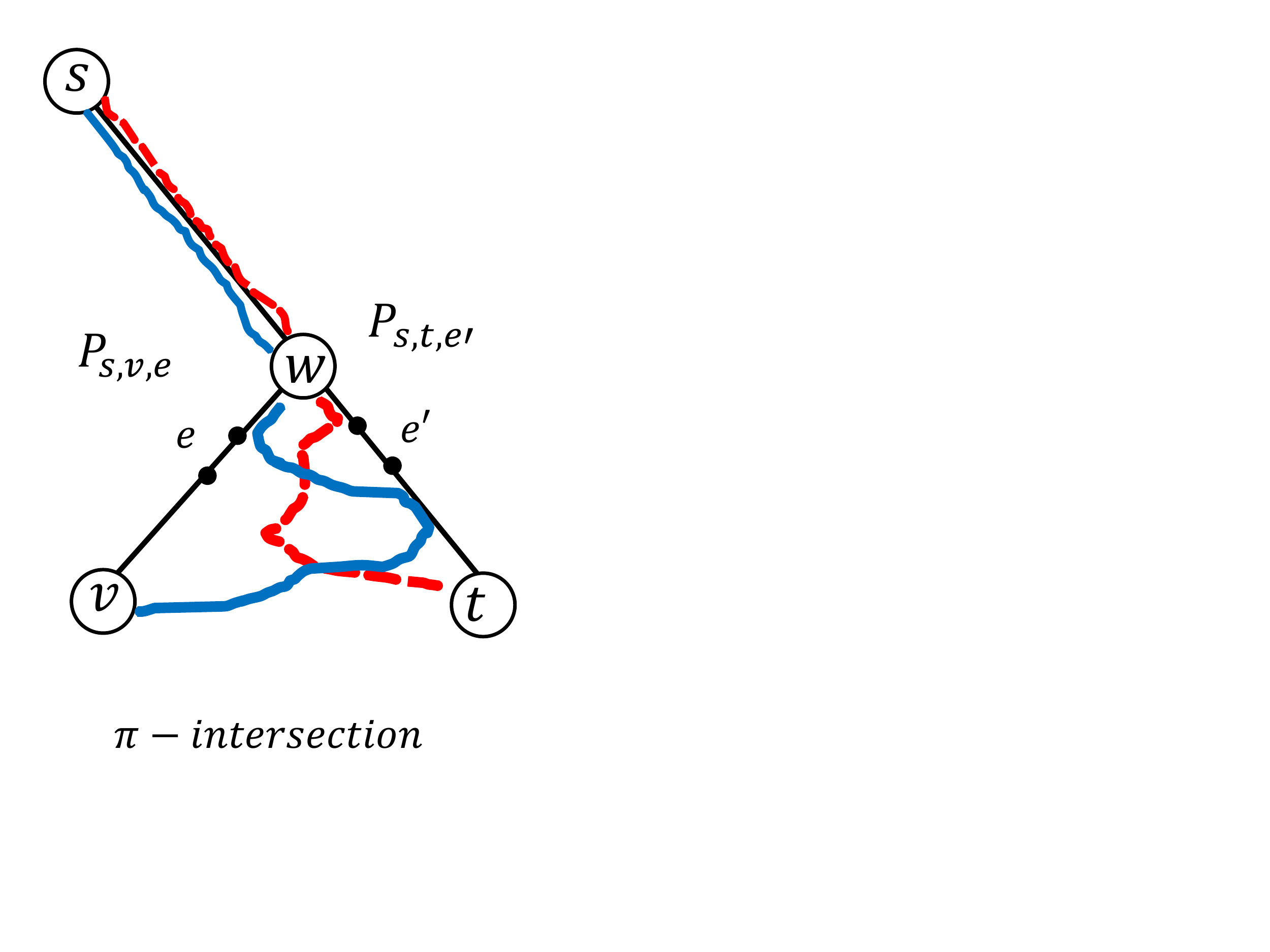}
\caption{ Illustration of $\pi$-intersection. The replacement-path $P_{v,e}$ (solid) $\pi$-intersects $P_{t,e'}\in \mathcal{I}^{\not\sim}(P_{v,e})$ (dashed).
\label{fig:piinter}}
\end{center}
\end{figure}

The next definition is important in this context.
For $v \neq t$, a path $P_{v,e}$ $\pi$-\emph{intersects} with path $P_{t,e'} \in \mathcal{I}^{\not\sim}(P_{v,e})$ if the detour of $P_{v,e}$ intersects at least one of the vertices of $\pi(\LCA(v,t), t)) \setminus \{\LCA(t,v)\}$, see Fig. \ref{fig:piinter}.
%\def\APPENDPIINTER{
%%%%%%%%%%%%%%%%%%%
%%%%%%%%%%%%%%%%%%%
%}%\APPENDPIINTER
Note that this property may \emph{not} be symmetric (unlike interference). That is, it might be the case that $P_{v,e}$ $\pi$-intersects $P_{t,e'}$ but not vice-versa.

The replacement paths of the uncovered pairs in some subset $\mathcal{P}_\ell \subseteq \mathcal{I}_1$ can be roughly classified into three types, termed A,B, and C with respect to $\mathcal{P}_\ell$.
%\\\textbf{MP: Should think on more informative notation.
%How about: Type A would be $(\not\sim,\pi)$-path, type B is $(\not\sim,\not\pi)$-path and type C is $(\sim)$-path?}\\
%\textbf{MP: should we classify the pairs according to a b C instead of the paths?}
A replacement-path $P_{v,e}$ for $\langle v,e \rangle \in \mathcal{P}_\ell$ is of \emph{type A with respect to} $\mathcal{P}_\ell$ if it $\pi$-intersects at least one path in $\mathcal{I}^{\not\sim}(\langle v,e \rangle)\cap \mathcal{P}_\ell$.
Let $\mathcal{P}^{A}_\ell \subseteq \mathcal{P}_\ell$ be the subset of all pairs whose paths is of type A, i.e.,
\begin{equation}
\label{eq:typea}
\mathcal{P}^{A}_\ell=\{ \langle v,e \rangle  \in \mathcal{P}_\ell \mid \exists \langle t,e' \rangle \in \mathcal{I}^{\not\sim}(\langle v,e \rangle)\cap \mathcal{P}_\ell ,~
%\mbox{~and~}
P_{v,e} \mbox{~$\pi$-intersects~} P_{t,e'}\}
~.
\end{equation}
%\textbf{MP: Note that in the analysis, we do not use the fact that $P$ interferes with $P'$ only the fact that it $\pi$-intersect with it and that $F(P)\not\sim F(P)$.}\\
A replacement-path $P_{v,e}$ for $\langle v,e \rangle \in \mathcal{P}_\ell$ is of \emph{type B with respect to} $\mathcal{P}_\ell$, if it is not of type A and it $(\not\sim)$-interferes with at least one path $P_{t,e'}$ for $\langle t,e' \rangle \in \mathcal{P}_\ell$ that is not of type A as well, i.e., $\langle t,e'\rangle \in \mathcal{P}_\ell \setminus \mathcal{P}^A_\ell$.
%Note that
In such a case, both $\langle v,e \rangle$ and $\langle t,e'\rangle$ are not in $\mathcal{P}^A_\ell$, and hence $P_{v,e}$ does not $\pi$-intersect $P_{t,e'}$ and vice-versa, implying that $P_{t,e'}$ is of type B as well.  Let $\mathcal{P}^{B}_\ell$  be the collection of the pairs whose corresponding path is of type B;
formally
%defined by
\begin{equation}
\label{eq:typeb}
\mathcal{P}^{B}_\ell=\{\langle v,e \rangle \in \mathcal{P}_\ell \setminus \mathcal{P}^{A}_\ell ~\mid~ \mathcal{I}^{\not\sim}(\langle v,e \rangle) \cap \left(\mathcal{P}_\ell \setminus \mathcal{P}^{A}_\ell \right) \neq \emptyset\}~.
\end{equation}
Finally, a replacement-path $P_{v,e} \in \mathcal{P}_\ell$ is of \emph{type C with respect to} $\mathcal{P}_\ell$ if it is not of type A or B. Note that such a path $P_{v,e}$ satisfies
that the intersection $\mathcal{I}^{\not\sim}(\langle v,e\rangle) \cap (\mathcal{P}_\ell \setminus \mathcal{P}^{A}_\ell)$ is empty. (This can happen either because $\mathcal{I}^{\not\sim}(\langle v,e\rangle) \cap \mathcal{P}_\ell=\emptyset$ or because $\left(\mathcal{I}^{\not\sim}(\langle v,e\rangle) \cap \mathcal{P}_\ell\right) \subseteq   \mathcal{P}^{A}_\ell$.) Let $\mathcal{P}^{C}_\ell=\mathcal{P}_\ell \setminus \left(\mathcal{P}^{A}_\ell \cup \mathcal{P}^{B}_\ell\right)$ be the set of pairs whose path is of type C.
Let
\begin{equation}
\label{eq:wideeps}
\keps=\lceil 1/\epsilon \rceil+2~.
\end{equation}
The uncovered pairs of $\mathcal{I}_1$ are now partitioned into $\keps+1$ subsets: $\keps$ $(\sim)$-sets $\mathcal{P}^C_1, \ldots,\mathcal{P}^C_{\keps}$ and a subset containing all the remaining pairs $\mathcal{I}'_1 =\mathcal{I}_1\setminus \bigcup_{i=1}^{\keps} \mathcal{P}^C_i$.
Essentially, the subset $\mathcal{I}'_1$ is ``implicit" and is not actually constructed by the algorithm; it consists of all $\mathcal{I}_1$ pairs $\langle v,e\rangle$ whose last edge of their path $P_{v,e}$ was added to $H$ during one of the $\keps$ iterations of Phase (S1). The analysis shows that the number of distinct last edges of the replacement paths of $\mathcal{I}'_1$ that were added into $H$ is bounded by $O(1/\epsilon \cdot n^{1+\epsilon})$.

\begin{figure}
\begin{center}
\includegraphics[scale=0.3]{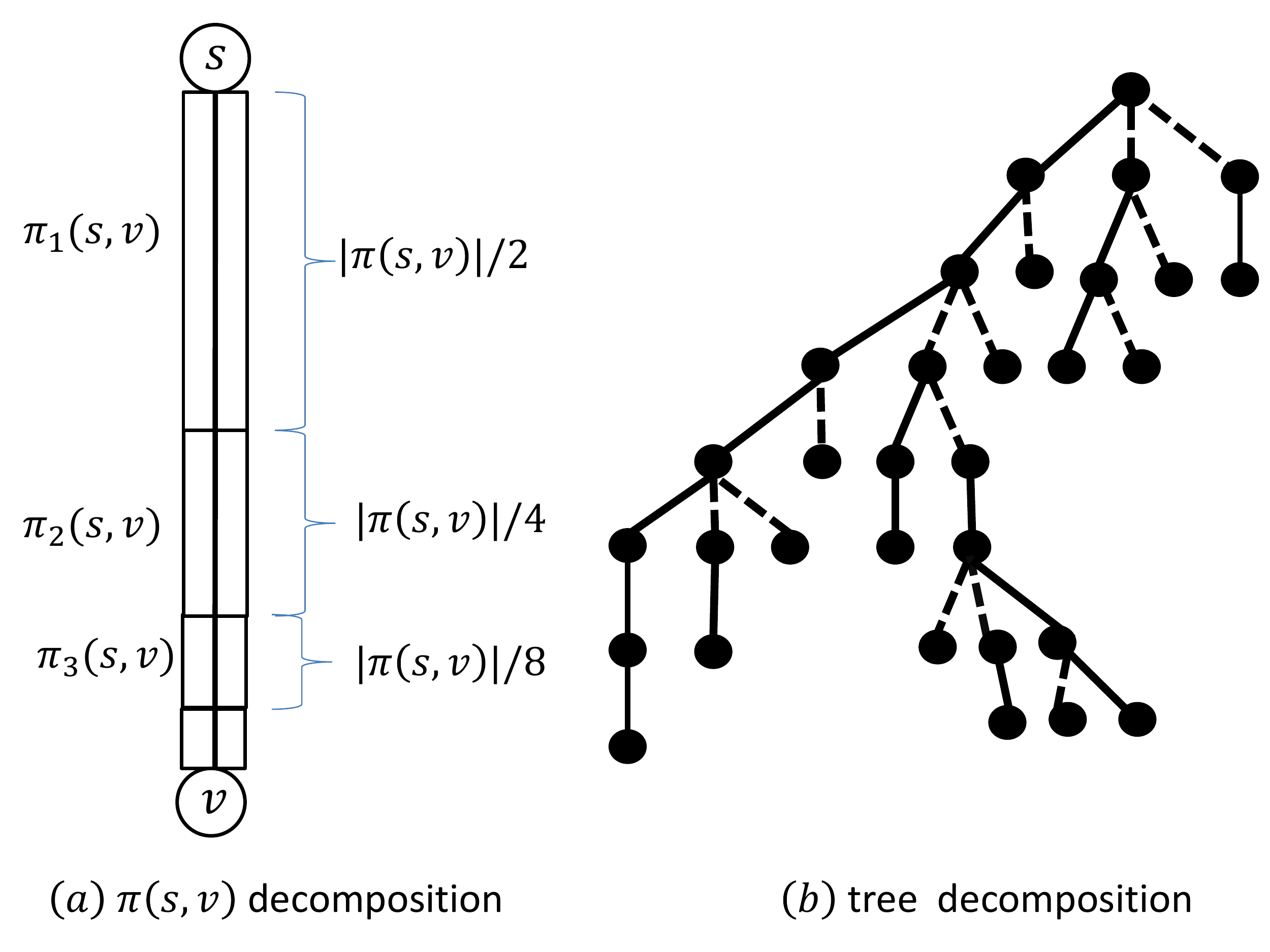}
\end{center}
\caption{\sf Decomposition of shortest-paths and trees.
(a) decomposition of the $s-v$ path $\pi(s,v)$ into $O(\log n)$ subsegment. The segment $i$, $\pi_i(s,v)$ contains roughly half of the remaining vertices. (b) Heavy-tree-decomposition $\mathcal{TD}$ on the BFS tree $T_0$ according to the algorithm of \cite{BS10}. The solid edges correspond to the paths of $\mathcal{TD}$.
}
\label{fig:treedecomp}
\end{figure}

The partition of $\mathcal{I}_1$ is conducted in $\keps$ iterations. At the end of each iteration, $O(n^{1+\epsilon})$ distinct last edges of the paths that correspond to the first $\langle v,e\rangle$ pairs from $\mathcal{I}_1$ (the paths protecting the deepest edges on $\pi(s,v)$) are added to $H$ (and intuitively, the pairs of these replacement paths join $\mathcal{I}'_1$).
Initially, let $\mathcal{P}_1=\mathcal{I}_1$. For every
$i =\{1, \ldots, \keps\}$, the next steps are performed:
\begin{itemize}
\item
Divide $\mathcal{P}_i$ into the subsets $\mathcal{P}^A_i, \mathcal{P}^B_i$ and $\mathcal{P}^C_i$ (according to Eq. (\ref{eq:typea},\ref{eq:typeb})).
\\(* Handling the paths of $\mathcal{P}^C_i$ is deferred to Phase (S2). The following steps attempt to handle the paths of $\mathcal{P}^A_i \cup \mathcal{P}^B_i$. *)
\item
Let $\overrightarrow{\mathcal{P}}^J_i(v)=\{\langle v,e_{i_{1}}\rangle, \ldots, \langle v,e_{i_{k_{J_v}}}\rangle\}$ be the ordered $\langle v, e\rangle$ uncovered pairs of $v$ in $\mathcal{P}^J_i$ for every $v \in V$ and $J \in \{A,B\}$ (in increasing distance of the failing edge $e_{i_{j}}$ from $v$).
\item
Add to $H$, the $\lceil n^{\epsilon} \rceil$ distinct last edges of the \emph{first} replacement-paths of the pairs in the ordering $\overrightarrow{\mathcal{P}}^J_i(v)$.
%, i.e., add $\LastE(P_{v,e_{i_{j}}})$ for every $J \in \{A,B\}$, $v \in V$ and $j \in \{1, \min\{\lceil n^{\epsilon} \rceil, k_v\}\}$.
\item
Set $\mathcal{P}_{i+1}=\{\langle v, e\rangle \in \mathcal{P}^A_{i}\cup \mathcal{P}^{B}_{i} ~\mid~ \LastE(P_{v,e}) \notin H\}$.
\end{itemize}
This completes
% the description of this
Phase (S1). Observe that a pair $\langle v,e \rangle \in \mathcal{P}_i$ that was classified as, say, type A in iteration $i$, but was not handled (i.e., its last edge was not added to $H$), joins $\mathcal{P}_{i+1}$ and is re-classified in iteration $i+1$, where it may be classified differently. In particular, if it gets classified into $\mathcal{P}^C_{i+1}$, then its handling will be deferred to Phase (S2). %In addition, the same path $P_{v,e_i}=P_{v,e_j}$ can be classified differently depending on its key $\langle e, v \rangle$.
\dnsparagraph{Phase (S2): Handling the remaining $(\sim)$-sets.}
The input for this step is a collection of  $(\sim)$ multi-sets
$\mathcal{S}=\{\mathcal{P}^{C}_{0}, \mathcal{P}^{C}_{1}, \ldots, \mathcal{P}^{C}_{\keps}\}$.
\dnsparagraph{Preprocessing Sub-Phase (S2.0): Building tree-decomposition for $T_0$.}
As a preprocessing step for handling the $(\sim)$-sets, the algorithm begins applying to the BFS tree $T_0$ the \emph{heavy-path-decomposition} technique presented by Sleator and Tarjan \cite{DCOMP81}
and slightly adapted by Baswana and Khanna \cite{BS10}.
Using this technique, the tree $T_0$ is broken into vertex disjoint paths $\mathcal{TD}=\{\psi_1, \ldots, \psi_t\}$ that satisfy some desired properties; for an illustration see Fig. \ref{fig:treedecomp}(b).
\begin{fact}
\label{fc:pathdecomp_prop}\cite{BS10}
There exists an $O(n)$ time algorithm for computing a path $\psi$ in $T_0$ whose removal splits $T_0$ into a set of disjoint subtrees $T_0(v_1), \ldots, T_0(v_j)$ s.t. for every
%such that for every
$1 \leq i \leq j$,
\begin{description}
\dnsitem{(1)}
$|T_0(v_i)| \leq n/2$ and $\psi \cap T_0(v_i)=\emptyset$, and
\dnsitem{(2)}
$T_0(v_i)$ is connected to $\psi$ through some edge, hereafter denoted $e(\psi,i)$ .
\end{description}
\end{fact}
The algorithm of Fact \ref{fc:pathdecomp_prop} is applied recursively on $T_0$. The output of this recursive procedure is a collection $\mathcal{TD}$ of paths $\psi \subseteq T_0$ plus a set of $T_0$ edges $e(\psi,i)$ that glue the paths $\psi$ to the tree $T_0$. Let $E^+(\mathcal{TD})=\bigcup_{\psi \in \mathcal{TD}}E(\psi)$ be the set of tree edges occurring on the paths of the decomposition and let $E^{-}(\mathcal{TD})=T_0 \setminus E^+(\mathcal{TD})$ be the collection of ``glue" edges.
In the next Sub-Phase, the algorithm iterates over all the vertices $v$ and add to the structure $H$ a collection of last edges of the replacement-paths protecting against the failing of the glue edges. (In the analysis it is shown that at most $O(n \log n)$ edges are added due to this step.)
\dnsparagraph{Sub-Phase (S2.1): Edge addition based on tree-decomposition [for fixed $v$].}
Define the last edges
of the new-ending replacement paths protecting the glue edges $E^{-}(\mathcal{TD}) \cap \psi(s,v)$ by
$$\widehat{E}(\mathcal{TD},v)=\{\LastE(P_{v,e}) ~\mid~ \langle v,e \rangle \in \UncoverPairs \mbox{~and~} e \in E^{-}(\mathcal{TD})\}.$$
Add $\widehat{E}(\mathcal{TD},v)$ to $H$.

We now turn to consider the main part of Phase S2.
The algorithm treats each $(\sim)$-set $\mathcal{P} \in \mathcal{S}$ separately, by adding into $H$ $O(n^{1+\epsilon} \cdot \log n)$ distinct last edges of  replacement paths carefully selected from the uncovered pairs of $\mathcal{P}$. In the analysis section, we then show that the total number of edges $e$ with a pair $\langle v,e\rangle$ in $\mathcal{P}$ that are unprotected by $H$ is bounded by $O(n^{1-\epsilon} \cdot \log n)$, and since there are $\keps+1=O(1/\epsilon)$ sets in $\mathcal{S}$ (see Eq. (\ref{eq:wideeps})), overall there are $O(1/\epsilon \cdot n^{1-\epsilon} \cdot \log n)$ edges in $T_0$ that are unprotected by $H$ (and will have to be reinforced).
\par The selection of the uncovered pairs $\langle v,e\rangle$ whose last edge of their replacement path $P_{v,e}$ is to be added into $H$ is performed in the following manner.
The algorithm iterates over every $(\sim)$-set $\mathcal{P} \in \mathcal{S}$ and every vertex $v \in V$, and selects a subset $\Add(\mathcal{P},v)$ of $v$'s uncovered pairs from $\mathcal{P}$, where the total number of last edges of their corresponding replacement paths in bounded by $O(n^{\epsilon} \cdot \log n)$, and then adds these last edges to $H$. The selection, for $\mathcal{P}$ and $v$, of pairs to be included in $\Add(\mathcal{P},v)$ is done in two main phases.
\dnsparagraph{Sub-Phase (S2.2)[for fixed $\mathcal{P},v$]: Covering pairs based on shortest-path decomposition into $O(\log n)$ fragments.}
The $s-v$ shortest-path $\pi(s,v)=[s=u_{0}, \ldots, u_{k}=v]$ is decomposed into $k'=\lfloor \log|\pi(s,v)| \rfloor$ subsegments of exponentially decreasing length, i.e., where each subsegment consists of the first half of the remaining $\pi(s,v)$ path. For an illustration see Fig. \ref{fig:treedecomp}(a). Formally, letting $u_{i_j}$ be the vertex at distance $\left\lceil \sum_{\ell=1}^j\left(|\pi(s,v)|/ 2^\ell \right)\right \rceil$ from $s$ on $\pi(s,v)$ for $j \in \{1, \ldots, k'\}$, and $u_{i_0}=u_0$, the $j$'th subsegment is given by $\pi_j(s,v)=\pi(u_{i_{j-1}},u_{i_{j}})$ for every $j \in \{1, \ldots, k'\}$. It then holds that
\begin{equation}
\label{eq:jseg}
|\pi_j(s,v)|\geq \left \lfloor |\pi(s,v)|/2^{j-1}| \right\rfloor
\mbox{~and~}
%\mbox{~~and~~}
\sum_{j'>j}|\pi_j(s,v)|\geq |\pi_j(s,v)|/2
%~.
\end{equation}
For each of the $k'$ subsegments $\pi_j(s,v)$,
let $\mathcal{P}_j(v)=\{\langle v,e \rangle \in \mathcal{P} ~\mid~ e \in \pi_j(s,v)\}$ be the set of $v$'s uncovered pairs from $\mathcal{P}$ whose paths protect the edges in $\pi_j(s,v)$; let $\mathcal{LE}_j(\mathcal{P},v)=\{\LastE(P_{v,e}) ~\mid~ \langle v,e \rangle \in \mathcal{P}_j(v)\}$ be the corresponding last edges of these replacement paths.

A subsegment $\pi_j(s,v)$ is \emph{heavy} with respect to $\mathcal{P}$ if $|\mathcal{LE}_j(\mathcal{P},v)|\geq \lceil n^{\epsilon}\rceil$, otherwise it is \emph{light}.
For every light subsegment $\pi_j(s,v)$, add $\mathcal{P}_j(v)$ to the collection of selected pairs $\Add(\mathcal{P},v)$ whose last edges (of their corresponding replacement paths) are later added to $H$. I.e., add to $\Add(\mathcal{P},v)$ the pairs $\bigcup \mathcal{P}_j(v)$, where the union is over all $j \in \{1, \ldots, k'\}$ s.t. $\pi_j(s,v)$ is light.
In addition, we move to $\Add(\mathcal{P},v)$ some additional pairs $\langle v, e^*_j\rangle$ as follows. For every $j \in \{1, \ldots, k'\}$, let $e^*_j$ be the first edge on $\pi_j(s,v)$ (closest to $s$) such that $\langle v,e^*_j\rangle$ is in $\mathcal{P}_j(v)$.
The algorithm adds $\bigcup_{j=1}^{k'}\langle v,e^*_j\rangle$ to $\Add(\mathcal{P},v)$. (This addition would ensure that the divergence point $d(P_{v,e})$ of the replacement paths $P_{v,e}$ protecting edges $e$ on the segment $\pi_j(s,v)$ and whose last edge was not added to the output structure $H$, is located inside the segment $\pi_j(s,v)$.)
%$$\Add(\mathcal{P},v)=\bigcup_{\{j \in \{1, \ldots, k'\} ~\mid~ \pi_j(s,v) \mbox{~is light}\}}\mathcal{P}_j(v).$$

\dnsparagraph{Sub-Phase (S2.3): Covering pairs depending on both the tree-decomposition and $\pi(s,v)$ decomposition [for fixed $\mathcal{P},\psi, v$].}

Define $E(\psi,\mathcal{P},v)=\{ e ~\mid~ \langle v,e \rangle \in \mathcal{P} \mbox{~~and~~} e \in \psi \cap \pi(s,v)\}$. Let $e^*$ be the upmost edge in $E(\psi,\mathcal{P},v)$ (i.e., closest to
s). Add $\langle v,e^* \rangle$ to $\Add(\mathcal{P},v)$.

\begin{figure}
\begin{center}
\includegraphics[scale=0.4]{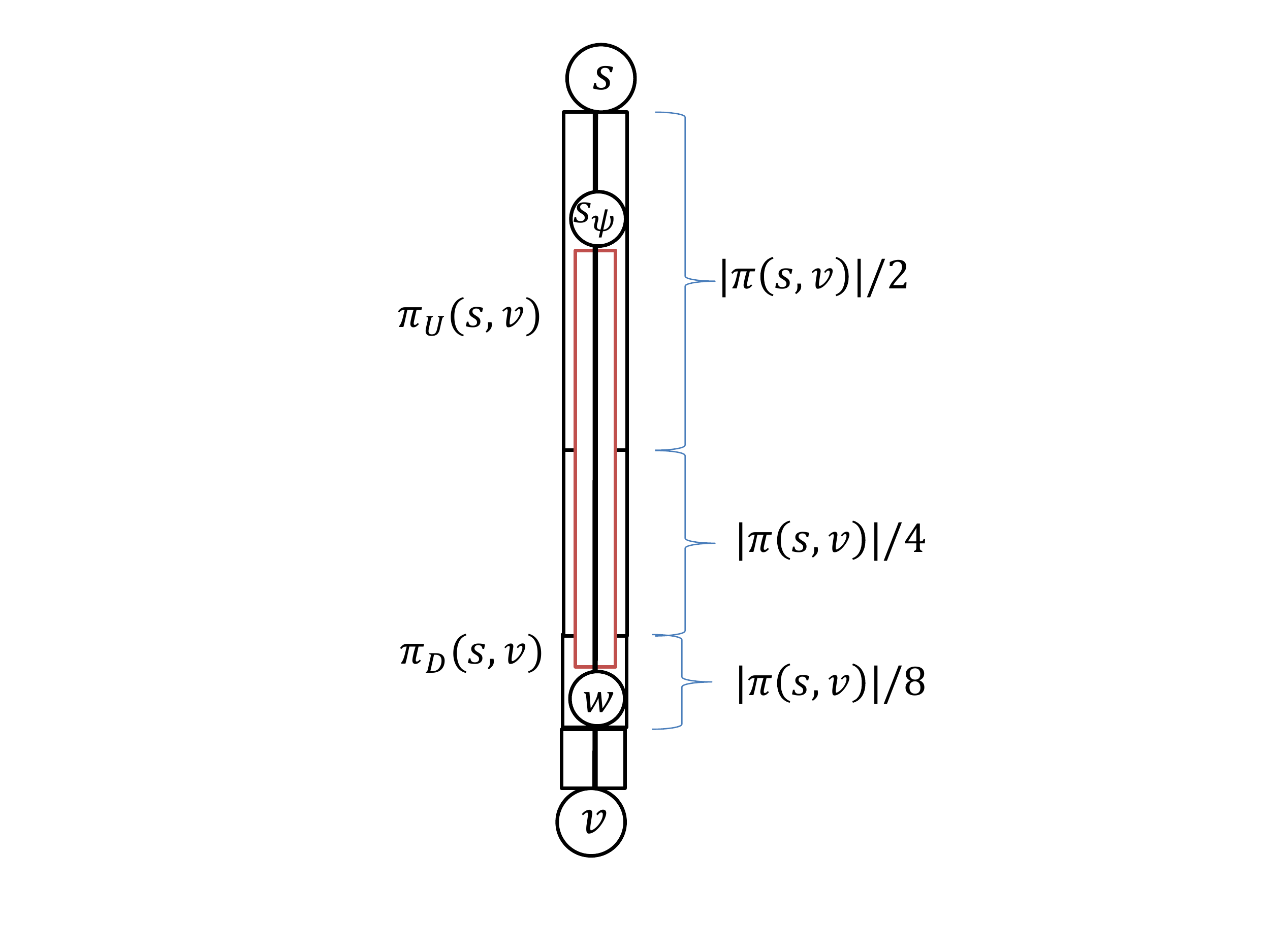}
\end{center}
\caption{\sf Illustration of the intersection between $s_{\psi}-t_{\psi}$ path $\psi$ in the tree decomposition $\mathcal{TD}$ and the shortest-path $\pi(s,v)$ decomposition. The vertex $w$ is the LCA of $t_{\psi}$ and $v$ in $T_0$. The intersection of the paths $\psi \cap \pi(s,v)$ is shown in red.
The upper (resp., lower) intersected segment is $\pi_U(s,v)$ (resp., $\pi_L(s,v)$). }
\label{fig:decompud}
\end{figure}

Next, consider the intersection of $\psi$ with $\pi(s,v)$.
Recall that in Sub-Phase (S2.1), the $s-v$ path $\pi(s,v)$ was decomposed into $k'=\lfloor \log|\pi(s,v)| \rfloor$ segments $\pi_1(s,v), \ldots, \pi_{k'}(s,v)$.
%
%Let $\pi_U(\psi,v)$ be the first (closest to $s$)
%(resp., last)
%heavy subsegment that intersects $\pi(s,v)$ such that $\pi_U(\psi,v) \nsubseteq \pi(s,v)$ and $\pi_U(\psi,v) \cap \pi(s,v) \neq \emptyset$.
%
Let $\pi_{U}(\psi,v)$ be the first, i.e., closest to $s$, subsegment of $\pi(s,v)$ that intersects $\psi$ such that $\pi_{U}(\psi,v) \nsubseteq \psi$ and $\pi_{U}(\psi,v) \cap \psi \neq \emptyset$ (if such exists).
Similarly, let $\pi_{L}(\psi,v)$ be the last, i.e., closest to $v$, subsegment of $\pi(s,v)$ that intersects $\psi$ such that $\pi_{L}(\psi,v) \nsubseteq \psi$ and $\pi_{L}(\psi,v) \cap \psi \neq \emptyset$.
See Fig. \ref{fig:decompud} for an illustration.

Let
$\mathcal{P}_{U}(\psi,v)=\{\langle v,e \rangle \in \mathcal{P} ~\mid~ e \in \pi_{U}(\psi,v)\cap \psi\}$ be the pairs in $\mathcal{P}$ whose replacement paths protect against the failing of the edges in the intersection $\pi_U(s,v)\cap \psi$ and let
$\mathcal{LE}_{U}(\mathcal{P},\psi,v)=\{\LastE(P_{ v, e}) ~\mid~ \langle v,e \rangle \in \mathcal{P}_{U}(\psi,v)\}$ be the last edges of the corresponding replacement paths.
If $|\mathcal{LE}_{U}(\mathcal{P},\psi,v)|\leq \lceil n^\epsilon \rceil$, then add $\mathcal{P}_{U}(\psi,v)$ to $\Add(\mathcal{P},v)$.
Finally, let $e^*_{U}$ be the upmost edge on $\pi(s,v)$ with a pair $\langle v,e^*_{U} \rangle\in \mathcal{P}_{U}(\psi,v)$. Then, add $\langle v,e^*_{U} \rangle$ to $\Add(\mathcal{P},v)$. The set $\mathcal{P}_{L}(\psi,v)$ is handled in the same manner as $\mathcal{P}_{U}(\psi,v)$.

Finally, for every $\mathcal{P}$ and $v$ and for every edge $e$ such that $\langle v,e\rangle \in \Add(\mathcal{P},v)$ add the last edge of $P_{v,e}$ to $H$. While the resulting sets $\Add(\mathcal{P},v)$ might contain many pairs, in the analysis section we show that the total number of new backup edges that will be added to $H$  as a result of these pairs will be at most $O(\log n\cdot n^{\epsilon})$ per vertex $v$ and $(\sim)$-set $\mathcal{P}$.
This completes the description of the algorithm.

\section{Analysis}
\subsection{Size Bound}
We start with size analysis and use the following fact.
\begin{fact}[\cite{BS10}]
\label{fc:td_logn}
For every node $v \in V$, \\
(a) $\pi(s,v) \cap E^{-}(\mathcal{TD})=O(\log n)$, and
(b) $\pi(s, v)$ intersects at most $O(\log n)$ paths in $\mathcal{TD}$ [Lemma 3.6 of \cite{BS10}].
\end{fact}

\begin{lemma}
\label{lem:sizeh}
$|E(H)|=\min\{O(1/\epsilon \cdot n^{1+\epsilon} \cdot \log n), n^{3/2}\}$.
\end{lemma}
\Proof
For $\epsilon \geq 1/2$, the claim trivially holds by \cite{PPFTBFS13}. From now on, consider $\epsilon \in (0,1/2)$.
By Fact \ref{fc:td_logn}(a), the set of edges $\widehat{E}(\mathcal{TD},v)$ that was added in Sub-Phase (S2.1)
contains $O(\log n)$ edges. We now focus on a specific vertex $v$ and $(\sim)$-set $\mathcal{P}$ and bound the number of new edges corresponding to the pairs of $\Add(\mathcal{P},v)$ that were collected in Sub-Phase (S2.2-3).

In Sub-Phase (S2.2), the algorithm adds the pairs of the light subsegments $\pi_j(s,v)$. Since there are $O(\log n)$ subsegments and as the number of last edges of replacement paths protecting the edges of a light subsegment is bounded by $O(n^{\epsilon})$ edges, overall $O(\log n \cdot n^{\epsilon})$ edges are added due to these pairs.

In Sub-Phase (S2.3) we restrict attention to a specific path $\psi \in \mathcal{TD}$ and consider the intersection of $\pi(s,v)$ and $\psi$. By Fact \ref{fc:td_logn}(b), every path $\pi(s,v)$ intersects with $O(\log n)$ paths $\psi$ in $\mathcal{TD}$. Since the algorithm adds the last edges of replacement paths protecting edges on $\pi_U(s,v)$ and $\pi_L(s,v)$ only if their number is bounded by $O(n^{\epsilon})$, overall $O(\log n \cdot n^{\epsilon})$ edges are added due to this sub-phase.
Finally, the total number of pairs $\langle v, e^*_j\rangle$ and
$\langle v, e^*_U\rangle,\langle v, e^*_L\rangle$ that are added in Sub-Phase (S2.2-3) is bounded by $O(\log n)$.
Altogether, we get that the pairs of $\Add(\mathcal{P},v)$ contributes $O(\log n\cdot n^{\epsilon})$ edges to $H$. The lemma follows by summing over all $n$ vertices and the $O(1/\epsilon)$ $(\sim)$ sets.
\QED
%}%\APPENDSIZE
We proceed by presenting some useful properties of the paths constructed
by Alg. \mbox{\tt Pcons}.
\subsection{Basic Replacement Path Properties}
\begin{lemma}
\label{lem:pcon_correct}
For every $v \in V$, $e \in \pi(s,v)$, it holds that $P_{v,e} \in SP(s, v, G \setminus \{e\})$.
\end{lemma}
\Proof
If the replacement path $P_{v,e}$ is not new-ending, i.e., $\LastE(P_{v,e}) \in T_0$, then the correctness follows immediately. The interesting remaining case is where the replacement-path $P_{v,e}$ had to include a new-edge that was not in $T_0$. We show that in this case, there exists an $s-v$ shortest path in $G \setminus \{e\}$ with a unique divergence point from $\pi(s,v)$ that occurs above the failing edge $e$. In particular, such a path can given by letting $P'=SP(s, v, G \setminus \{e\},W)$. To see this, assume towards contradiction that the divergence point of $P'$ from $\pi(s,v)$ is not unique. Let $w_1$ (resp., $w_2$) be the first (resp., second) divergence point of $P'$ from $\pi(s,v)$. There are two cases. If $e \in \pi(w_1,w_2)$, then $P'[w_2,v]=SP(w_2,v, G \setminus\{e\},W)=\pi(w_2,v)$, in contradiction to the fact that $P_{v,e}$ is new-ending. Otherwise, $e \in \pi(w_2,v)$ and $P'[w_1,w_2]=SP(w_1,w_2, G \setminus\{e\},W)=\pi(w_1,w_2)$, contradicting the fact that $w_1$ is a divergence point from $\pi(s,v)$. This establishes the claim that there exists a replacement-path with a unique divergence point. Since the algorithm picks the replacement-path whose unique divergence point from $\pi(s,v)$ is as close to $s$ as possible, the correctness follows.
\QED
%}%\APPENDPCONCOR
Recall that for a new-ending path $P_{v,e}$ (i.e., $\langle v,e \rangle \in \UncoverPairs$), $d(P_{v,e})$ is the first divergence point of $P_{v,e}$ from $\pi(s,v)$.
By the construction of the new-ending paths, we have the following.
\begin{claim}
\label{cl:newend_claim}
For every new-ending path $P_{v,e}$:
\begin{description}
\item{(1)}
the divergence point $d(P_{v,e})$ is unique;
\dnsitem{(2)}
there exists no $s-v$ replacement path in $G \setminus \{e\}$ whose unique divergence point is \emph{above} $d(P_{v,e})$ on $\pi(s,v)$ (i.e., closer to $s$).
%\item{(3)}
%the last vertex of the detour $E(P)\setminus E(\pi(s,v))$ is $v$.
\end{description}
\end{claim}
\Proof
Begin with (1).
Consider a new-ending path $P_{v,e}$ and let $\pi(s,v)=[u_0=s, u_1, \ldots, u_k=v]$. Recall that
for every $j \in \{0, \ldots, k-1\}$, $G_j(v)=G \setminus V(\pi(u_j,u_k))\cup \{u_j,u_k\}$. Let $j^*$ be the minimal index $j$ satisfying that $\dist(s,v, G_j(v)\setminus \{e\})=\dist(s,v, G \setminus \{e\})$. Alg.  \mbox{\tt Pcons} defines $P_{v,e}=SP(s,v, G_{j*}(v)\setminus \{e\},W)$.
We now show that $u_{j^*}$ is the unique divergence point by showing that $D(P_{v,e})=P_{v,e}[u_{j^*},v]$ and $\pi(s,v)$ are vertex disjoint except for the common endpoints $u_{j^*}$ and $v$.
Assume towards contradiction otherwise, and let $u_{\ell} \in \left(V(D(P_{v,e})) \cap V(\pi(s,v))\right) \setminus \{u_{j^*},v\}$ be the last vertex (closest to $v$) on $D(P_{v,e})$ that occurs on $\pi(s,v)\setminus \{v\}$. By the definition of $P_{v,e}$, $u_{\ell}$ occurs above $u_{j^*}$ on $\pi(s,v)$ and hence also above failing edge $e$ on $\pi(s,v)$.
Consider the path $P'=\pi(s,u_{\ell}) \circ P_{v,e}[u_{\ell},v]$.
By the selection of $u_{\ell}$, $P' \in SP(s, v, G \setminus \{e\})$ and $u_{\ell}$ is the unique divergence point of $P'$ and $\pi(s,v)$. In particular, $P' \subseteq G_{\ell}$ where $\ell<j^*$. Contradiction to the selection of $j^*$. Part (1) follows.  Part (2) follows immediately by Part (1) and the definition of the paths by Alg. \mbox{\tt Pcons}.
\QED
\begin{claim}
\label{cl:dpointbtw}
Consider two new-ending $s-v$ replacement paths $P_{v,e_{i_1}}, P_{v,e_{i_2}}$ such that $\LastE(P_{v,e_{i_1}})\neq \LastE(P_{v,e_{i_2}})$ where without loss of generality $e_{i_1}=(x_{i_1},y_{i_1})$ is above (closer to $s$) $e_{i_2}=(x_{i_2},y_{i_2})$ on $\pi(s,v)$.
Then, $d(P_{v,e_{i_2}}) \in \pi(y_{i_1}, x_{i_2})$.
\end{claim}
\Proof
Towards contradiction, assume otherwise. It then holds that $d(P_{v,e_{i_2}}) \in \pi(s, x_{i_1})$. Since the detour segments are edge disjoint with $\pi(s,v)$ (see Cl. \ref{cl:newend_claim}(1)), we get that there
are two $s-v$ paths in $G \setminus \{e_{i_1},e_{i_2}\}$ given by $P_{v,e_{i_1}}, P_{v,e_{i_2}}$, and the optimality of these paths implies $|P_{v,e_{i_1}}|=|P_{v,e_{i_2}}|$.
There are three cases. \\Case 1: $d(P_{v,e_{i_1}})=d(P_{v,e_{i_2}})$. In this case, by the uniqueness of the weight assignment $W$, we get that $P_{v,e_{i_1}}=P_{v,e_{i_2}}$, in contradiction to the fact that $\LastE(P_{v,e_{i_1}})\neq \LastE(P_{v,e_{i_1}})$.
Case 2: $d(P_{v,e_{i_1}})$ is above $d(P_{v,e_{i_2}})$.
In this case, we get a contradiction to Cl. \ref{cl:newend_claim}(2) with respect $P_{v,e_{i_2}}$.
Case 3: $d(P_{v,e_{i_2}})$ is above $d(P_{v,e_{i_1}})$.
In this case, we get a contradiction to Cl. \ref{cl:newend_claim}(2) with respect $P_{v,e_{i_1}}$. The claim follows.
\QED
%}%\APPENDDPBTW
\begin{claim}
\label{cl:detourvertex_disjoint}
For every $P=P_{v,e}$ such that $\langle v,e\rangle \in \UncoverPairs(v)$,
\begin{description}
\dnsitem{(1)}
$|D(P)|=\Omega(\dist(e,v,\pi(s,v)))$.
\dnsitem{(2)}
For every $P'=P_{v,e'}$ $\langle v,e\rangle \in \UncoverPairs(v)$ satisfying that $\LastE(P)\neq \LastE(P')$ it holds that $D(P') \cap D(P)=\{v\}$.
\end{description}
\end{claim}
\Proof
%Without loss of generality, assume that $P$ was constructed by Alg. \mbox{\tt Pcons} \emph{before} $P'$.
Since $d(P)$ is above the edge $e$ on $\pi(s,v)$, it holds that $|D(P)|=|P(d(P),v)|\geq \dist(e,v,\pi(s,v))$. Consider (2) and assume, towards contradiction, that there exists a mutual vertex $w \in \left(D(P') \cap D(P)\right)\setminus \{v\}$.
Since $d(P)$ and $d(P')$ are unique divergence points, it holds that
$P[w,v]\cap V(\pi(s,v))=\{v\}$ and $P'[w,v]\cap V(\pi(s,v))=\{v\}$.
Hence,
$$P[w,v]=SP(w, v, (G \setminus V(\pi(s,v)))\cup \{v\},W)=P'[w,v],$$
in contradiction to the fact that $\LastE(P)\neq \LastE(P')$. \QED
%}%\APPENDDDIS
\subsection{Bounding the number of $T_0$ edges unprotected by $H$}
Throughout, we consider the final structure $H$ (obtained by the end of Phase (S2)) and denote the path $P_{v,e}$ as $H$-\emph{new-ending} if $\LastE(P_{v,e}) \notin H$. Let $\UncoverPairs(H)=\{\langle v,e \rangle ~\mid~ \LastE(P_{v,e})\notin H\}$ be the uncovered pairs
in the final structure $H$.

For every $(\sim)$-set $\mathcal{P} \in \mathcal{S}$, let $\mathcal{P}_{miss}=\{\langle v,e \rangle\in \mathcal{P} \cap\UncoverPairs(H)\}$ be the pairs of $\mathcal{P}$ that are uncovered by $H$. %Let $E_{miss}(\mathcal{P})=\{F(P) ~\mid~ P \in \mathcal{P}_{miss}\}$ be the set of last-unprotected edges in $H$ corresponding to the paths of $\mathcal{P}_{miss}$.
Let  $E_{miss}(\mathcal{P},v)=\{e ~\mid~ \langle v,e \rangle \in \mathcal{P}_{miss}\}$ be the set of edges on $\pi(s,v)$ such that the last edge of the replacement paths of $\mathcal{P}$ pairs  were not added to $H$. Let
\begin{equation}
\label{eq:emiss}
E_{miss}(\mathcal{P})=\bigcup_{v \in V}E_{miss}(\mathcal{P},v) \end{equation}
be the collection of $T_0$ edges unprotected by $H$, corresponding to the paths of $\mathcal{P}$ and let $E_{miss}(H)=\{e ~\mid~ \exists v \mbox{~s.t~}\langle v,e \rangle \in \UncoverPairs(H)\}$ be the set of $T_0$ edges that are unprotected by $H$.
Toward the end of this section, we show that
\begin{lemma}
\label{lem:newh_final}
$|E_{miss}(H)|=O(1/\epsilon \cdot n^{1-\epsilon} \cdot \log n)$.
\end{lemma}
The analysis proceeds in two steps.
Let $\mathcal{P}^{C}=\bigcup_{i=1}^{\keps}\mathcal{P}^{C}_{i}$ be the collection of pairs whose corresponding paths are of type C defined in Phase (S1). First, we show that due to Phase (S1), $\mathcal{I}_1 \setminus \mathcal{P}^{C}$ contains \emph{no} uncovered pair in $H$, i.e., there is no pair $\langle v,e \rangle \in \mathcal{I}_1 \setminus \mathcal{P}^{C}$ such that $P_{v,e}$ is $H$-new-ending path.
This implies that it suffices to consider the uncovered pairs of $\mathcal{P}^C_i$, since $\UncoverPairs(H)=\bigcup_{\mathcal{P} \in \mathcal{S}}\mathcal{P}_{miss}$.
In the second
step,
%part of the analysis,
we complete the argument by showing that for each of the $O(1/\epsilon)$ $(\sim)$-sets $\mathcal{P}$, the cardinality of $E_{miss}(\mathcal{P})$, the set of $T_0$ edges that are unprotected by $H$, is bounded by
$O(n^{1-\epsilon} \cdot \log n)$. Since there are $O(1/\epsilon)$ such sets, overall, we get that
$$|E_{miss}(H)|=|\bigcup_{\mathcal{P} \in \mathcal{S}} E_{miss}(\mathcal{P})|=O(1/\epsilon \cdot n^{1-\epsilon} \cdot \log n)$$
as desired. We now describe the analysis in detail.

\subsubsection{Analysis of Phase (S1)}
\label{sub:ana1}
\begin{figure}[h!]
\begin{center}
\includegraphics[scale=0.31]{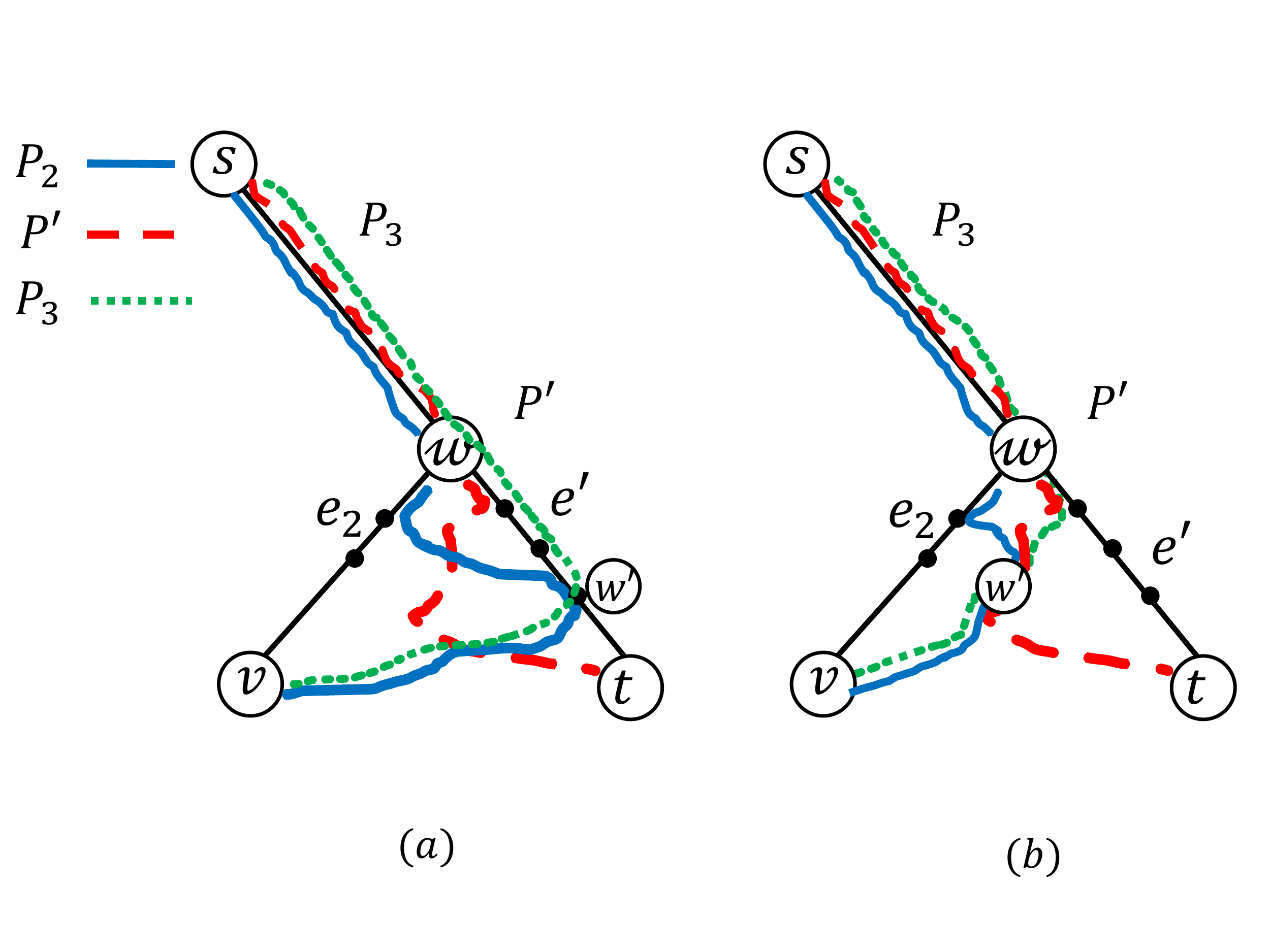}
\caption{ Schematic Illustration for Lemma \ref{lem:inter_edge}:
(a) $P_2$ of type A. (b) $P_2$ of type B that does not intersect $\pi(w, t)$.
\label{fig:lemmainteredge}}
\end{center}
\end{figure}

We begin by establishing a property that holds for every two pairs $\langle v,e_1 \rangle, \langle v,e_2 \rangle \in \mathcal{P}^{J}_i$ for $J \in \{A,B\}$ such that $\LastE(P_{v,e_1})\neq  \LastE(P_{v,e_2})$. This property plays a key role in our analysis and justifies the classification of the paths of $\mathcal{P}_i$ pairs into the three types.
\begin{lemma}
\label{lem:inter_edge}
Let $P_1=P_{v,e_1},P_2=P_{v,e_2}$ be such that $e_1=(x_1,y_1)$ is above $e_2=(x_2,y_2)$ on $\pi(s,v)$, $\LastE(P_1)\neq \LastE(P_2)$,
$\langle v,e_2\rangle\in \mathcal{P}^{J}_i$ for some $J \in \{A,B\}$ and $i \in \{2, \ldots, \keps\}$.
Then there exist a vertex $t$ and an edge $e' \in \pi(s,t)$ satisfying (see Fig. \ref{fig:lemmainteredge})
\begin{description}
\dnsitem{(a)}
$\langle t,e' \rangle \in \mathcal{P}_{i}$ and hence also $\langle t,e' \rangle \in \mathcal{P}^A_{i-1} \cup \mathcal{P}^B_{i-1}$,
\dnsitem{(b)}
$\LCA(t,v) \in \pi(y_1,x_2)$.
\end{description}
\end{lemma}
\Proof
Let $\mathcal{I}_i(\langle v,e_2 \rangle)=\mathcal{I}^{\not\sim}(\langle v,e_2 \rangle) \cap \mathcal{P}_{i}$.
Since $\langle v,e_2 \rangle \in \mathcal{P}^A_{i}\cup \mathcal{P}^{B}_i$, by Eq. (\ref{eq:typea}) and (\ref{eq:typeb}), we get that
$\mathcal{I}_i(\langle v,e_2 \rangle) \neq \emptyset$ and $\mathcal{I}_i(\langle v,e_2 \rangle) \subseteq \mathcal{P}^A_{i-1} \cup \mathcal{P}^B_{i-1}$.

To identify the path $P'=P_{t,e'}$ where $\langle t,e'\rangle\in \mathcal{I}_i(\langle v,e_2 \rangle)$, consider two cases depending on the type of the path $P_2$ with respect to $\mathcal{P}_{i}$.
\emph{Case 1}: $\langle v,e_2 \rangle \in \mathcal{P}^A_{i}$ (i.e., $P_2$ is of type A).
Let $\langle t,e' \rangle \in \mathcal{I}_i(\langle v,e_2\rangle)$ be such that $P$ $\pi$-intersects $P'=P_{t,e'}$. By Eq. (\ref{eq:typea}) such $\langle t,e' \rangle$ exists.
\emph{Case 2}: $\langle v,e_2 \rangle\in \mathcal{P}^B_{i}$ (i.e., $P_2$ is of type B). Let $P'=P_{t,e'}$ be some type B path for $\langle t,e'\rangle\in\mathcal{I}_i(\langle v,e_2\rangle) \setminus \mathcal{P}^A_{i}$. By Eq. (\ref{eq:typeb}) such a pair $\langle t,e'\rangle$ exists. By the definition of type B, $P_2$ does not $\pi$-intersect $P'$ and vice-versa.
Note that in either case, $P'$ satisfies part (a) of the lemma. To prove part (b), let $w=\LCA(v,t)$.
Since $\langle t,e'\rangle \in \mathcal{I}^{\not\sim}(\langle v,e_2\rangle)$ (i.e., $e' \not\sim e_2$), it holds that $w$ is not below $x_2$. In addition, since $P_1$ and $P_2$ are new-ending $s-v$ paths ending with a distinct edge, by Cl. \ref{cl:dpointbtw}, it holds that $d_2$, the unique divergence point of $P_2$ and $\pi(s,v)$, occurs on the segment $\pi(y_1,x_2)$.

\begin{claim}
\label{cl:rp_internalw}
There exists an $s-v$ replacement-path protecting against $e_2$, $P_3 \subseteq G \setminus \{e_2\}$, whose unique divergence point from $\pi(s,v)$ is not below $w$ (see Fig. \ref{fig:lemmainteredge}).
\end{claim}
\Proof
First consider the case where $\langle v,e_2\rangle \in \mathcal{P}^A_{i}$, see Fig. \ref{fig:lemmainteredge}(a). In this case, by the selection of $P'$, it holds that $P_2$ $\pi$-intersects $P'$.
Let $w' \in (V(\pi(w, t)) \setminus \{w\}) \cap V(P_2)$ and define $P_3=\pi(s,w') \circ P_2[w',v]$. First, observe that $w$ is the unique divergence point of $P_3$ and $\pi(s,v)$ since $P_2[w',v]\subseteq D(P_2)$. Next, observe that
$e_2\notin P_3$. This holds since $\pi(s,w')=\pi(s,w)\circ \pi(w,w')$. Since $e_2 \in \pi(w,v)$ and $E(\pi(w,w'))\cap E(\pi(w,v))=\emptyset$, indeed the failing edge is not on $P_3$.
Finally, by the optimality of the BFS tree $T_0$, $|P_3|=|P_2|$. Hence, the path $P_3$ satisfies the desired property as it diverges from $\pi(s,v)$ at $w$.

It remains to consider the case where $\langle v,e_2\rangle \in \mathcal{P}^{B}_i$.
See Fig. \ref{fig:lemmainteredge}(b). Since both $P_2$ and $P'$ are of type B, $P_2$ does not $\pi$-intersect $P'$ and vice-versa, and hence
\begin{equation}
\label{eq:typeinter}
V(\pi(w, v) \cap P')\setminus \{w\}=\emptyset  \mbox{~~and also~~}  V((\pi(w, t) \cap P_2))\setminus \{w\}=\emptyset~.
\end{equation}
Let $w' \notin \{d(P_2),d(P'),v,t\}$ be a common point of the detours $D(P_2)$ and $D(P')$.
Since $\langle t,e' \rangle \in \mathcal{I}^{\not\sim}(P_2)$, by Eq. (\ref{eq:interfere}), such vertex $w'$ exists.
Let $P_3=P'[s,w']\circ P_2[w',v]$.
We first claim that $P_3$ has a unique divergence point from $\pi(s,v)$ which is not below $w$.
Let $d(P')$ be the unique divergence point of $P'$ from $\pi(s,t)$ (which exists by Cl. \ref{cl:newend_claim}(1)).
Clearly, $P'[s,w']=\pi(s,d(P')) \circ P'[d(P'),w']$.
Since $P'[d(P'),w'] \subseteq D(P')$, it holds that $\left(P'[d(P'),w']\cap \pi(s,w) \right)\setminus\{d(P')\}=\emptyset$. Since $P'$ does not $\pi$-intersect with $P_2$, by Eq. (\ref{eq:typeinter}), it also holds that $\left(P'[d(P'),w']\cap \pi(w,v)\right) \setminus \{d(P'),w\}=\emptyset$, and since $P_2[w',v]\subseteq D(P_2)$, overall it holds that $V(P_3[d(P'),v]) \cap V(\pi(s,v))\setminus \{d(P'),v\}=\emptyset$.

Note that the last point common to $P'$ and $\pi(s,v)$ is not below $w$ and hence the unique divergence point of $P_3$ and $\pi(s,v)$ is not below $w$.
In addition, observe that $P_3 \subseteq G \setminus \{e_2\}$ since $e_2 \in \pi(w, v)$ and $P'$ does not intersect $\pi(w,v)\setminus \{w\}$. It remains to bound the length of $P_3$. By Eq. (\ref{eq:typeinter}), $P'[s,w'], P_2[s,w'] \subseteq G \setminus \{e',e_2\}$, and by the optimality of $P'$ and $P_2$, it holds that $|P'[s,w']|=|P_2[s,w']|$.
The claim follows.
\QED
Since Algorithm \mbox{\tt Pcons} attempt to select the replacement-path whose divergence point is as close to $s$ as possible, (see Cl. \ref{cl:newend_claim}(2)), it holds that $d_2$ is not below $w$. Altogether, $w$ is above $e_2$ but not above $d_2$, implying that $w \in \pi(y_1,x_2)$ as well, thus proving part (b) of the lemma.
\QED
%\def\APPENDFIGKEYLEMMA{
%%%%%%%%%%%%%%%%%%%
%%%%%%%%%%%%%%%%%%%
%}%\APPENDFIGKEYLEMMA

We conclude the analysis of Phase (S1) by showing that $\LastE(P_{v,e}) \in H$ for every $\langle v,e \rangle \in \mathcal{I}_1 \setminus \mathcal{P}^{C}$. The high level idea of the proof is to use Lemma \ref{lem:inter_edge} to show that the existence of at least one $H$-new-ending path $P_{v,e}$ where $\langle v,e \rangle \in \mathcal{I}_1 \setminus \mathcal{P}^{C}$ implies that $T_0$ has expansion at least $n^{\epsilon}$, so after  $O(1/\epsilon)$ steps of expansion, it covers more than $n$ vertices, leading to contradiction.
\begin{lemma}
\label{lem:i1_ana}
$\left(\mathcal{I}_1 \setminus \mathcal{P}^{C} \right) \cap \UncoverPairs(H)=\emptyset$.
\end{lemma}
\Proof
Assume, towards contradiction, that there exists at least one uncovered pair $\langle v,e \rangle \in \mathcal{I}_1 \setminus \mathcal{P}^{C}$ such that  $\LastE(P_{v,e})\notin H$.
Let $\mathcal{O}$ be the collection of all ordered pairs of vertices $(x,y)$ where $x$ is an ancestor of $y$ in $T_0$,
that is, $\mathcal{O}=\{(x,y) \in V \times V ~\mid~ \LCA(x,y)=x\}$.
For every pair of vertices $(x,y) \in \mathcal{O}$ and index $t \in \{1, \ldots, \keps\}$, define the collection of $s-y$ replacement paths in $\mathcal{P}^A_t \cup \mathcal{P}^B_t$ protecting the edges on $\pi(x,y)$ by
\begin{equation}
\label{eq:pathptnsim}
\mathcal{P}_{t}(x,y)=\{\langle y,e \rangle \in \mathcal{P}^A_t \cup \mathcal{P}^B_t ~\mid~ e \in \pi(x,y)\}~.
\end{equation}

The structure of our reasoning is as follows. For every $i \in \{1, \ldots,\keps\}$, we define a collection $\widehat{\Pi}_i \subseteq \mathcal{O}$ of $\Omega(n^{\epsilon(i-1)})$ ordered vertex-pairs $(x,y)$, such that the paths $\pi(x,y)$ are of length $\Omega(n^{\epsilon})$, and
the internal segments of the paths $\pi(x,y)$ and $\pi(x',y')$ are vertex disjoint for every two distinct pairs $(x,y),(x',y') \in  \widehat{\Pi}_i$. Hence, overall, the total number of vertices occupied by the tree-paths connecting the pairs of $\widehat{\Pi}_i$ is $\Omega(n^{\epsilon \cdot i})$.
Solving for $i=\keps$, we get that the graph contains $\Omega(n^{1+\epsilon})$ distinct vertices and hence leading to contradiction.

We next define the sets $\widehat{\Pi}_i$ and show that $\widehat{\Pi}_i$ satisfies the following properties for every $i \in \{1, \ldots, \keps\}$.
\begin{description}
\dnsitem{(Q1)}
$|\widehat{\Pi}_i|=\Omega(n^{\epsilon \cdot (i-1)})$.
\dnsitem{(Q2)}
$T_0(z)$ and $T_0(z')$ are vertex-disjoint,
where $z,z'$ is the second vertex on $\pi(x,y)$ and $\pi(x',y')$ respectively, for every $(x,y),(x',y') \in \widehat{\Pi}_i$ and $T_0(z),T_0(z') \subseteq T_0$ are the subtrees of $T_0$ rooted at $z,z'$ respectively.
\dnsitem{(Q3)}
For every $(x,y) \in \widehat{\Pi}_i$,
$\mathcal{P}_{\keps-i+1}(x,y)$ contains at least $n^{\epsilon}$ pairs whose corresponding $s-y$ replacement paths end with a distinct last edge.
\dnsitem{(Q4)}
$|\pi(x,y)|\geq n^{\epsilon}$ for every $(x,y) \in \widehat{\Pi}_i$.
\end{description}
We now construct $\widehat{\Pi}_i$ inductively and show by induction that it satisfies these properties. For $i=1$, let $\widehat{\Pi}_1=\{(s,v)\}$ where $v$ is the vertex satisfying that there exists an $e \in \pi(s,v)$ such that $\langle v,e \rangle  \in \mathcal{I}_1 \setminus \mathcal{P}^{C}$ and $P_{v,e}$ is $H$-new-ending (i.e., $\langle v,e \rangle \in  \UncoverPairs(H) \cap \left(\mathcal{I}_1 \setminus \mathcal{P}^{C} \right)$.)

Properties (Q1-Q2) hold vacuously as $\widehat{\Pi}_1$ contains (only) one pair. We now verify (Q3), that is, we show that $\mathcal{P}_{\keps}(x,y)$ contains pairs corresponding to at least $n^{\epsilon}$  replacement paths whose last edges are distinct.

Since $\langle v,e \rangle \in \mathcal{I}_1 \setminus \mathcal{P}^C$, it follows that $\langle v,e \rangle \notin \mathcal{P}^C_{\keps}$ and hence $P_{v,e}$ is of type A or B with respect to the collection $\mathcal{P}_{\keps}$.
Let $J \in \{A,B\}$, be such that $\langle v,e \rangle \in  \mathcal{P}^{J}_{\keps}$ (so, $P_{v,e}$ is of type (J)). Recall that $\overrightarrow{\mathcal{P}}^J_{\keps}(v)$ is the collection of all $v$'s pairs in $\mathcal{P}^J_{\keps}$ ordered in increasing distance of their failing edge from $v$.
Recall that the algorithm adds $\lceil n^{\epsilon} \rceil$ distinct last edges of the paths corresponding to the \emph{first} pairs in this ordering into $H$. Hence, by the fact that $\LastE(P_{v,e})$ was not added to $H$, it follows that the paths of the pairs of $\mathcal{P}^J_{\keps}(v)$ end with more than $\lceil n^{\epsilon} \rceil$ distinct edges, hence (Q3) holds.
\par Finally, to see (Q4), observe that the replacement path of every pair $\langle v,e \rangle \in \mathcal{P}_{\keps}(s,v)$ protects a different edge on $\pi(s,v)$, i.e., $e\neq e'$ for every $\langle v,e \rangle,\langle v,e' \rangle \in \mathcal{P}_{\keps}(s,v)$.
In addition, $e_j \in \pi(s,v)$ for every pair $\langle v,e_j\rangle \in \mathcal{P}_{\keps}(s,v)$. Hence,
$|\pi(s,v)|\geq |\mathcal{P}_{\keps}(s,v)| > n^{\epsilon},$ as required, and (Q4) holds.

\begin{figure}[h!]
\begin{center}
\includegraphics[scale=0.35]{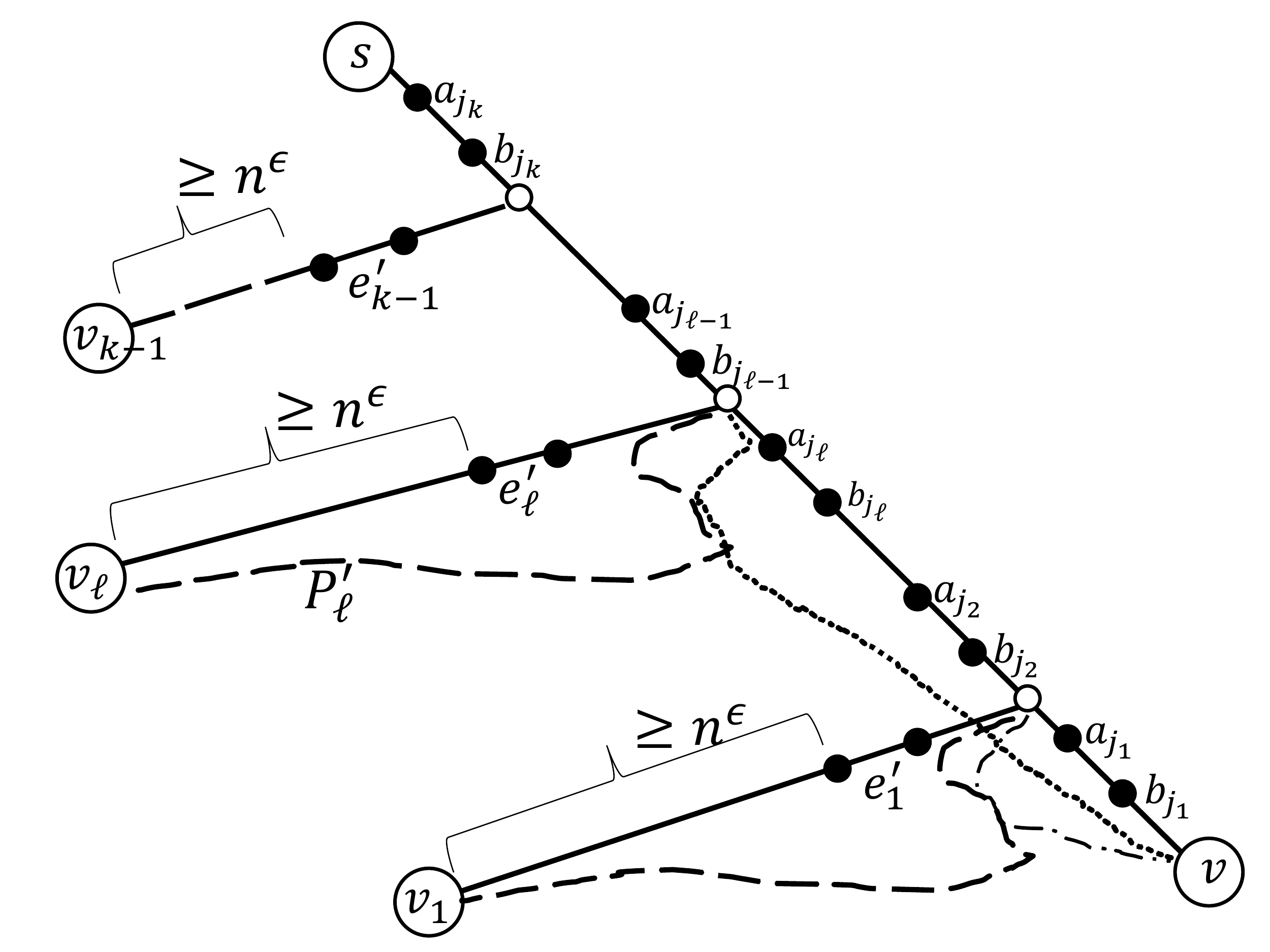}
\caption{ Illustration for Lemma \ref{lem:i1_ana}.
Shown is the construction of $\widehat{\Pi}_2$ where $\widehat{\Pi}_1=\{(s,v)\}$. By the induction base, $\pi(s,v)$ contains at least $n^{\epsilon}$ edges protected by the paths of type A and B whose corresponding pairs are in $\mathcal{P}_{\keps}(s,v)$.
These edges are sorted in increasing distance from $v$. By Lemma
\ref{lem:inter_edge}, between any two consecutive edges $(b_{j_{\ell}},b_{j_{\ell}})$  and $(a_{j_{\ell+1}},b_{j_{\ell+1}})$, there is a least common ancestor $\LCA(v,v_\ell)$. The length of the segment $\pi(\LCA(v,v_\ell),v_\ell)$ is at least $n^{\epsilon}$ since the last edges of the paths protecting the $n^{\epsilon}$ edges closest to $v_\ell$ were taken into $H$ in step 2 of Phase (S1), but the last edge of the path $P_{s,v,e'_\ell}$ was not taken. Hence, after $i=2$, the subtree rooted at $v$ contains $\Omega(n^{2\cdot \epsilon})$ distinct vertices.
\label{fig:i1_ana}}
\end{center}
\end{figure}

For the inductive step, assume that the collection $\widehat{\Pi}_{i-1}$ is given and satisfies (Q1-Q4).
We now describe the construction of $\widehat{\Pi}_i$ and show that it satisfies the properties as well. To do that, every pair $(x,y) \in \widehat{\Pi}_{i-1}$ is used to produce $\Omega(n^{\epsilon})$ new pairs $(x_1,y_1), \ldots, (x_k,y_k)$ for $k\geq \lceil n^{\epsilon} \rceil$ in the following manner.
%By induction assumption, $\widehat{\Pi}_{i-1}$ satisfies property (Q3) and hence $|\mathcal{P}_{\keps-i+2}(x,y)|\geq n^{1/\epsilon}$.
Let $\mathcal{P}^{UN}_{\keps-i+2}(x,y)$ be a maximum collection of replacement \emph{paths} corresponding to the pairs of $\mathcal{P}_{\keps-i+2}(x,y)$ each ending with a distinct last edge.
By the induction assumption for Property (Q3), $|\mathcal{P}^{UN}_{\keps-i+2}(x,y)|\geq n^{\epsilon}$.
Let $E_{i-1}(x,y)=\{ e \in \pi(x,y) ~\mid~ P_{v,e}\in \mathcal{P}^{UN}_{\keps-i+2}(x,y)\}$ be the set of edges on $\pi(x,y)$ protected by the paths of $\mathcal{P}^{UN}_{\keps-i+2}(x,y)$.
Let $\overrightarrow{E}_{i-1}(x,y)=\{e_{j_1}, \ldots, e_{j_k}\}$ be the set of $E_{i-1}(x,y)$ ordered in increasing distance from $y$.
As $|E_{i-1}(x,y)|=|\mathcal{P}^{UN}_{\keps-i+2}(x,y)|$, it follows that
$k \geq \lceil n^{\epsilon} \rceil$.
Let $e_{j_\ell}=(a_{j_\ell},b_{j_\ell})$ for every $\ell \in \{1,\ldots, k\}$.
Since $\LastE(P_1)\neq \LastE(P_2)$ for every $P_1,P_2 \in \mathcal{P}^{UN}_{\keps-i+2}(x,y)$, we can safely apply
Lemma \ref{lem:inter_edge}. By this lemma, for every $\ell \in \{1,\ldots, k-1\}$, there exists a vertex $v_{\ell}$ and an $s-v_{\ell}$ replacement-path $P'_{\ell}=P_{v_{\ell},e'_{\ell}}$ for  $\langle v_{\ell},e'_{\ell} \rangle\in \mathcal{P}^A_{\keps-i+1} \cup \mathcal{P}^B_{\keps-i+1}$ one level up, such that $\LCA(v_{\ell},y)$ is located on $\pi(s,y)$ between the two failing edges $e_{j_\ell}, e_{j_{\ell+1}} \in \pi(x,y)$, that is,
$\LCA(y, v_{\ell}) \in \pi(a_{j_\ell}, b_{j_{\ell+1}})$.
Let $\widehat{\Pi}(x,y)=\{(\LCA(y, v_{\ell}), v_{\ell}) ~\mid~ 1\leq \ell\leq k\}$ and define
$\widehat{\Pi}_{i}=\bigcup \widehat{\Pi}(x,y)$, where the union is over all
$(x,y) \in \widehat{\Pi}_{i-1}$. See Fig. \ref{fig:i1_ana} for an illustration for the case where $i=2$.
%$$\widehat{\Pi}_{i}=\bigcup_{(x,y) \in \widehat{\Pi}_{i-1}} \widehat{\Pi}(x,y).$$
\par We now show that $\widehat{\Pi}_{i}$ satisfies (Q1-Q4).
By induction assumption (Q1) for $i-1$, $\widehat{\Pi}_{i-1}$ contains $\Omega(n^{\epsilon \cdot (i-2)})$ pairs, and by the construction, each pair $(x,y)$ gives raise to a collection of pairs $\widehat{\Pi}(x,y)$ of size $\Omega(n^{\epsilon})$. We now show that these pairs are distinct, and hence the claim holds.
By construction, $z_1 \neq z'_1$ for every $(z_1,z_2),(z'_1,z'_2) \in \widehat{\Pi}(x,y)$, as each such vertex is located between two consecutive failing edges on $\pi(x,y)$.
Note that for every $(z_1,z_2) \in \widehat{\Pi}(x,y)$, it holds that $z_1$ occurs on $\pi(x,y)$ strictly below the first edge on $\pi(x,y)$ (i.e., $z_1$ is between two failing edges on $\pi(x,y)$). Thus $z_1 \in T_0(z)$ where $z$ is the second vertex on $\pi(x,y)$. By Property (Q2) for $\widehat{\Pi}_{i-1}$, also $z_1 \neq z'_1$ for every $(z_1,z_2) \in \widehat{\Pi}(x,y)$ and $(z'_1,z'_2) \in \widehat{\Pi}(x',y')$ for every $(x,y), (x',y') \in \widehat{\Pi}_{i-1}$.
Hence, Property (Q1) holds. By a similar argument, (Q2) holds as well. In particular, by (Q2) for $\widehat{\Pi}_{i-1}$ the claim holds for $(z_1,z_2) \in \widehat{\Pi}(x,y)$ and $(z'_1,z'_2) \in \widehat{\Pi}(x',y')$ corresponding to distinct pairs
$(x,y), (x',y') \in \widehat{\Pi}_{i-1}$. In addition, by the definition of $\widehat{\Pi}(x,y)$, the first vertex $z_1$ of each pair $(z_1,z_2) \in \widehat{\Pi}(x,y)$ is located between two different consecutive edges on $\pi(x,y)$.

We now turn to consider (Q3).
By the construction of $\widehat{\Pi}(x,y)$ for $(x,y) \in \widehat{\Pi}_{i-1}$, the first vertex of every pair $(z_{j_1}, z_{j_2}) \in \widehat{\Pi}(x,y)$ is $z_{j_1}=\LCA(y,z_{j_2})$ and there exists a pair of edges $e \in \pi(x,y)$ and $e' \in \pi(s, z_{j_2})$ satisfying that $e \not\sim e'$ and $\langle z_{j_2},e' \rangle \in \mathcal{P}^A_{\keps-i+1}\cup \mathcal{P}^B_{\keps-i+1}$. Since $e \not\sim e'$, it follows that $e' \in \pi(z_{j_1}, z_{j_2})$ (i.e., $e'$ occurs below $\LCA(y,z_{j_2})$).
Let $J \in \{A,B\}$ be such that $\langle z_{j_2},e' \rangle\in \mathcal{P}_{\keps-i+2} \cap \mathcal{P}^{J}_{\keps-i+1}$.

Recall that the algorithm adds, at the end of step $\keps-i+1$, $\lceil n^{\epsilon} \rceil$ last distinct edges of $s-z_{j_2}$ replacement paths corresponding to the first pairs in the ordered set
$\overrightarrow{\mathcal{P}}^{J}_{\keps-i+1}(z_{j_2})$. Hence, the fact that $\langle z_{j_2},e'\rangle \in \mathcal{P}^{J}_{\keps-i+2}$, implies that the last edge of
$P_{z_{j_2},e'}$ was \emph{not} taken into $H$, and thus there are at least $\lceil n^{\epsilon} \rceil$ pairs such that each their corresponding replacement paths ending with a distinct last edge, and these pairs precede it in the ordering, i.e., their corresponding paths protect edges on $\pi(s,z_{j_2})$ \emph{below} $e'$.
Since all the protected edges of the pairs of $\mathcal{P}_{\keps-i+1}(z_{j_1},z_{j_2})$ belong to the segment $\pi(z_{j_1},z_{j_2})$,
%it holds that $\{P_{z_{j_2}, e_{t_\ell}}~\mid~ 1\leq \ell \leq \lceil n^{\epsilon} \rceil\} \subseteq \mathcal{P}_{\keps-i+1}(z_{j_1},z_{j_2})$,
property (Q3) follows. Finally, since each replacement-path corresponds to a pair in $\mathcal{P}_{\keps-i+1}(z_{j_1},z_{j_2})$ protecting against the failing of a distinct edge on $\pi(z_{j_1},z_{j_2}) $, Property (Q4) holds as well. The induction step holds. 
\par We now complete the proof. By Property (Q1), $\widehat{\Pi}_{\keps}$ contains $\Omega(n^{\epsilon(\keps-1)})$ pairs. By (Q2), the paths $\pi(x,y)$ and $\pi(x',y')$ are vertex-disjoint for every $(x,y), (x',y') \in \widehat{\Pi}_{\keps}$. By (Q4), the length of each path $\pi(x,y)$ is $\Omega(n^{\epsilon})$ for every $(x,y) \in \widehat{\Pi}_{\keps}$, so overall there are $\Omega(n^{\epsilon\cdot \keps})=\Omega(n^{1+\epsilon})$ vertices in these paths, in contradiction to the fact that the number of vertices in $T_0$ is bounded by $n$. The claim follows.
\QED

\subsubsection{Analysis of Phase (S2)}
\label{subsec:ana2}
We begin by showing the following.
%$(\sim)$-set.
\begin{observation}
\label{obs:typec_sim}
$\mathcal{P}^{C}_{i}$ is a $(\sim)$-set for every $i \in \{1, \ldots, \keps\}$.
\end{observation}
\Proof
To prove this, we consider two pairs $\langle v,e\rangle, \langle t,e'\rangle$ in $\mathcal{P}^{C}_{i}$  for some $i \in \{1, \ldots, \widehat{\epsilon}\}$ and show that $\langle t,e'\rangle \notin \mathcal{I}^{\not\sim}(\langle v,e\rangle)$.
By definition, $\langle v,e\rangle, \langle t,e'\rangle \in \mathcal{P}^C_i\subseteq \mathcal{P}_i$.
If $\mathcal{I}^{\not\sim}(\langle v,e\rangle) \cap \mathcal{P}_i=\emptyset$ then the claim holds vacuously. So consider the remaining case. Since $P_1=P_{v,e}$ is of type C with respect to $\mathcal{P}_i$, Eq. (\ref{eq:typea}) and (\ref{eq:typeb}) imply that $\mathcal{I}^{\not\sim}(\langle v,e\rangle) \cap \mathcal{P}_i \subseteq \mathcal{P}^{A}_{i}$.  Since $\langle t,e'\rangle \notin  \mathcal{P}^{A}_{i}$, it holds that $\langle t,e'\rangle \notin \mathcal{I}^{\not\sim}(\langle v,e\rangle) \cap \mathcal{P}_i$, and as $\langle t,e'\rangle \in \mathcal{P}_i$, we conclude that $\langle t,e'\rangle \notin \mathcal{I}^{\not\sim}(\langle v,e\rangle)$ (by symmetry, $\langle v,e\rangle \notin \mathcal{I}^{\not\sim}(\langle t,e'\rangle)$ holds as well).  The observation follows.
\QED
%We begin by proving Obs. \ref{obs:typec_sim}, which claims that every subset $\mathcal{P}^{C}_{i}$ is a
%$(\sim)$-set.
%
%The following definitions are useful in our reasoning.
Recall that $E^{-}(\mathcal{TD})$ is the collection of glue edges, namely, $T_0$ edges that do not appear by the paths $\psi$ of the tree-decomposition $\mathcal{TD}$. Sub-Phase (S2.2) and Obs. \ref{obs:lastedge} imply:
%, we have the following.
\begin{claim}
\label{obs:tree_uncov}
Every glue edge $e \in E^{-}(\mathcal{TD})$ is protected by $H$.
\end{claim}
Hence, it remains to bound the number of unprotected edges on the paths of $\mathcal{TD}$. The following definitions are useful in our reasoning. For a vertex
$v$ and an $s_{\psi}-t_{\psi}$  path $\psi \in \mathcal{TD}$, define
\begin{eqnarray*}
\mathcal{P}_{miss}(\psi,v)&=&\{\langle v,e \rangle \in \mathcal{P}_{miss} ~\mid~ e \in \pi(s,v) \cap \psi\},
\end{eqnarray*}
as the set of uncovered pairs in $H$ that belong to $\mathcal{P}$ and whose corresponding replacement paths protect against the failure of the edges on $\pi(s,v) \cap \psi$.
Let $E_{miss}(\mathcal{P}, \psi,v)=\{e ~\mid~ \langle v,e \rangle \in \mathcal{P}_{miss}(\psi,v)\}$ be the corresponding last-unprotected edges by $H$ on $\pi(s,v) \cap \psi$ and let
$E_{miss}(\mathcal{P}, \psi)=\bigcup_{v \in V} E_{miss}(\mathcal{P}, \psi,v)$. By Cl. \ref{obs:tree_uncov},
$$E_{miss}(\mathcal{P})=\bigcup_{\psi \in \mathcal{TD}}E_{miss}(\mathcal{P},\psi).$$

Note that the replacement paths of the pairs of $\mathcal{P}_{miss}(\psi,v)$ may end with the same last edge.
We now identify a set $\mathcal{P}^{\sf UN}_{miss}(\mathcal{P}, \psi,v)$
of unique representatives for each last edge as follows.
For every edge $e'$ that has several replacement paths $P_{v,e}$ for  $\langle v,e \rangle \in \mathcal{P}_{miss}(\psi,v)$ whose last edge $\LastE(P_{v,e})=e'$, we pick one representative pair  $\langle v,e^* \rangle$ corresponding to the path $P_{v,e^*}$ whose failing edge $e^*$ is closest to $s$ among all other candidates. Formally, let $\mathcal{LE}_{miss}(\mathcal{P}, \psi,v)=\{\LastE(P_{v,e})~\mid~ \langle v,e \rangle \in \mathcal{P}_{miss}(\psi,v)\}$ be the last edges of the replacement paths of the pairs in $\mathcal{P}_{miss}(\psi,v)$.
For every $e' \in \mathcal{LE}_{miss}(\mathcal{P}, \psi,v)$, let $\mathcal{P}(e',\mathcal{P}, \psi,v)=\{
\langle v,e \rangle \in \mathcal{P}_{miss}(\psi,v) ~\mid~ \LastE(P_{v,e})=e'\}$.
The representative pair for the last edge $e'$ denoted by $\widehat{P}(e')=\langle v,e^* \rangle$ for $\langle v,e^* \rangle\in \mathcal{P}(e',\mathcal{P}, \psi,v)$ satisfying that  $\dist(s,e^*,\pi(s,v))<\dist(s,e'',\pi(s,v))$ for every $e'' \neq e^*$ and $\langle v,e''\rangle\in \mathcal{P}(e',\mathcal{P}, \psi,v)$.
Finally, define
$\mathcal{P}^{\sf UN}_{miss}(\mathcal{P}, \psi,v)=\{\widehat{P}(e') ~\mid~ e' \in \mathcal{LE}_{miss}(\mathcal{P}, \psi,v)\}$ and
$$E^{\sf UN}_{miss}(\mathcal{P}, \psi,v)=\{e_j ~\mid~ \langle v,e_{j} \rangle \in \mathcal{P}^{\sf UN}_{miss}(\mathcal{P}, \psi,v)\}.$$
We proceed
by showing that $E^{\sf UN}_{miss}(\mathcal{P}, \psi,v)$ is either \emph{empty} or sufficiently \emph{large}.
\begin{lemma}
\label{lem:emiss_sufflarge}
For every $v \in V$ and every  $\psi \in \mathcal{TD}$, if $E^{\sf UN}_{miss}(\mathcal{P}, \psi,v) \neq \emptyset$,
then:\\
(a) $|E^{\sf UN}_{miss}(\mathcal{P}, \psi,v)|\geq \lceil n^{\epsilon}\rceil$, and
(b) the first $\lceil n^{\epsilon}\rceil$ edges in $E^{\sf UN}_{miss}(\mathcal{P}, \psi,v)$ are contained in $\pi_{j^*}(s,v)\subseteq \pi(s,v)$, which is the highest \emph{heavy} subsegment that intersects $\psi$ with respect to $\mathcal{P}$.
\end{lemma}
\Proof
Recall that in Sub-Phase (S2.1), the $s-v$ shortest-path $\pi(s, v)$ was partitioned into
$k'=\lfloor \log|\pi(s,v)| \rfloor$ segments where the $j$'th segment $\pi_j(s, v)$ is given by $\pi_j(s, v) = \pi(u_{i_{j-1}},u_{i_j})$
for every $j  \in \{ 1, \ldots, k'\}$.
Let $e_1 \in \pi(s,v)$ be the closest edge to $s$ in $E_{miss}(\mathcal{P}, \psi,v)$ and $e_1 \in E^{\sf UN}_{miss}(\mathcal{P}, \psi,v)$. Since $e_1 \in E_{miss}(\mathcal{P}, \psi,v)$, it implies that $e_1$ belongs to a heavy subsegment and in particular, $e_1 \in \pi_{j^*}(s,v)$.
First, consider the case where this
subsegment is fully contained in $\psi$ , i.e., that $\pi_{j^*}(s,v) \subseteq \psi$. Recall that $\mathcal{P}_{j^*}(v)=\{
\langle v,e \rangle \in \mathcal{P} ~\mid~ e \in \pi_{j^*}(s,v)\}$
is the collection of $s-v$ replacement paths in $\mathcal{P}$ that protect the edges on $\pi_{j^*}(s,v)$ and $\mathcal{LE}_{j^*}(v)$ is the corresponding last edges of these paths. Since $e_1 \in E_{miss}(\mathcal{P}, \psi,v)$,  it follows that $\pi_{j^*}(s,v)$ is heavy with respect to $\mathcal{P}$, i.e., that $|\mathcal{LE}_{j^*}(v)|
>\lceil n^{\epsilon}\rceil$. Hence, the pairs of $\mathcal{P}_{j^*}(v)$
correspond to at least $\lceil n^{\epsilon}\rceil$ replacement paths that end with distinct last edges, implying that $|E^{\sf UN}_{miss}(\mathcal{P}, \psi,v)|\geq \lceil n^{\epsilon}\rceil$.

%Since $\mathcal{P}_j(v) \subseteq \mathcal{P}_{miss}(\psi,v)$, it holds that $$|E_{miss}(\mathcal{P}, \psi,v)|=|\mathcal{P}_{miss}(\psi,v)|=\Omega(n^{\epsilon}).$$

Next, consider the complementary
case where $\pi_{j^*}(s,v) \nsubseteq \psi$ . This implies that $e_1$ belongs to the subsegment $\pi_U(\psi,v)$ or $\pi_L(\psi,v)$ that intersects with $\psi$. Assume first that $e_1 \in \pi_U(\psi,v)$. Since $e_1 \in E_{miss}(\mathcal{P}, \psi,v)$, it follows that
$|\mathcal{LE}_{U}(\mathcal{P},\psi,v)|\geq \lceil n^{\epsilon}\rceil$ and hence $\mathcal{P}_{U}(\psi,v)$ contains pairs corresponding to at least $\lceil n^{\epsilon}\rceil$ replacement paths that end with a distinct last edge.
The case where $e_1 \in \pi_L(\psi,v)$ is analogous.
\QED
Note that if $E^{\sf UN}_{miss}(\mathcal{P}, \psi,v)=\emptyset$, then also $E_{miss}(\mathcal{P}, \psi,v)=\emptyset$, so this set needs not concern us anymore.  Hence hereafter we concentrate on vertices $v$
with a large set $E^{\sf UN}_{miss}(\mathcal{P}, \psi,v)$. For such a vertex $v$, let $$\overrightarrow{E}^{\sf UN}_{miss}(\mathcal{P}, \psi,v)=\{e_{i_1}, \ldots, e_{i_\ell}\}$$ be the edges of $E^{\sf UN}_{miss}(\mathcal{P}, \psi,v)$ ordered in \emph{increasing} distance from $s$. By Lemma \ref{lem:emiss_sufflarge}, $\ell \geq \lceil n^{\epsilon}\rceil$.
Define $\mathcal{D}(\mathcal{P},\psi,v)=\{D(P_{v,e_{i_j}}) ~\mid~ j \in \{1, \ldots, \lceil n^{\epsilon}\rceil\}\}$ as the collection of the detours protecting against the failure of the first $\lceil n^{\epsilon}\rceil$ ordered edges in the ordering $\overrightarrow{E}^{\sf UN}_{miss}(\mathcal{P}, \psi,v)$. Note that by the definition of $E^{\sf UN}_{miss}(\mathcal{P}, \psi,v)$, each of the detours in $\mathcal{D}(\mathcal{P},\psi,v)$ ends with a distinct last edge (in particular, by Cl. \ref{cl:detourvertex_disjoint}, these detours are vertex disjoint, except for the terminal $v$).

\begin{figure}[h!]
\begin{center}
\includegraphics[scale=0.35]{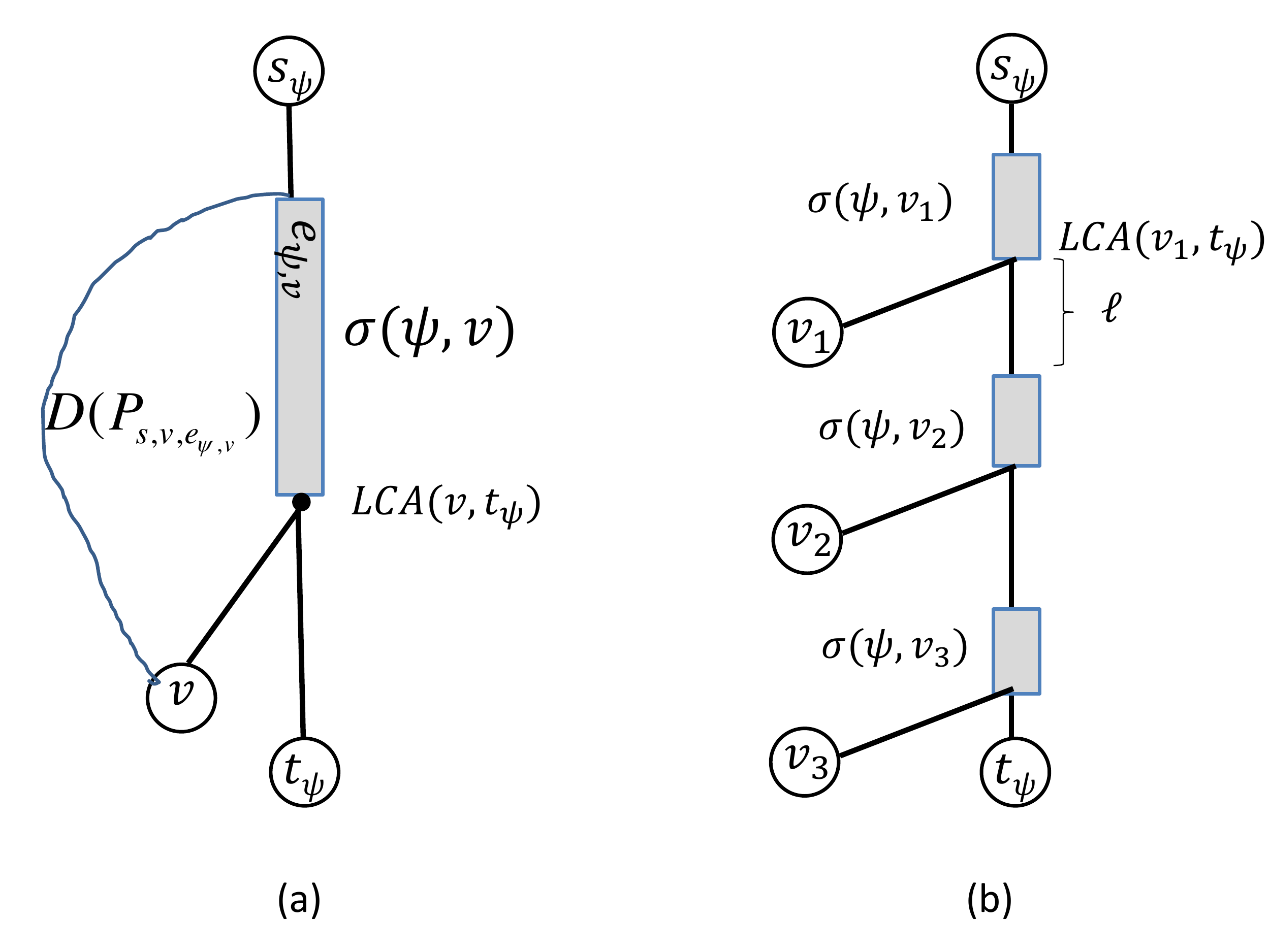}
\caption{ Illustration of the segments $\sigma(\mathcal{P},\psi , v)$.
For brevity, we omit $\mathcal{P}$ and simply write $\sigma(\psi , v)$.
Shown is an $s_{\psi}-t_{\psi}$  path  $\psi \in \mathcal{TD}$ and (a) a segment $\sigma(\psi , v)$ defined by the divergence point of the replacement-path $P_{v,e^*(\mathcal{P},\psi,v)}$ protecting
the highest unprotected edge on $\pi(s, v) \cap \psi$ and the LCA of $v$ and $t_{\psi}$; (b) the independent
segments $\mathcal{SG}_{IS}(\mathcal{P},\psi)$, where
$\ell=\max\{ |\sigma(\psi , v_1)|,|\sigma(\psi , v_2)|\}$
 is the minimum spacing between
two segments.
\label{fig:segment}}
\end{center}
\end{figure}

In addition, for a vertex $v$ with a large set $E^{\sf UN}_{miss}(\mathcal{P}, \psi,v)$, Let $e^*(\mathcal{P},\psi,v) \in E_{miss}(\mathcal{P}, \psi,v)$ be the closest edge to $s$ on $\pi(s,v) \cap \psi$ among all edges in $E_{miss}(\mathcal{P}, \psi,v)$. Hence, $e^*(\mathcal{P},\psi,v) \in E^{\sf UN}_{miss}(\mathcal{P}, \psi,v)$.
Note that by the end of the Sub-Phases (S2.2.1-2) and by Cl. \ref{cl:dpointbtw}, the divergence point of $P_{v,e^*(\mathcal{P},\psi,v)}$ from $\pi(s,v)$ must occur on $\psi$, i.e.,
$d(P_{v,e^*(\mathcal{P},\psi,v)}) \in \pi(s,v) \cap \psi$. (This is because  the last edges of the new ending paths protecting the first failing edges on each subsegment $\pi_j(s,v)$ and the intersected segments $\pi_U(s,v), \pi_L(s,v)$ were added into $H$, the divergence point of the $H$-new-ending paths protecting the other edges on these segments are internal to their segments.)
Define the segments
$$\sigma(\mathcal{P},\psi,v)=\pi(d(P_{v,e^*(\mathcal{P},\psi,v)}), \LCA(v,t_{\psi})),$$
and the segment collection
\begin{equation}
\label{eq:lca_seg}
\mathcal{SG}(\mathcal{P},v)=\{\sigma(\mathcal{P},\psi,v) ~\mid~ E_{miss}(\mathcal{P}, \psi,v) \neq \emptyset\}~,
\end{equation}
where $\psi$ is an $s_{\psi}-t_{\psi}$ path, for illustration see Fig. \ref{fig:segment}.
%\def\APPENDSEGMENT{
%%%%%%%%%%%%%%%%%%%
%%%%%%%%%%%%%%%%%%%
%}%\APPENDSEGMENT

We next claim that each of the detours of $\mathcal{D}(\mathcal{P},\psi,v)$ is sufficiently long.
\begin{lemma}
\label{lem:suff_long}
If $\mathcal{D}(\mathcal{P},\psi,v)$ is nonempty, then $|D_i|\geq |\sigma(\mathcal{P},\psi,v)|/4$ for every $D_i \in \mathcal{D}(\mathcal{P},\psi,v)$.
\end{lemma}
\Proof
By Cl. \ref{lem:emiss_sufflarge}(b),
the first $\lceil n^{\epsilon}\rceil$ edges in $E^{\sf UN}_{miss}(\mathcal{P}, \psi,v)$ are contained in $\pi_{j^*}(s,v)\subseteq \pi(s,v)$, which is the highest \emph{heavy} subsegment with respect to $\mathcal{P}$ and $v$ that intersects $\pi(s,v)$. Hence letting $d^*=d(P_{v,e^*(\mathcal{P},\psi,v)})$, we get that $d^*$ occurs on $\pi_{j^*}(s,v)$  and
the detours of a nonempty set $\mathcal{D}(\mathcal{P},\psi,v)$ protect the failing of edges on $\pi_{j^*}(s,v)$.

By the right inequality of Eq. (\ref{eq:jseg}), for every $D_i \in \mathcal{D}(\mathcal{P},\psi,v)$,
\begin{eqnarray}
\label{eq:det_leng_int}
|V(D_i)|&\geq& \sum_{j' >j^*} |\pi_{j'}(s,v)|\geq |\pi_{j^*}(s,v)|/2 \geq \sum_{j' \geq j^*} |\pi_{j'}(s,v)|/4 \nonumber
\\&\geq &
|\pi(d^*,v)|/4 =|\sigma(\mathcal{P},\psi,v)|/4~.
\end{eqnarray}
The lemma follows.
\QED
%}%\APPENDSUFFLODET
\begin{observation}
\label{obs:emiss_in}
$E_{miss}(\mathcal{P},\psi)\subseteq \bigcup_{\sigma \in \mathcal{SG}(\mathcal{P},\psi)} \sigma$.
\end{observation}
\Proof
Assume towards contradiction that there exists an edge $e \in E_{miss}(\mathcal{P},\psi)\setminus  \bigcup_{\sigma \in \mathcal{SG}(\mathcal{P},\psi)} \sigma$.
Let $v \in V$ be such that $\LastE(P_{v,e})\notin H$, i.e., $e \in  E_{miss}(\mathcal{P},\psi,v)$.

\begin{figure}[h!]
\begin{center}
\includegraphics[scale=0.35]{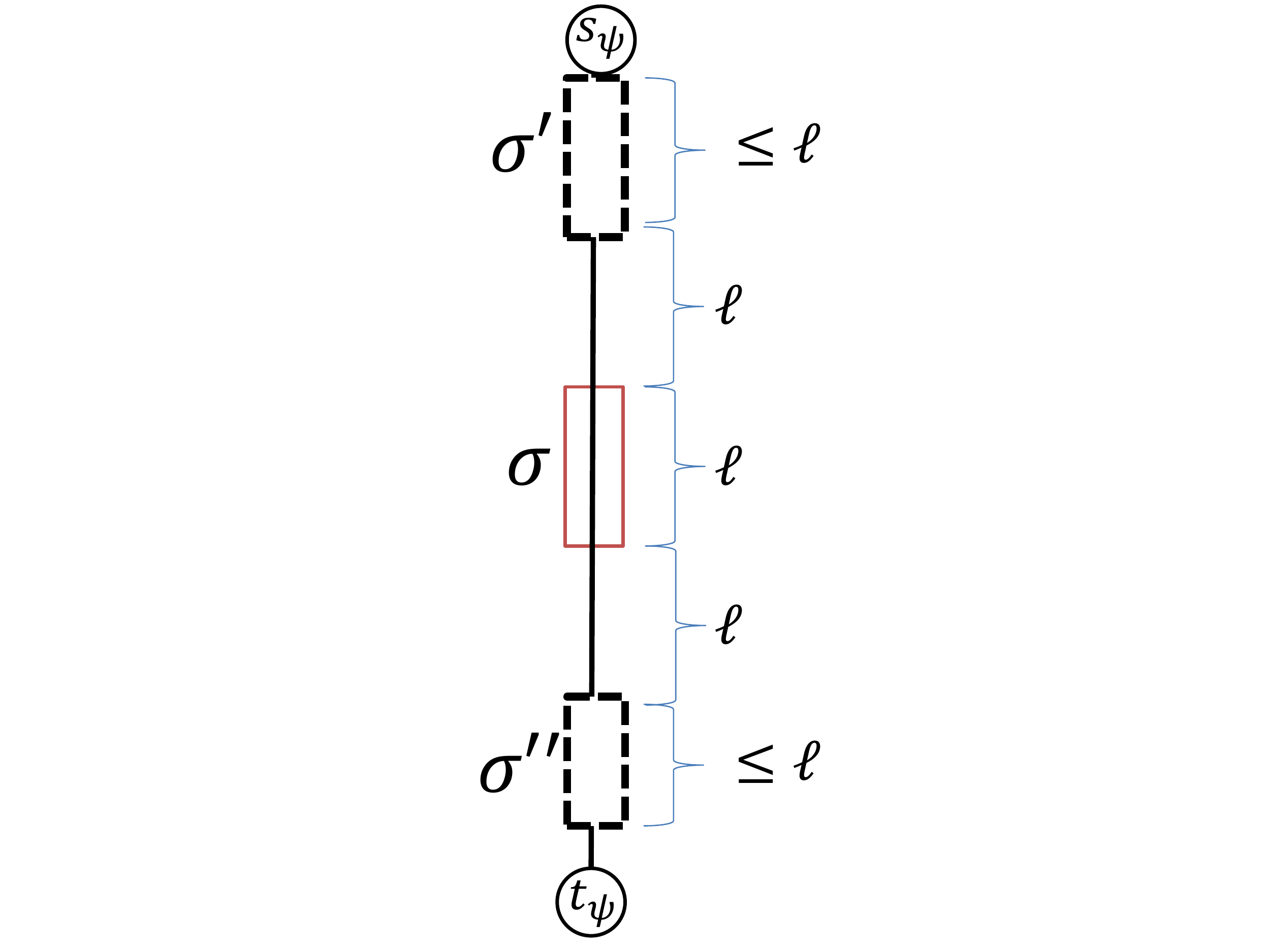}
\caption{ Illustration for Claim \ref{obs:is_many}.
Shown is an $s_{\psi}-t_{\psi}$ path $\psi \in \mathcal{TD}$ and the segment $\sigma$ of length $|\sigma|=\ell$ that was taken into $\mathcal{SG}_{IS}(\mathcal{P},\psi)$. The dependent segments $\Delta(\sigma)$ that were discarded from $\mathcal{SG}'(\mathcal{P},\psi)$ when adding $\sigma$ occupied at most $5|\sigma|$ vertices from $\psi$. To see this, observe that in the most extreme scenario, $\sigma'$ and $\sigma''$ shown in the figure are in $\Delta(\sigma)$. By the maximality of the length of $\sigma$ at the time it was taken into $\mathcal{SG}_{IS}(\mathcal{P},\psi)$, it holds that $|\sigma'|,|\sigma''|\leq \ell$.
\label{fig:claim5fig}}
\end{center}
\end{figure}

Recall that $e^*(\mathcal{P},\psi,v)$ is the
closest edge to $s$ in $E_{miss}(\mathcal{P},\psi,v)$, hence
$e^*(\mathcal{P},\psi,v) \in E^{\sf UN}_{miss}(\mathcal{P},\psi,v)$,
and $e$ is not above $e^*(\mathcal{P},\psi,v)$ on $\pi(s, v)$. In addition, since $e \in \pi(s,v) \cap \psi$, it holds that $e$ is above $LCA(v, t_{\psi})$. Altogether, we get that $e \in \sigma(\mathcal{P},\psi,v)$, contradiction. The observation follows.
\QED
%}%\APPENDEMISSIN
From now on, we focus on a particular path $\psi$  in the tree decomposition $\mathcal{TD}$.
We proceed by defining a notion of independence between two segments $\sigma_i = \sigma(\mathcal{P}, \psi, v_i)$ and $\sigma_j = \sigma(\mathcal{P}, \psi, v_j)$ in
$\mathcal{SG}(\mathcal{P},\psi)$ (see Eq. (\ref{eq:lca_seg}) for the definition of $\mathcal{SG}(\mathcal{P},\psi)$).
Let $x_i$ (resp. $x_j$) be the first vertex of $\sigma_i$ (resp., $\sigma_j$) and let $y_i = \LCA(v_i, t_{\psi})$ (resp.,
$y_j = \LCA(v_j, t_{\psi})$) be the last vertex of $\sigma_i,\sigma_j$.
%\textbf{MP: the notion of independence is not
%good here, because we already have this notion for detours. In addition, in this
%context, segments in $\mathcal{SG}(\mathcal{P},\psi)$ are independent if they are sufficiently distant on the path $\psi$.
%Maybe we should call it ``distant segments"? or "distant independent segments"?}
\begin{definition}[Independent Segments]
Let $\sigma_i=\pi(x_i, y_i),\sigma_j=\pi(x_j, y_j)\in \mathcal{SG}(\mathcal{P},\psi)$ be such that $\dist(s,x_i,G) \leq  \dist(s,x_j,G)$ and let $\ell=\max\{|\sigma_i|,|\sigma_j|\}$.
Then, $\sigma_i$ and $\sigma_j$ are \emph{independent} if $\dist(s,x_j,G)-\dist(s,y_i,G)\geq \ell$, otherwise they are \emph{dependent}.
\end{definition}
%For every segment $\sigma(\mathcal{P},\psi,v)$, define its value by
%$$\Val(\sigma(\mathcal{P},\psi,v))=|\sigma(\mathcal{P},\psi,v)|.$$
By Lemma \ref{lem:emiss_sufflarge}, we have the following.
\begin{observation}
\label{obs:valnesp}
For a vertex $v$ with $E^{\sf UN}_{miss}(\mathcal{P}, \psi,v)\neq \emptyset$, we have
$|\sigma(\mathcal{P},\psi,v)|=\Omega(n^{\epsilon})$.
\end{observation}
Set $\mathcal{SG}'(\mathcal{P},\psi)\gets \mathcal{SG}(\mathcal{P},\psi)$. We now compute a collection of maximal weighted independent
set $\mathcal{SG}_{IS}(\mathcal{P},\psi)$ greedily by adding to $\mathcal{SG}_{IS}(\mathcal{P},\psi)$ at each step the segment $\sigma(\mathcal{P},\psi,v) \in \mathcal{SG}'(\mathcal{P},\psi)$ whose length is maximal among all remaining segments $\mathcal{SG}'(\mathcal{P},\psi)$ and removing from it the segments $\sigma(\mathcal{P},\psi,v'')$ that are \emph{dependent} with $\sigma(\mathcal{P},\psi,v)$. For illustration see Fig. \ref{fig:segment}(b).
The next observation shows that the total length of the independent set $\mathcal{SG}_{IS}(\mathcal{P},\psi)$ is of the same order as the original set $\mathcal{SG}(\mathcal{P},\psi)$.
\begin{claim}
\label{obs:is_many}
$\displaystyle \sum_{\sigma \in \mathcal{SG}_{IS}(\mathcal{P},\psi)}|\sigma|\geq |E_{miss}(\mathcal{P},\psi)|/5$ (see Fig. \ref{fig:claim5fig}).
%\begin{equation*}
%\sum_{\sigma \in \mathcal{SG}_{IS}(\mathcal{P},\psi)}\Val(\sigma)\geq |E_{miss}(\mathcal{P},\psi)|/6~.
%\end{equation*}
\end{claim}
\Proof
Note that by the definition of independence, $\sigma \cap \sigma'=\emptyset$ for every $\sigma,\sigma' \in \mathcal{SG}_{IS}(\mathcal{P},\psi)$.
By Obs. \ref{obs:emiss_in}, $|\bigcup_{\sigma \in \mathcal{SG}(\mathcal{P},\psi)} \sigma|\geq |E_{miss}(\mathcal{P},\psi)|$.
Let $\Delta(\sigma)$ be the collection of segments discarded from $\mathcal{SG}'(\mathcal{P},\psi)$ before adding $\sigma$ into $\mathcal{SG}_{IS}(\mathcal{P},\psi)$. We now show that
$$\left|\bigcup_{\sigma' \in \Delta(\sigma)} \sigma' \right|~~\leq~~ 5 \cdot |\sigma|.$$
Note that every $\sigma'$ in $\Delta(\sigma)$ is dependent with respect to $\sigma$. The inequality follows by the maximality of $\sigma$ in $\Delta(\sigma)$ and by the definition of independence. The claim follows by summing over all independent segments $\sigma$ in $\mathcal{SG}_{IS}(\mathcal{P},\psi)$.
\QED
%}%\APPENDISMANY

We next show the following (for the given $\psi$ and $\mathcal{P}$).
\begin{lemma}
\label{lem:interimsig}
$n^{\epsilon} \cdot \sum_{\sigma \in \mathcal{SG}_{IS}(\mathcal{P},\psi)}|\sigma|\leq |\bigcup_{P\in\mathcal{P}_{miss}(\mathcal{P},\psi)} V(D(P))|$.
\end{lemma}
To prove the lemma, we consider an iterative process on the set
$VS=\{v ~\mid~ \sigma(\mathcal{P},\psi,v) \in \mathcal{SG}_{IS}(\mathcal{P},\psi)\}$, the set of vertices whose segment is in the independent set $\mathcal{SG}_{IS}(\mathcal{P},\psi)$. In this process, the detours
of these vertices $v'$ are added in decreasing distance of $\LCA(v',t_{\psi})$ and $s$. (Note that the
order is strictly decreasing since the segments are independent and hence also vertex disjoint.)
Formally, let $\overrightarrow{VS}=\{v_{i_1}, \ldots, v_{i_k}\}$
be the collection of $VS$ vertices sorted in decreasing distance of $\LCA(v_{i_j},t_{\psi})$ and $s$, i.e., $\dist(s, \LCA(v_{i_1},t_{\psi}))>\ldots>\dist(s, \LCA(v_{i_k},t_{\psi}))$.
%(Note that since the segments of $\mathcal{SG}_{IS}(\mathcal{P},\psi)$ are independent, every vertex has at most \emph{one}
%segment in this set.)
Starting with $G'_1=\emptyset$, at step $\tau\geq 1$, let
$$G'_{\tau+1}=G'_{\tau}\cup \bigcup_{D' \in \mathcal{D}(\mathcal{P},\psi, v_{i_{\tau}})}D'~.$$
Let $\widehat{G}=G'_k$ be the final subgraph. Hence,
\begin{equation}
\label{eq:final_det_graph}
\widehat{G} \subseteq \bigcup_{\langle v,e \rangle \in \mathcal{P}} D(P_{v,e})~.
\end{equation}
\begin{lemma}
\label{lemma:grad_addition}
For every $\tau \in \{1, \ldots, k\}$,
$|V(G'_\tau)\setminus V(G'_{\tau-1})|\geq n^{\epsilon}/4 \cdot |\sigma(\mathcal{P},\psi, v_{i_\tau}))|$.
\end{lemma}
\Proof
Consider a specific detour $\widetilde{D} \in \mathcal{D}(\mathcal{P},\psi,v_{i_\tau})$ of a path $P_{v_{i_\tau},e}$ for $e \in E_{miss}(\mathcal{P}, \psi,v_{i_\tau})$.
First note that since each of the detours in $\mathcal{D}(\mathcal{P},\psi,v_{i_\tau})$ ends with a distinct edge, by Cl. \ref{cl:detourvertex_disjoint}(2), these detours are vertex disjoint (besides the common endpoint $v_{i_\tau}$). We distinguish between two cases.
Case (1): the internal segment of the detour $\widetilde{D}$ does not intersect with the vertices of $G'_{\tau-1}$.
In this case, by Lemma \ref{lem:suff_long},
$|V(\widetilde{D})|\geq |\sigma(\mathcal{P},\psi, v_{i_\tau})|/4$.
\\Case (2): the internal segment of $\widetilde{D}$ intersects at least one vertex of $G'_{\tau-1}$. Let $w$ be the first internal vertex of
$\widetilde{D}$ that occurs in some $D'$, i.e., $w \in V(\widetilde{D})\setminus \{d,v_{i_\tau}\}$ where $d$ (resp., $v_{i_\tau}$) is the first (resp., last) vertex of $\widetilde{D}$.
That is, $\widetilde{D}[d,w]\setminus \{w,d\}$ is vertex disjoint with $G'_{\tau-1}$.
%(Recall that all replacement paths of the pairs in $\mathcal{P}$ are new-ending and hence the detour collides with the shortest-path $\pi(s,v_{i_{\tau}})$ at the last  vertex $v_{i_{\tau}}$, see Obs. \ref{obs:detour_exist}.)
Note that by the definition of the segments, $d \in \sigma(\mathcal{P},\psi, v_{i_\tau})$.
We now show that $|\widetilde{D}(d,w)|\geq |\sigma(\mathcal{P},\psi,v_{i_\tau})|$.
By Cl. \ref{cl:detourvertex_disjoint}(2) and by the ordering $\overrightarrow{VS}$, $D'$ is a detour of some $P_{v_{i_j},e'}$ path for $v_{i_j}$ such that $j<\tau$.
Let $\sigma_\tau=\sigma(\mathcal{P},\psi, v_{i_\tau})=\pi(x_\tau,y_\tau)$ and
$\sigma_j=\sigma(\mathcal{P},\psi, v_{i_j})=\pi(x_j,y_j)$ (this definition is consistent, as the segments $\sigma$ are subpaths on the tree $T_0$).

\begin{figure}[h!]
\begin{center}
\includegraphics[scale=0.35]{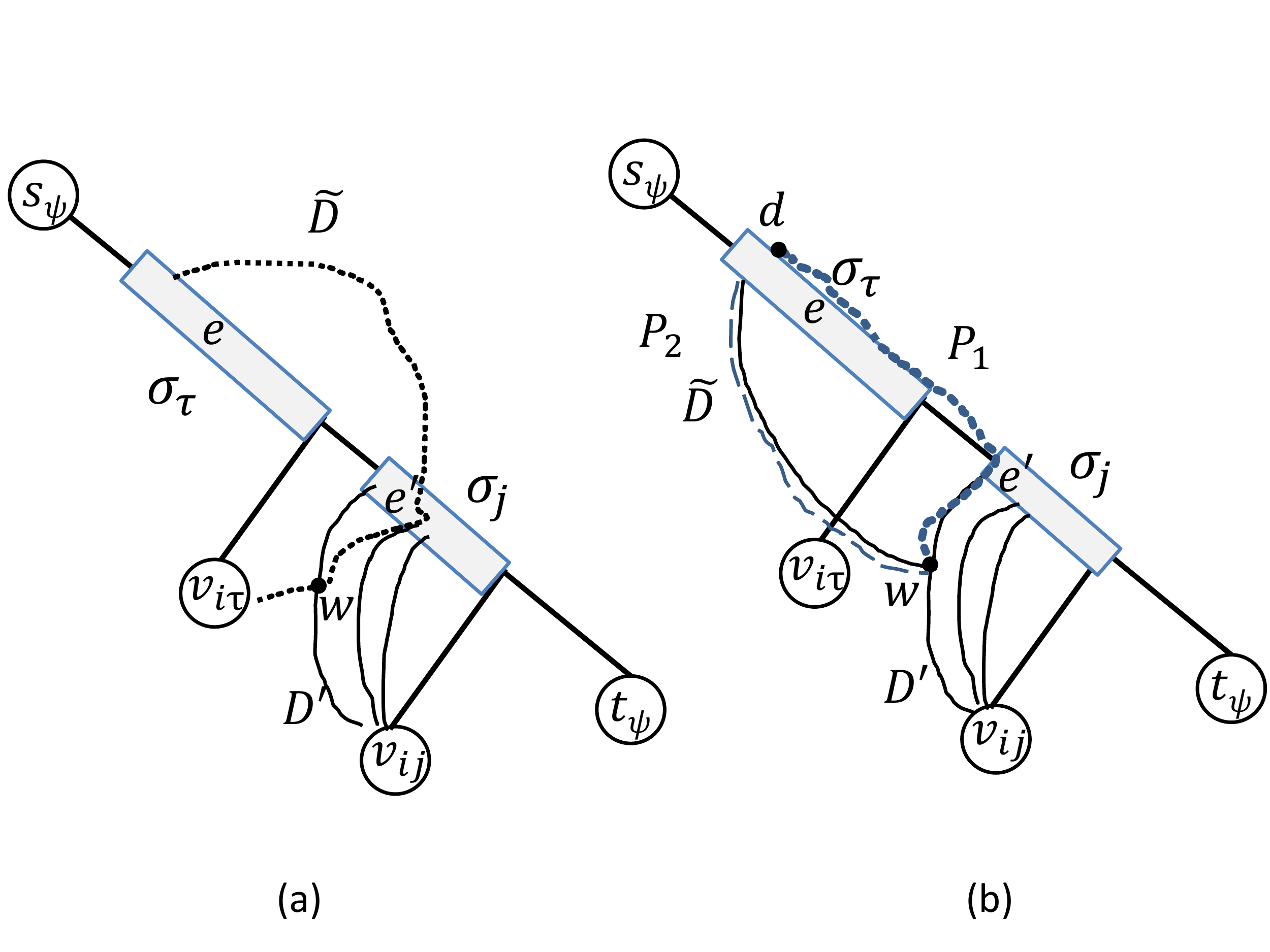}
\caption{ Illustration for Lemma \ref{lemma:grad_addition}.
Shown is an $s_{\psi}-t_{\psi}$ path $\psi \in \mathcal{TD}$ and two independent
segments $\sigma_\tau, \sigma_j$ in $\mathcal{SG}_{IS}(\mathcal{P},\psi)$. The detour $\widetilde{D}$ was added to $\widehat{G}$ at step $\tau$ and it intersects $D'$, which was added to $\widehat{G}$ at step $j<\tau$.
(a) The detour $\widetilde{D}$
traverses $e'$, the failing edge protected
by $D'$, before the first mutual vertex $w$.
(b) $e' \notin \widetilde{D}[d,w]$
(where $d$ is the first vertex of $D$).
\label{fig:lemmafig}}
\end{center}
\end{figure}

By the ordering of the insertion into $\widehat{G}$, it holds that
$\dist(s, y_\tau,G)<\dist(s, x_j,G)$. In addition, since $\sigma_j$ and $\sigma_\tau$ are independent (i.e., in $\mathcal{SG}_{IS}(\mathcal{P},\psi)$),
\begin{equation}
\label{eq:ind_leng_is}
|\pi(y_\tau,x_j)|\geq \max\{|\sigma_\tau|, |\sigma_j|\}~.
\end{equation}
%In addition, $e' \in E_{miss}(\mathcal{P},v_{i_j})$.
This case is further divided into two cases depending on
whether or not the failing edge $e'$ (protected by the detour $D'$) occurs on $\widetilde{D}[d,w]$.
First, consider the case where $e'=(x',y') \in \widetilde{D}[d,w]$, that is, $e'$ occurs on $\widetilde{D}$ \emph{before} the first common vertex. For illustration, see Fig. \ref{fig:lemmafig}(a). By definition, $e' \in \sigma_j$ and by the independence of the subsegments, it holds that
%=\LCA(v_{i_\tau} , t_{\psi} )
$e' \notin \sigma_\tau$. By the ordering, $y_\tau$ occurs on $\psi$ not below $d'$ where $d'$ is the first vertex of the detour $D'$.
We get that
$$|\widetilde{D}[d,w]|\geq |\widetilde{D}[d,y']|\geq |\pi(d,y')|\geq |\pi(y_\tau,x_j)|\geq |\sigma_\tau|~,$$
where the last inequality holds by Eq. (\ref{eq:ind_leng_is}).
Finally, we turn to consider the complementary case where $e' \notin \widetilde{D}[d,w]$. Then there are two $d-w$ shortest paths in $G\setminus \{e'\}$ given by $P_1=\pi(d,d') \circ D'[d',w]$ and $P_2= \widetilde{D}[d,w]$ where $P_1\subseteq P_{v_{i_j},e'}$ and $P_2\subseteq P_{v_{i_\tau},e}$. As $P_1$ is optimal in $G \setminus \{e'\}$,
$$|P_2|\geq |P_1|\geq |\pi(d,d')|\geq |\pi(y_\tau,x_j)|\geq |\sigma_\tau|~,$$
where the last inequality follows again by Eq. (\ref{eq:ind_leng_is}).
\QED
%}%\APPENDGRAD

%\def\APPENDFIGINTERTYPE{
%%%%%%%%%%%%%%%%%%%

%%%%%%%%%%%%%%%%%%%
%}%\APPENDFIGINTERTYPE
%Let $E_{miss}(\mathcal{P})=\{e ~\mid~ P_{v,e} \in \mathcal{P} \mbox{~~and~~} \LastE(P_{v,e})\notin H\}$ be the set of last unprotected edges in $H$ protected by the replacement paths of $\mathcal{P}$.
By the last two lemmas we get:
\begin{lemma}
\label{lem:manyindetor}
$|\bigcup_{P\in\mathcal{P}_{miss}(\mathcal{P},\psi)} V(D(P))|=\Omega(n^{\epsilon} \cdot |E_{miss}(\mathcal{P},\psi)|)$.
\end{lemma}
%We are now ready to complete the proof of Lemma \ref{lem:manyindetor}.
%\Proof [Lemma \ref{lem:manyindetor}]
%By Lemma \ref{lemma:grad_addition},
%$|V(\widehat{G})|=\Omega(n^{\epsilon}\cdot \sum_{\sigma \in \mathcal{SG}_{IS}(\mathcal{P},\psi)}|\sigma|$. By Cl. \ref{obs:is_many}, we get that $|V(\widehat{G})|=\Omega(n^{\epsilon} \cdot %|E_{miss}(\mathcal{P},\psi)|)$. The lemma follows by
% combining with
%Eq. (\ref{eq:final_det_graph}).
%\QED
%We have the following.
\begin{corollary}
\label{lem:emissp}
$|E_{miss}(\mathcal{P})|=O(n^{1-\epsilon}\cdot \log n)$ for every $\mathcal{P} \in \{\mathcal{I}_2, \ldots, \mathcal{P}^{C}_{1}, \ldots, \mathcal{P}^{C}_{\keps}\}$.
\end{corollary}
\Proof
Recall that by Cl. \ref{obs:tree_uncov}, the edges that are not covered by the paths of the tree-decomposition are protected by $H$. Hence, the set of unprotected edges is given
by
\begin{equation}
\label{eq:emiss_all}
E_{miss}(\mathcal{P})=\bigcup_{\psi \in \mathcal{TD}}E_{miss}(\mathcal{P},\psi)~.
\end{equation}
The recursive tree-decomposition algorithm of \cite{BS10} (Procedure Partition therein) consists of $O(\log n)$ levels. Let $\psi_{i_1}, \ldots, \psi_{i_\ell} \in \mathcal{TD}$ be a collection of paths constructed in the same recursion level. Let $v_{i_1}, \ldots, v_{i_\ell}$ be the first vertices of these paths respectively (the path endpoint that is closer to
$s$). It can be shown that the subtrees
$T(v_{i_1}), \ldots, T(v_{i_\ell}) \subseteq T_0$  are vertex disjoint, and hence $e \not\sim e'$ for every $e\in \psi_{i_j}$ and $e'\in \psi_{i_{j'}}$.
By Lemma \ref{lem:manyindetor}, the number of vertices occupied by the detour segments protecting the edges of $E_{miss}(\mathcal{P},\psi_{i_j})$ is $\Omega(n^{\epsilon}\cdot |E_{miss}(\mathcal{P},\psi_{i_j})|)$. Since $\mathcal{P}$ is a $(\sim)$-set, the internal detour segments protecting the edges on $\psi_{i_j}$ and on  $\psi_{i_{j'}}$ are vertex-disjoint.
Note that the definition of independence between detours, refers to empty intersection of
their internal segments (excluding the first and the last vertices). Since the length of each detour is $\Omega(n^{\epsilon})$ (see Obs. \ref{obs:valnesp} and the proof of Lemma \ref{lemma:grad_addition}), this is negligible. Overall, we get that the number of vertices occupied by these detours is bounded by $\Omega(n^{\epsilon} \cdot \sum_{j=1}^k|E_{miss}(\mathcal{P}, \psi_{i_j})|)$.
As the number of vertices is bounded by $n$, we get that $\sum_{j=1}^k|E_{miss}(\mathcal{P}, \psi_{i_j})|=O(n^{1-\epsilon})$. Summing over all $O(\log n)$ recursion levels, and combining with Eq. (\ref{eq:emiss_all}),
$$|E_{miss}(\mathcal{P})|=\sum_{\psi \in \mathcal{TD}}|E_{miss}(\mathcal{P},\psi)|=O(n^{1-\epsilon}\cdot \log n).$$
The lemma follows.
\QED
%}%\APPENDEMISSP
We are now ready to complete the proof of Thm. \ref{thm:square}.
\Proof[Thm. \ref{thm:square}]
%Let $\mathcal{E}_{\New}=\{e ~\mid~ \exists v \mbox{~such that~}\langle v,e \rangle \in \UncoverPairs\}$ be the total set of $T_0$ edges that are unprotected by $T_0$ (i.e., that has some new-ending $P_{v,e}$ paths).
Recall that $\UncoverPairs=\mathcal{I}_1 \cup \bigcup_{i=0}^{\keps}\mathcal{P}^C_i$.
By Lemma \ref{lem:i1_ana}, the uncovered pairs in $H$ (i.e., pairs that correspond to $H$-new-ending paths) are in $\bigcup_{i=0}^{\keps}\mathcal{P}^C_i$.
Finally, using Cor. \ref{lem:emissp} and summing over all $O(1/\epsilon)$ $(\sim)$-sets $\mathcal{P}$
yields
Lemma \ref{lem:newh_final}, which bounds the number of unprotected edges (that need to be reinforced) as in the theorem. Lemma \ref{lem:sizeh} bounds the size of $H$, hence the number $b(n)$ of backup edges. 
Consequently the theorem follows.
\QED

% are established.

%
%
%
%\paragraph{Upper Bound}
%For every source $s \in S$, add $O(n \log n)$ edges to protect against paths $P$ with nonempty $I^{\not\sim}(P)$.
%
%In addition, for every vertex $v$ and source $s \in S$ add $O(\log n \cdot (n/\NSource)^{\epsilon})$ edges.
%
%Our goal is to show that there are at most $O(\sigma^{\epsilon} \cdot n^{1-\epsilon})$ unprotected edges in $H$.
%To see this, we show that for at least half of the sources, $S' \subseteq S$, $|S'|\geq \sigma/2$, it holds that
%there are at most $2(n/\NSource)^{1-\epsilon}$ unprotected edges in the BFS trees.
%
%By the same reasoning as for the single source, if we assume otherwise, every source with at least $y=2(n/\NSource)^{1-\epsilon}$, requires at $(n/\NSource)^{\epsilon}\cdot y=n/\sigma$ vertices.
%
%
\section{Lower Bound}
In this section, we establish lower bounds on the size of the $\epsilon$ \FTBFS\ structures. These bounds match the upper bound of Sec. \ref{sec:uni_upb} up to logarithmic factors in both the number of reinforced edges and the size of the construct. These lower bound constructions are generalizations of \cite{PPFTBFS13}. We first consider  the single source case. Note that for $\epsilon \in [1/2,1]$, by the lower bound \FTBFS\ in \cite{PPFTBFS13} $\Omega(n^{3/2})$ edges are required. Hence, it remains to establish the lower bound for $\epsilon \in (0,1/2)$.
\begin{theorem}
\label{thm:lowerbound_edgeonef}
For every $\epsilon \in (0,1/2)$,
there exists an $n$-vertex graph $G(V, E)$ and a source node $s \in V$ such that any $\epsilon$ \FTBFS\ tree rooted at $s$ with at most $\lfloor n^{1-\epsilon}/6 \rfloor$ reinforced edges has $\Omega(n^{1+\epsilon})$ edges. In other words, there exists a graph for which any $(b(n),r(n))$ \FTBFS\ structure for $r(n)=\Omega(n^{1-\epsilon})$ requires $\Omega(\min\{n^{1+\epsilon}, n^{3/2}\})$ backup edges.
\end{theorem}
\Proof
Let us first describe the structure of
$G=(V,E)$. Set $d_{\epsilon}=\lfloor n^{\epsilon}/4 \rfloor$ and $k_{\epsilon}=\lfloor n^{1-2\epsilon} \rfloor$.
We first describe the structure of a subgraph $G_{\epsilon}$ which provides the basic building block of the construction.
In particular, the final graph $G$ consists of $k_{\epsilon}$ copies of the graph $G_{\epsilon}$ denoted by $G_{\epsilon,1}, \ldots, G_{\epsilon, k_{\epsilon}}$ that are connected to the source vertex $s$ as will be described later.

We begin by describing the structure of the $i$th copy $G_{\epsilon,i}$ (the copies are  identical), which consists of four main components.
The first is a path $$\pi_i=[s_i=v^i_1, \ldots, v^{i}_{d_{\epsilon}+1}=v^*_i]$$ of length $d_\epsilon$.
The second component consists of a node set $Z_i=\{z^i_1,\ldots,z^i_{d_\epsilon}\}$
and a collection of $d_{\epsilon}$ disjoint paths of deceasing length, $P^i_1, \ldots, P^i_{d_{\epsilon}}$,
where $P^i_j=[v^i_j=p^{i}_{j,1}, \ldots, p^{i}_{j,t_j}=z^i_j]$ connects $v^i_j$ on $\pi_i$ with $z^i_j$ and its length is $t_j=|P^i_j|=6+2(d_{\epsilon}-j)$, for every $j \in \{1,\ldots,d_{\epsilon}\}$.
Altogether, the set of nodes in these paths, \\ $Q_i=\bigcup_{j=1}^{d_{\epsilon}} V(P^i_j)$, is of size $|Q_i| = d_{\epsilon}^2+5d_{\epsilon}=\Theta(n^{2\epsilon})$.
The third component is a set of nodes $X_i$ of size $n-1-k_{\epsilon}(|\pi_i|+|Q_i|+|Z_i|)=\Theta(n^{2\epsilon})$,
%$c \cdot n$, for some constant $c$ to be determined later.
all connected to the terminal node $v^*_i$.
The last component is a \emph{complete} bipartite graph $B_i=(X_i, Z_i)$ connecting every vertex in $X_i$ to every vertex in $Z_i$.

So far, $V(G_{\epsilon,i})=V(\pi_i) \cup Z_i \cup X_i \cup Q_i$ and $$E(G_{\epsilon,i})= E(\pi_i) \cup \{(v^*_i, x^i_t) ~\mid~ x^i_t \in X_i\} \cup E(B_i) \cup \bigcup_{j=1}^{d_{\epsilon}} E(P^i_j).$$

\begin{figure}
\begin{center}
\includegraphics[scale=0.3]{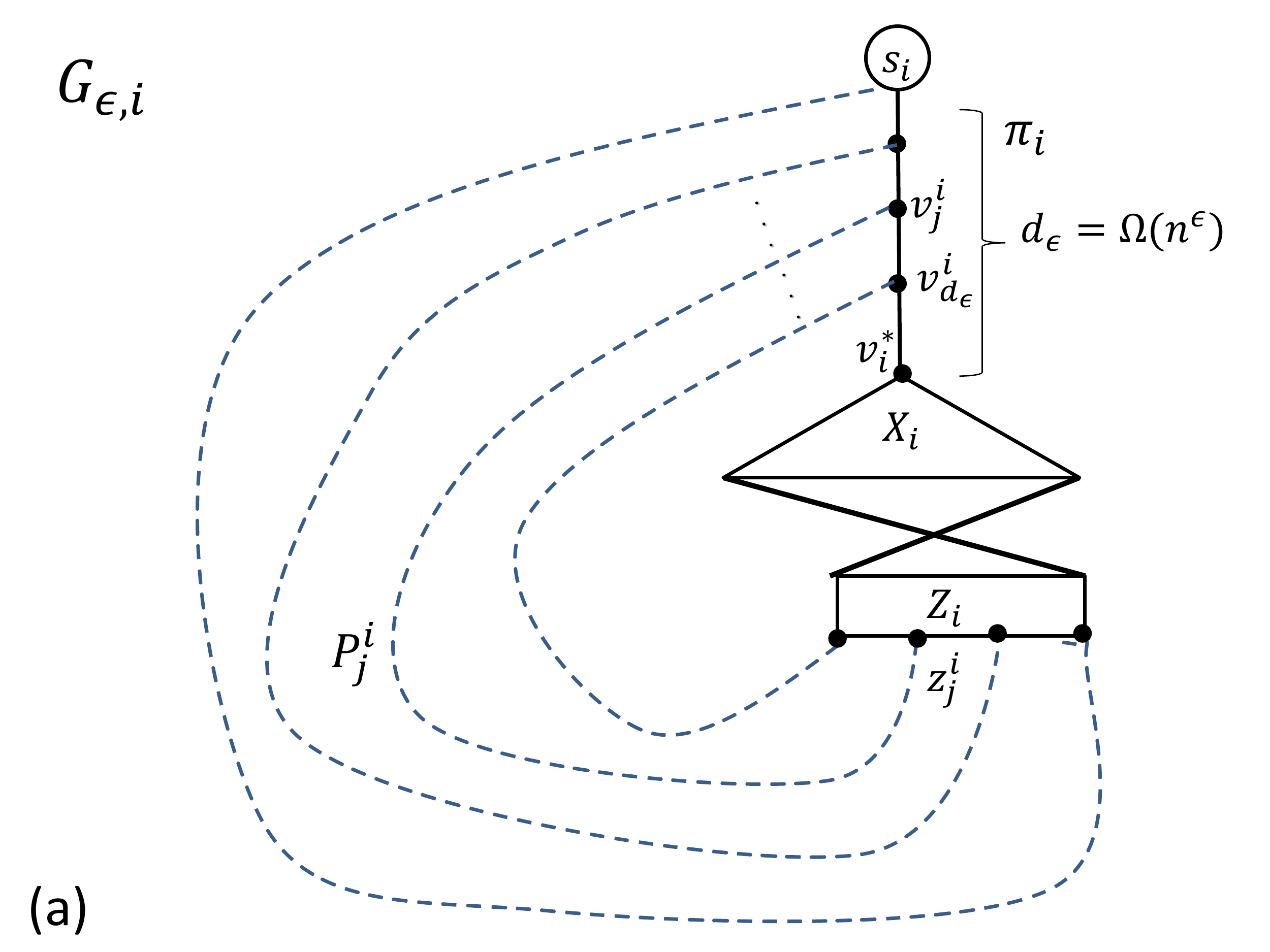}
\hfill
\includegraphics[scale=0.3]{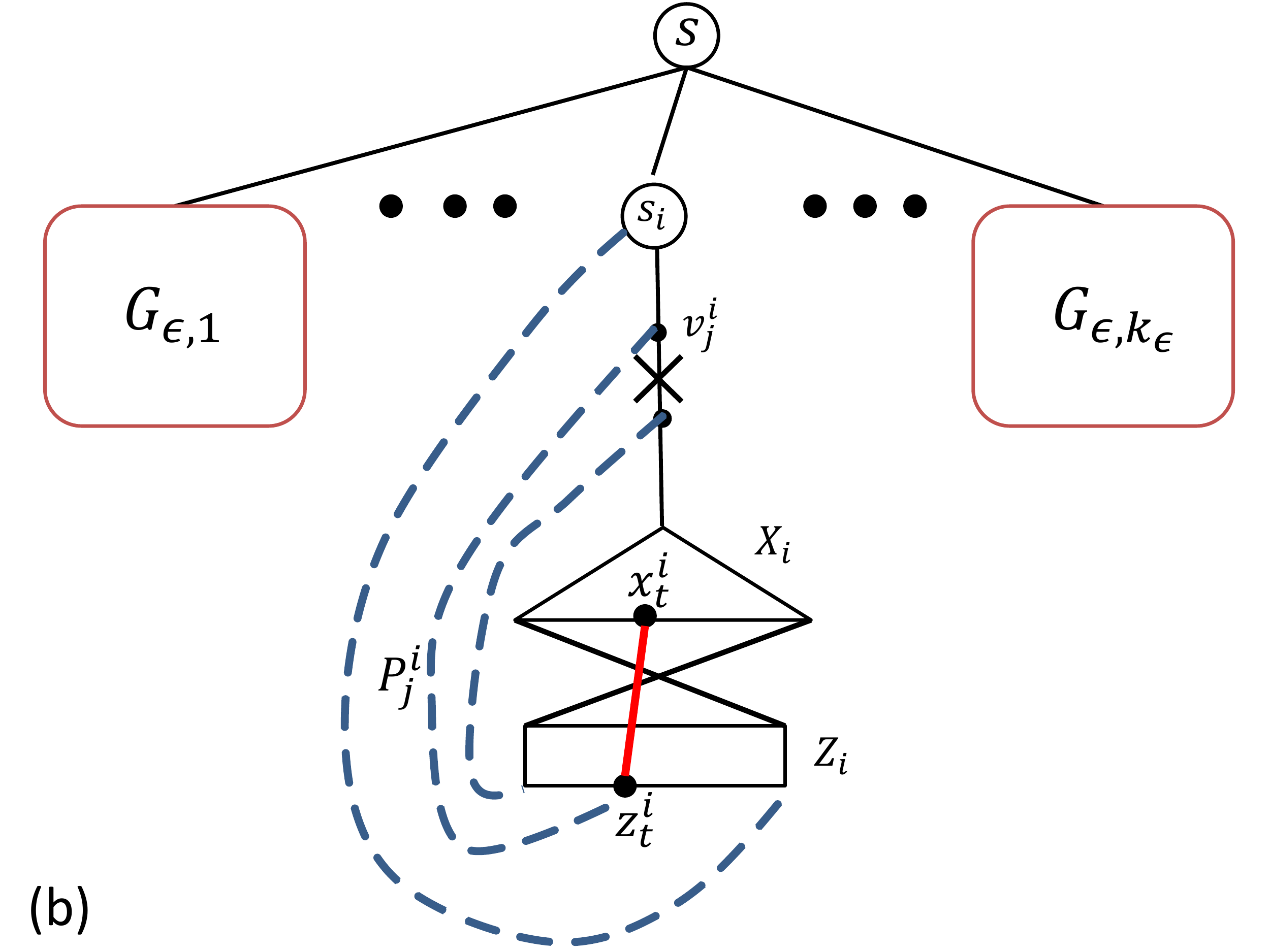}
\end{center}
\caption{\sf Schematic description of the lower bound construction for $\epsilon$ \FTBFS.
(a) The graph $G_{\epsilon_i}$. The dashed lines corresponds to the collection monotone increasing lengths of paths $P^i_j$. The set of vertices $X_i$ contains $\Omega(n^{2\epsilon})$ vertices that is fully connected to the collection of $\Omega(n^{\epsilon})$ vertices $Z_i$.
(b) The graph $G=(V,E)$ consists of $k_{\epsilon}=\Omega(n^{1-2\epsilon})$ copies of the graph $G_{\epsilon,i}$ connected to the source vertex $s$.
The bold dashed red edge $(x^i_t, z^i_j)$ is required
upon failure of the edge $e^i_j$.}
\label{fig:lowerbound1f}
\end{figure}

Finally, $s$ is connected via a star to the first vertex $s_i$ of each path $\pi_i$ for every $i \in \{1, \ldots, k_{\epsilon}\}$, see Fig. \ref{fig:lowerbound1f}(b) for illustration.

Overall, $V(G)=\{s\} \cup \bigcup_{i=1}^{k_{\epsilon}} V(G_{\epsilon,i})$ and
$E(G)=E(s) \cup \bigcup_{i=1}^{k_{\epsilon}} V(G_{\epsilon,i}),$
where $E(s)=\{ (s,s_i) ~\mid~ i \in \{1, \ldots, k_{\epsilon}\}\}$.

A BFS tree $T_0$ rooted at $s$ for this $G$
%(illustrated by the solid edges in the figure)
%(illustrated in Fig. \ref{fig:lowerbound1f})
is given by
\vspace{-6pt}
\begin{eqnarray*}
E(T_0) &=& E(s) \cup \bigcup_{i=1}^{k_{\epsilon}} \left(\pi_i \cup \bigcup_{x \in X_i}(v^*_i, x) \cup \bigcup_{t=1}^{k_{\epsilon}}(x^i_t, z^i_t) \right.
\\&&
\left. \cup
\bigcup_{j=1}^{d_{\epsilon}} E(P^i_j) \setminus \{(p^{i}_{j,\ell_j}, p^i_{j,\ell_j-1})\}\right),
\end{eqnarray*}
where $\ell_j=t_j-(d_{\epsilon}-j)$ for every $j \in \{1, \ldots, d_{\epsilon}\}$.

%\vspace{-2pt}
%For an illustration of the construction, see Fig. \ref{fig:lowerbound1f}.
Let $B=\bigcup_{i=1}^{k_{\epsilon}}B_i$ be the collection of edges on the $k_{\epsilon}$ complete bipartite graphs $B_i$ and let
$\Pi=\bigcup_{i=1}^{k_{\epsilon}}\pi_i$.
\begin{observation}
\label{obs:lb_eps_ss_size}
\begin{description}
\item{(a)}
$|V(G_{\epsilon,i})|=n/k_{\epsilon}-1$ and hence $|V(G)|=n$.
\item{(b)}
$|E(G)|=\Omega(|EB|)=\Omega(n^{1+\epsilon})$.
\item{(c)}
$|E(\Pi)|=k_{\epsilon}\cdot d_{\epsilon}\geq \lfloor n^{1-\epsilon}/5 \rfloor$.
\end{description}
\end{observation}
We now show that every $\epsilon$ \FTBFS\ structure $H$ must
contain a constant fraction of the edges in  $B=\bigcup_{i=1}^{k_{\epsilon}}B_i$, namely, the edges $e^i_{t,j}=(x^i_t, z^i_j)$ (the thick edges in Fig. \ref{fig:lowerbound1f}(b)).

%Let $\Pi=\bigcup_{i=1}^{k_{\epsilon}}\pi_i$.
Note that the edges of $\Pi$ are ``costly" in the sense that to protecting against their failure, many edges connecting $X_i$ and $Z_i$ should be introduced into the fault-tolerant structure $H$. To provide a succinct structure, one can choose to fortify these edges, however, since $|E(\Pi)|=\Omega(n^{1-\epsilon})$, upon setting the fortification budget as in the statement of Thm. \ref{thm:lowerbound_edgeonef}, a constant fraction of these edges could not be reinforced, resulting eventually in a dense structure. We now formalize this intuition.

For ease of analysis, consider a partition the edges of $B$ into $d_{\epsilon} \cdot k_{\epsilon}$ disjoint subsets corresponding to the number of edges in $\Pi$. Let $e^i_j=(v^i_j,v^i_{j+1}) \in \pi_i$ be the $j$'th edge on the path $\pi_i \subseteq G_{\epsilon,i}$ and define $E^i_j=\{(x^i_t,z^i_j) ~\mid~ x^i_t \in X_i\}$. Observe that $E^i_j$ and $E^{i'}_{j'}$ are disjoint for every pair $(i,j)\neq (i',j')$ for every $i \in \{1, \ldots, k_{\epsilon}\}$ and every $j \in \{1, \ldots, d_{\epsilon}\}$ and in addition, $B=\bigcup_{i=1}^{k_{\epsilon}} \bigcup_{j=1}^{d_{\epsilon}} E^i_j$.

For every $\epsilon$ \FTBFS\ structure $H \subseteq G$, let $E'(H)$ be the set of at most $\lfloor n^{1-\epsilon}/6 \rfloor= O(n^{\epsilon})$ reinforced edges in $H$ (which is the maximum allowed number of reinforced edges by the statement of the theorem).
\begin{claim}
\label{cl:lb_edgeimp}
$E^{i}_{j} \subseteq H$ for every $e^i_j \in \Pi\setminus E'(H)$.
\end{claim}
\Proof
Assume, towards contradiction, that $H$ does not contain $e'=(x^i_t, z^i_j) \in E^{i}_{j}$ for some $x^i_t \in X_i$
(the bold dashed edge $(x^i_t, z^i_j)$ in Fig. \ref{fig:lowerbound1f}(b)).
Note that upon failure of the edge $e^i_j=(v^i_j,v^i_{j+1}) \in \pi_i$,
the unique $s-x^i_t$ shortest path connecting $s$ and $x^i_t$ in
$G \setminus \{e^i_j\}$ is $P'_{i,j}= \pi[s,v^i_j] \circ P^i_{j}\circ e'$,
and all other alternatives are strictly longer.
% the following are not the only ones...
%
%the paths connecting $s$ and $x_i$ are given by
%$P'_1=P_1 \circ (z_1, x_i)$,
%$P'_2=\pi[v_1,v_2] \circ P_2 \circ (z_2, x_i)$,
%$\ldots, P'_{j}= \pi[v_1,v_j] \circ P_{j}\circ [z_{j}, x_i]$.
%Since $|P_1| > |P_2| \ldots >|P_{j}|$, it also holds that
%$|P'_1| > |P'_2| \ldots >|P'_{j}|$. We get that $P'_j$ is the unique
%$s-x_i$ shortest-path in $G \setminus \{e_j\}$.
Since $e' \notin H$, also $P'_{i,j} \nsubseteq H$, and therefore
$\dist(s, x^i_t, G \setminus \{e^i_j\})< \dist(s, x^i_t, H \setminus \{e^i_j\})$,
in contradiction to the fact that $H$ is an $\epsilon$ \FTBFS\ structure. It follows $E^i_j \subseteq H$.
\QED
We are now ready to complete of Theorem
\ref{thm:lowerbound_edgeonef}.
%
%\Proof[Theorem \ref{thm:lowerbound_edgeonef}]
By Obs. \ref{obs:lb_eps_ss_size}(c), $|E(\Pi)|\geq \lfloor n^{1-\epsilon}/5 \rfloor$. By the statement of
Thm. \ref{thm:lowerbound_edgeonef}, the $\epsilon$ \FTBFS\ structure contains at most $\lfloor n^{\epsilon}/6 \rfloor$ reinforced edges $E'(H)$. Hence, even if all those edges are taken from the edges of $\Pi$ (i.e., $E'(H)\subseteq E(\Pi)$), there are still $|E(\Pi)\setminus E'(H)|=\Omega(n^{1-\epsilon})$ edges in $\Pi$ that remain unenforced.
By Cl. \ref{cl:lb_edgeimp}, each such edge requires that $E^{i}_{j}$ would be added to $H$.
Note that $|E^i_j|=|X_i|=\Omega(n^{2\epsilon})$.
By the disjointness of the $E^i_j$ sets, $|E(H)|\geq \Omega(n^{2\epsilon}) \cdot |\Pi \setminus E'(H)|=\Omega(n^{1+\epsilon})$, as required.
\QED
%}%\APPENDLBsingle
%-------------------------
%\def\APPENDFIGLB{
%}%\APPENDFIGLB
\paragraph{Multiple Sources.}
In this subsection, we consider an intermediate setting where it is necessary to construct an $\epsilon$ fault tolerant subgraph containing several $\epsilon$ \FTBFS\ trees in parallel,
one for each source  $s \in S$, for some $S\subseteq V$.

For a given subset of $\NSource$ sources $S \subseteq V$, a subgraph $H \subseteq G$ is an $\epsilon$ \FTMBFS\ structure if there exists a subset $E' \subseteq E$ of $O(\NSource^{\epsilon} \cdot n^{1-\epsilon})$ edges such that $\dist(s,v, H \setminus \{e\})=\dist(s,v, G \setminus \{e\})$ for every $s, v \in S \times V$ and every $e \in E\setminus E'$.
Towards the end of this section, we show the following.
\begin{theorem}
\label{thm:lowerbound_f_multisource}
For every real $\epsilon \in (0,1/2]$ and $\NSource \in \{1,\ldots, n\}$, there exists a graph $G=(V,E)$, a subset of $\NSource$ sources $S \subseteq V$, such that any $\epsilon$ \FTMBFS\ structure with at most $\lfloor (\NSource^{\epsilon}\cdot n^{1-\epsilon})/6 \rfloor$ reinforced edges, contains
$\Omega(\NSource^{1-\epsilon} \cdot n^{1+\epsilon})$ edges.
\end{theorem}

Set $d_{\epsilon,\NSource}=\lfloor (n/4\NSource)^{\epsilon} \rfloor$ and $k_{\epsilon,\NSource}=\lfloor (n/\NSource)^{1-2\epsilon} \rfloor$.

We first describe the structure of a subgraph $G_{\epsilon,\NSource}$ which is a subgraph of $G_{\epsilon}$ in the construction for the single source case.
The final constructs uses $\NSource \cdot k_{\epsilon,\NSource}$ copies of this subgraph: $k_{\epsilon,\NSource}$ copies per source $s \in S$.
Consider now the $(i,j)'$th copy $G^{i,j}_{\epsilon,\NSource}$ for $i \in \{1, \ldots, \NSource\}$ and $j \in \{1, \ldots, k_{\epsilon,\NSource}\}$ (all copies are identical).
This subgraph consists of three main components. The first is a path $\pi_{i,j}=[s_{i,j}=v^{i,j}_1, \ldots, v^{i,j}_{d_{\epsilon,\NSource}+1}=v^*_{i,j}]$ of length $d_{\epsilon,\NSource}$.
The second component consists of a node set $Z_{i,j}=\{z^{i,j}_1,\ldots,z^{i,j}_{d_\epsilon,\NSource}\}$
and a collection of $d_{\epsilon,\NSource}$ disjoint paths of deceasing length, $P^{i,j}_1, \ldots, P^{i,j}_{d_{\epsilon,\NSource}}$,
where $P^{i,j}_\ell=[v^{i,j}_\ell=p^{i,j}_{\ell,1}, \ldots, p^{i,j}_{\ell,t_\ell}=z^{i,j}_\ell]$ connects $v^{i,j}_\ell$ on $\pi_{i,j}$ with $z^{i,j}_\ell$ and its length is $t_\ell=|P^{i,j}_\ell|=6+2(d_{\epsilon,\NSource}-\ell)$, for every $\ell \in \{1,\ldots,d_{\epsilon,\NSource}\}$.

Altogether, the set of nodes in these paths, \\ $Q_{i,j}=\bigcup_{\ell=1}^{d_{\epsilon,\NSource}} V(P^{i,j}_\ell)$, is of size $|Q_{i,j}| = d_{\epsilon,\NSource}^2+5d_{\epsilon,\NSource}=\Theta((n/\NSource)^{2\epsilon})$.
This completes the description of the $G^{i,j}_{\epsilon,\NSource}$ subgraph.

For every $j \in \{1, \ldots, k_{\epsilon,\NSource}\}$, let $X_j$ be a set of $\Omega(n/k_{\epsilon,\NSource})$ vertices.
The $X_j$ vertices are connected via a star to a vertex $\widetilde{v}_j$. The latter is connected to the terminal node $v^*_{i,j} \in \pi_{i,j}$ for every $i \in \{1, \ldots, \NSource\}$.
Formally, these edges are defined by
$\widehat{E}(X_j)=\{(\widetilde{v}_j,x) ~\mid~ x \in X_j\} \cup \{( \widetilde{v}_j, v^*_{i,j}) ~\mid~ i \in \{1, \ldots, \NSource\}\}$.
In addition, the $X_j$ vertices are fully connected to each of the vertices in $Z_j=\bigcup_{i=1}^{\NSource} Z_{i,j}$.
Formally, $G$ contains $k_{\epsilon,\NSource}$ complete bipartite graphs $B_j=B(X_j, Z_j)$ for every $j \in \{1, \ldots, k_{\epsilon,\NSource}\}$.
Finally, every source vertex $s_i \in S$ is connected to $s_{i,j}$ for every $j \in \{1, \ldots, k_{\epsilon,\NSource}\}$.
Overall, the vertices of $G=(V,E)$ are given by
$$V(G)=\bigcup_{i=1}^{\NSource}\bigcup_{j=1}^{k_{\epsilon,\NSource}} V(G^{i,j}_{\epsilon,\NSource}) \cup \bigcup_{j=1}^{k_{\epsilon,\NSource}} \left(X_j \cup \{\widetilde{v}_j\} \right) \cup S$$
and the edges are
$$E(G)=\bigcup_{i=1}^{\NSource}\bigcup_{j=1}^{k_{\epsilon,\NSource}} \left( E(G^{i,j}_{\epsilon,\NSource}) \cup \{(s_i,s_{i,j})\} \right)\cup \bigcup_{j=1}^{k_{\epsilon,\NSource}} \left(E(B_j) \cup \widehat{E}(X_j) \right).$$
%See Fig. \ref{XX} for illustration.
\begin{observation}
\label{obs:lb_multi_source}
\begin{description}
\item{(a)}
$|V(G^{i,j}_{\epsilon,\NSource})|=\Theta((n/\NSource)^{2\epsilon})$.
\item{(b)}
The constants can be set precisely, so that $|V(G)|=n$ and $|X_j|=\Theta(n/k_{\epsilon,\NSource})$.
\item{(c)}
$|E(G)|\geq |EB|=\Omega(\NSource^{1-\epsilon} \cdot n^{1+\epsilon})$.
\end{description}
\end{observation}
Let $\Pi=\bigcup_{i=1}^{\NSource}\bigcup_{j=1}^{k_{\epsilon,\sigma}}E(\pi_{i,j})$.
Since each $\pi_{i,j}$ is of length $d_{\epsilon,\sigma}$, overall \begin{equation}
\label{eq:pilb_ms}
|E(\Pi)|\geq \lfloor \NSource^{\epsilon} \cdot n^{1-\epsilon} /5\rfloor~.
\end{equation}
Recall that
$e^{i,j}_{\ell}=(v^{i,j}_{\ell},v^{i,j}_{\ell+1})$
is the $\ell$'th edge on $\pi_{i,j}$.

The reasoning goes in a very similar way to the single source case. In particular, we show that the edges of $\Pi$ are costly in the sense that unless they are reinforced in $H$, they require the introduction of many edges into $H$.

For ease of analysis, the edges of the bipartite graphs $B=\bigcup_{j=1}^{k_{\epsilon, \NSource}}B_j$ are partitioned into $|\Pi|$ disjoint sets $E^{i,j}_{\ell}$, each corresponds to one particular  edge $e^{i,j}_{\ell} \in \Pi$.
Let $E^{i,j}_{\ell}=\{(x^{j}_t, z^{i,j}_{\ell}) ~\mid~x^{j}_t \in X_j\}$, hence $|E^{i,j}_{\ell}|=|X_j|=\Omega(n/k_{\epsilon,\NSource})$ and $EB=\bigcup_{i=1}^{\NSource}\bigcup_{j=1}^{k_{\epsilon,\sigma}}\bigcup_{\ell=1}^{d_{\epsilon,\sigma}}E^{i,j}_{\ell}$.
For every $\epsilon$ \FTMBFS\ structure $H \subseteq G$, let $E'(H)$ be the set of at most $\lfloor \NSource^{\epsilon} \cdot n^{1-\epsilon}/6 \rfloor= O(\NSource^{\epsilon} \cdot n^{1-\epsilon})$ reinforced edges in $H$ (which is the maximum allowed number of reinforced edges by the statement of the theorem).
\begin{claim}
\label{cl:lb_edgeimp_multi}
For every $e^{i,j}_{\ell} \in \Pi\setminus E'(H)$, it holds that $E^{i,j}_{\ell} \subseteq H$.
\end{claim}
\Proof
Assume, towards contradiction, that $H$ does not contain $e'=(x^j_t, z^{i,j}_{\ell}) \in E^{i,j}_{\ell}$ for some $x^j_t \in X_j$.
Note that upon the failure of the edge $e^{i,j}_{\ell}=(v^{i,j}_\ell,v^{i,j}_{\ell+1}) \in \pi_{i,j}$,
the unique $s_i-x^j_t$ shortest path connecting $s_i$ and $x^j_t$ in $G \setminus \{e^{i,j}_{\ell}\}$ is $P'_{i,j,\ell}= \pi[s_i,v^{i,j}_\ell] \circ P^{i,j}_{\ell}\circ e'$,
and all other alternatives are strictly longer.
Note that the paths from the other sources $s_{i'} \in S \setminus \{s_i\}$ are not helpful as well.
% the following are not the only ones...
%
%the paths connecting $s$ and $x_i$ are given by
%$P'_1=P_1 \circ (z_1, x_i)$,
%$P'_2=\pi[v_1,v_2] \circ P_2 \circ (z_2, x_i)$,
%$\ldots, P'_{j}= \pi[v_1,v_j] \circ P_{j}\circ [z_{j}, x_i]$.
%Since $|P_1| > |P_2| \ldots >|P_{j}|$, it also holds that
%$|P'_1| > |P'_2| \ldots >|P'_{j}|$. We get that $P'_j$ is the unique
%$s-x_i$ shortest-path in $G \setminus \{e_j\}$.
Since $e' \notin H$, also $P'_{i,j,\ell} \nsubseteq H$, and therefore
$\dist(s_i, x^j_t, G \setminus \{e^{i,j}_{\ell}\})< \dist(s_i, x^j_t, H \setminus \{e^{i,j}_{\ell}\})$,
in contradiction to the fact that $H$ is an $\epsilon$ \FTMBFS\ structure. It follows $E^{i,j}_{\ell} \subseteq H$.
\QED
We are now ready to complete of Theorem
\ref{thm:lowerbound_f_multisource}.
%
%\Proof[Theorem \ref{thm:lowerbound_edgeonef}]
Consider any $\epsilon$ \FTMBFS\ structure $H$.
By Eq. (\ref{eq:pilb_ms}) and by the bound on the number of reinforced edges as stated in
Thm. \ref{thm:lowerbound_f_multisource}, even if all reinforced edges $E'(H)$ are taken from the edges of $\Pi$ (i.e., $E'(H)\subseteq E(\Pi)$), there are still $|E(\Pi)\setminus E'(H)|=\Omega(\NSource^{\epsilon}\cdot n^{1-\epsilon})$ edges in $\Pi$ that remain unreinforced.
By Cl. \ref{cl:lb_edgeimp_multi}, each such edge $e^{i,j}_{\ell}$ requires that a subset of $\Omega(n/k_{\epsilon,\NSource})$ edges $E^{i,j}_{\ell}$ to be included in $H$.
It then holds that $|E(H)|\geq \Omega(n/k_{\epsilon,\NSource}) \cdot |\Pi \setminus E'(H)|=\Omega(\NSource^{1-\epsilon} \cdot n^{1+\epsilon})$ as required. The theorem follows.
\QED

{\small
\bibliographystyle{abbrv}

} %\small

\end{document}